\documentclass[opre, nonblindrev]{informs3}

\OneAndAHalfSpacedXI 

\usepackage{endnotes}
\let\footnote=\endnote

\usepackage{amsmath}
\usepackage{booktabs}
\allowdisplaybreaks
\usepackage{dsfont, pifont, mathrsfs}
\usepackage{multirow}
\usepackage{setspace}
\usepackage[hidelinks, colorlinks=true, citecolor=blue]{hyperref}
\usepackage[ruled,vlined,linesnumbered]{algorithm2e}

\usepackage{natbib}
 \bibpunct[, ]{(}{)}{,}{a}{}{,}%
 \def\bibfont{\small}%
\TheoremsNumberedThrough     
\ECRepeatTheorems

\EquationsNumberedThrough    

\begin{document}

\RUNAUTHOR{Lin and Warren (2026)}
\RUNTITLE{Spatial Modeling in Multilevel Areal Data}

\TITLE{On the Need for Spatial Random Effects in Bayesian Regression Models for Multilevel Areal Data}

\ARTICLEAUTHORS{%
\AUTHOR{Shuqi Lin\textsuperscript{$1$} \quad\quad Joshua L. Warren\textsuperscript{$\ast, 1$}}
\AFF{\textsuperscript{1}Department of Biostatistics, Yale School of Public Health, New Haven, CT, USA\texorpdfstring{\\}{}%
\EMAIL{$^\ast$\texttt{joshua.warren@yale.edu}}}
} 

\ABSTRACT{%
Although spatial models for areal data are widely used in multilevel settings, the conditions under which spatial and nonspatial random effects yield similar inference for regression coefficients are not well understood. We address this question within a hierarchical Bayesian regression framework and show that the answer is primarily governed by the number of observations per geographic area. We show that the relative difference in the variances of the full conditional distributions vanishes at a rate inversely proportional to within-area sample size, and that the difference in the means is of smaller order than their standard deviations. This gives a closed-form replication threshold below which spatial modelling materially affects inference and above which a nonspatial model is effectively equivalent. The threshold depends on the spatial correlation, the ratio of between-area to within-area variance, and the alignment between the covariate and spatial patterns in the data. When the covariate does not vary within areas, spatial modelling remains necessary regardless of replication. The threshold can inform study design from pilot data or prior knowledge, and we propose a two-stage procedure to extend its use to analysis. Simulations confirm the threshold, and an analysis of Minnesota radon data shows that the need for spatial random effects can differ between coefficients in the same model.
}%


\maketitle

\section{Introduction}\label{sec:introduction}
Spatially referenced data arise routinely across the health, environmental, and social sciences \citep{banerjee2003hierarchical, cressie2015statistics}. Observations from nearby locations are often correlated due to shared environmental conditions, social interactions, or other unmeasured factors. Ignoring spatial dependence can lead to underestimated measures of uncertainty (e.g., standard errors, posterior standard deviations), and therefore, inflated type I error rates \citep{legendre1993spatial, dormann2007methods}. 

This has motivated a large body of methods development work on statistical modeling that explicitly incorporates spatial dependence structure \citep{banerjee2003hierarchical, cressie2015statistics}. Typically, these models introduce location-specific random effects with a correlation structure defined broadly by distances between the spatial units, where the meaning of distance differs based on the type of spatial data (e.g., point-referenced, areal/lattice). This structure allows for information sharing across spatially proximate random effects, leading to more stable local estimates that would otherwise be too noisy to support reliable inference on their own. As a result, spatial modeling has become a default analytical choice for spatially referenced data.

The majority of spatial models for areal data were developed in settings where each region contributes a single observation, as in disease mapping studies where region-level disease counts are the primary outcome \citep{besag1991bayesian, leroux2000estimation}. However, when working with multilevel data, multiple observations are nested within the same spatial region and the model is introduced at the level of the individual. Random effects are then included primarily to account for clustering among individuals from the same region. In this setting, it is not clear what role a spatially structured random effect plays relative to a simpler nonspatial random effect. 

In the absence of formal guidance, current practice has been somewhat inconsistent. Several multilevel studies have incorporated spatially structured area-level effects using conditional autoregressive (CAR) random effects \citep{dong2016spatial,clark2021community,asare2022spatiotemporal, bravo2022air,djeudeu2022multilevel,cars2025environmental,li2025association}, and alternative regularization strategies have also been considered \citep{choi2022simultaneous}, while others analyzing data with similar structure have used independent and identically distributed (iid) random effects \citep{wu2020air,barry2022ability,broen2023burkitt,woodward2023combining,burridge2025association}. In the case of \citet{wu2020air}, county-level latitude and longitude were additionally included as fixed-effect predictors to evaluate whether residual spatial correlation may have influenced the findings, acknowledging that unmodeled spatial structure could be a concern in this setting. In none of these cases was the choice formally justified, and to our knowledge no general guidance exists to inform this decision.

Despite these inconsistencies, a few prior studies have suggested that spatial structure becomes less consequential for accurate statistical inference in multilevel settings as within-area sample size increases. \citet{savitz20095} noted that spatial borrowing is most beneficial when local sample sizes are small and that estimators based on no pooling, iid random effects, and spatially structured random effects converge as within-area sample size grows. \citet{xu2014comparing} and \citet{dasgupta2014comparing} empirically observed that spatial and nonspatial models produced similar fixed effect estimates in individual-level datasets, though neither identified the sample size as the mechanism driving convergence nor attempted a theoretical comparison. While these works observed or remarked upon the convergence phenomenon, a formal assessment remains absent.

A related but distinct line of work concerns spatial confounding, in which the inclusion of spatial random effects can affect inference on regression coefficients \citep{ReichHodgesZadnik2006, KhanCalder2022, ZimmermanVerHoef2022, khan2026rethinking}. Both literatures are concerned with how spatial random effects influence coefficient estimation, but the underlying mechanisms are fundamentally different. Spatial confounding is used to describe collinearity between covariates and spatial random effects in the fitted model and confounding by unmeasured spatial variables associated with the covariates and outcome \citep{khan2026rethinking}. Spectral and related approaches characterize how spatial scale, covariate structure, and smoothing induce coefficient bias \citep{paciorek2010importance, nobre2021effects, guan2023spectral, dupont2026demystifying} and establish conditions for consistent estimation \citep{gilbert2025consistency, datta2026infill}. Our work addresses a separate question entirely: assuming a correctly specified model in which spatial random effects represent residual dependence and are independent of the covariates, how much within-area replication is needed before spatial and nonspatial models yield equivalent inference on regression coefficients?

We formally investigate this question in Bayesian multilevel models for areal data. Using the Leroux CAR distribution as a representative spatial prior \citep{leroux2000estimation}, we make three contributions. First, we show that sufficient within-area replication makes inference under spatial and independent random effects models negligibly different. Specifically, for the full conditional distributions of the regression coefficient (i.e., the distributions needed to implement a Markov chain Monte Carlo (MCMC) posterior sampling algorithm), the absolute difference in means and the absolute relative difference in variances are $O_p\left(m^{-1}\right)$ and $O\left(m^{-1}\right)$ in the within-area sample size $m$, respectively. Simulations confirm that these results extend to the marginal posterior distributions of primary interest. Second, we derive a closed-form threshold giving the minimum within-area sample size required, for a prespecified tolerance, before the two models yield effectively equivalent inference, as a function of interpretable quantities including the spatial correlation parameter, the variance ratio, and properties of the covariate. Third, we develop a practical procedure for applying this criterion when variance parameters are unknown, and validate it via simulation. We consider balanced and unbalanced designs with one or more covariates, and further investigate non-Gaussian outcomes through simulation. An analysis of Minnesota radon data illustrates how the need for spatial modelling can differ between coefficients in the same model.  Proofs, computational details, and additional simulation results are provided in the appendices. The accompanying R package \texttt{SpThreshold} is available at \url{https://github.com/ShuqiLinn/SpThreshold}.

\section{Multilevel model for areal spatial data} \label{sec:model}
We consider a linear regression model for multilevel areal spatial data. Our primary objective is accurate statistical inference for the regression parameter $\beta_1$ defined in~\eqref{eq:model}, learning the relationship between the predictor and the outcome while properly adjusting for correlation in the data. Specifically, let $Y_{ij}$ denote the response for the $j^{\text{th}}$ observation within spatial unit $i$, where $i = 1, \ldots, n$ and $j = 1, \ldots, m$. We present the model with balanced replication and a single covariate; the results extend to unequal within-area sample sizes and to multiple covariates (Appendix~\ref{sec:supp_additional_settings}). The model is given as \begin{equation} Y_{ij} = \beta_0 + \beta_1 \text{x}_{ij} + \theta_i + \varepsilon_{ij}, \quad \varepsilon_{ij} \mid \sigma^2 \overset{\text{iid}}{\sim} \mathcal{N}\left(0, \sigma^2\right), \label{eq:model}\end{equation} where $\beta_0$ is an intercept parameter; $\beta_1$ is the coefficient for covariate $\text{x}_{ij}$; $\theta_i$ is the location-specific random effect shared by all observations in unit $i$ which induces correlation among observations from the same spatial unit and is assumed to be independent of the covariate and $\varepsilon_{ij}$; and $\varepsilon_{ij}$ is an observation-level residual error term.

In addition to the clustering correlation induced by $\theta_i$, it is common to account for spatial correlation among these parameters by introducing a spatial model. CAR models are widely used for areal data in the Bayesian setting due to the computational convenience of their conditional distributions, with several specifications available \citep{lee2011comparison}. We consider the CAR distribution of \citet{leroux2000estimation} where conditionally \[\theta_i \mid \boldsymbol{\theta}_{-i}, \tau^2_{sp}, \rho \stackrel{\text{ind}}{\sim} \mathcal{N}\left(\frac{\rho \sum_{i'=1}^n \text{w}_{ii'} \theta_{i'}} {\rho \sum_{i'=1}^n \text{w}_{ii'} + \left(1-\rho\right)}, \frac{\tau^2_{sp}} {\rho \sum_{i'=1}^n \text{w}_{ii'} + \left(1-\rho\right)} \right), \quad i=1,\hdots,n,\] with $\boldsymbol{\theta}_{-i}^\top = \left(\theta_1, \hdots, \theta_{i-1}, \theta_{i+1}, \hdots, \theta_n\right)$. This model allows for the borrowing of information spatially by specifying that \textit{a priori} spatial unit $i$'s random effect is normally distributed and centered at a weighted average of values from neighboring units, where $\text{w}_{ii'}$ is a binary indicator describing whether units $i$ and $i'$ are neighbors (e.g., share a common border or vertex). Unlike the intrinsic CAR model \citep{besag1974spatial}, this version represents a proper prior distribution for the vector of random effect parameters due to the inclusion of $\rho \in \left[0,1\right)$. This parameter defines the amount of spatial borrowing that occurs; when $\rho = 0$, $\theta_i \mid \tau^2_{sp} \stackrel{\text{iid}}{\sim} \mathcal{N}\left(0,\tau^2_{sp}\right)$, and as $\rho$ approaches one, the original intrinsic CAR model is obtained.

Jointly, $\boldsymbol{\theta}^\top = \left(\theta_1, \hdots, \theta_n\right)$ has a multivariate Gaussian distribution such that \begin{equation*} \boldsymbol{\theta} \mid \tau^2_{sp}, \rho \sim \mathcal{N}_n\bigl( \boldsymbol{0}_n, \tau^2_{sp} Q(\rho)^{-1} \bigr), \quad Q(\rho) = \rho \text{L} + \left(1-\rho\right) I_n, \end{equation*} where $\boldsymbol{0}_n$ is an $n$-length column vector with each entry equal to zero; $\text{L} = \operatorname{diag}(\text{w}_{1\cdot}, \ldots, \text{w}_{n\cdot}) - \text{W}$ is the graph Laplacian with $\text{W}$ the $n \times n$ adjacency matrix with entries $\text{w}_{ii'}$, and $\text{w}_{i\cdot} = \sum_{i'=1}^n \text{w}_{ii'}$; and $I_n$ is the $n$ by $n$ identity matrix. This multivariate Gaussian form allows for a convenient description of the joint distribution of the outcome vector and facilitates marginalizing out the $\boldsymbol{\theta}$ parameters, which we leverage in Section~\ref{sec:spatial} when exploring the connection between $m$ and the need for spatial modeling.

In matrix form, the model from~\eqref{eq:model} can be written as $$\boldsymbol{Y} = \beta_0 \boldsymbol{1}_{nm} + \beta_1 \textbf{x} + \text{Z} \boldsymbol{\theta} + \boldsymbol{\varepsilon},$$ where $\boldsymbol{Y}^\top = \left(\boldsymbol{Y}_{1.}^\top, \hdots, \boldsymbol{Y}_{n.}^\top\right)$ and $\boldsymbol{Y}_{i.}^\top = \left(Y_{i1}, \hdots, Y_{im}\right)$; $\boldsymbol{1}_{nm}$ is an $nm$-length vector with each entry equal to one; $\textbf{x}$ is a vector containing the $\text{x}_{ij}$ covariates and is sorted in the same way as $\boldsymbol{Y}$; $\text{Z} = I_n \otimes \boldsymbol{1}_m$ is a sparse binary matrix that maps each individual to their respective areal unit, where $\otimes$ represents the Kronecker product; and $\boldsymbol{\varepsilon}$ is a vector containing the $\varepsilon_{ij}$ parameters and is sorted in the same way as $\boldsymbol{Y}$. 

We compare this spatial model with a nonspatial version in which $\theta_i \mid \tau^2_{ns} \overset{\mathrm{iid}}{\sim}\mathcal{N}(0,\tau^2_{ns})$. The variance parameters have different interpretations: $\tau^2_{ns}$ is the marginal variance of the random effects under the nonspatial model, whereas the marginal variance under the spatial model for unit $i$ is $\tau^2_{sp}\left[Q\left(\rho\right)^{-1}\right]_{ii}$ and can differ across locations. Since directly equating the two variance parameters would therefore conflate quantities with fundamentally different interpretations, for the comparisons in Section~\ref{sec:spatial} we instead place the models on equal footing by matching the average (across all spatial units) spatial marginal variance to $\tau^2_{ns}$ and using a common residual variance $\sigma^2$ such as \begin{equation} \tau_{ns}^2 = \frac{1}{n}\sum_{i=1}^{n}\operatorname{Var}\left(\theta_i\mid\tau_{sp}^2,\rho\right) = \frac{\tau_{sp}^2}{n}
\operatorname{tr}\left\{Q\left(\rho\right)^{-1}\right\}. \label{eq:spatial_variance_scale} \end{equation}
Under~\eqref{eq:spatial_variance_scale}, both models imply the same average marginal variance, $\tau_{ns}^2 + \sigma^2$, for the outcomes. When $\rho = 0$, $Q\left(\rho\right) = I_n$, so~\eqref{eq:spatial_variance_scale} gives $\tau^2_{sp} = \tau^2_{ns}$, confirming that the two parameterizations coincide in the absence of spatial correlation.

To complete the Bayesian model specification, we assign priors to each model parameter. We assign flat priors to the regression coefficients, $f\left(\beta_0\right), f\left(\beta_1\right) \propto 1$; inverse-gamma priors to the variance parameters, $\sigma^2, \tau^2_{sp}, \tau^2_{ns} \overset{\text{iid}}{\sim} \text{IG}\left(a,b\right)$; and a uniform prior to the correlation parameter, $\rho \sim \text{Uniform}\left(0, 1\right)$. The nonspatial model uses the same priors, with $\tau^2_{ns}$ replacing $\tau^2_{sp}$ and $\rho$ fixed at zero.

In Appendix~\ref{sec:supp_fc}, we detail the full conditional distributions needed to fit the model in~\eqref{eq:model} using standard MCMC methods based on two versions of the likelihood function; conditional on $\boldsymbol{\theta}$ and marginalizing over $\boldsymbol{\theta}$.

\section{When to model spatial correlation} \label{sec:spatial}

\subsection{Full conditional distribution for \texorpdfstring{$\beta_1$}{beta1}}
\label{sec:conditional}
Our primary goal is to understand how marginal posterior inference for $\beta_1$ under the model in~\eqref{eq:model} is impacted as the within-unit sample size $m$ increases, and specifically how it compares to inference under the nonspatial version of~\eqref{eq:model} where $\rho=0$ (i.e., $\theta_i \mid \tau^2_{ns} \stackrel{\text{iid}}{\sim} \mathcal{N}\left(0,\tau^2_{ns}\right)$). To explore these relationships, we work with the full conditional distribution for $\beta_1$ based on the likelihood function where $\boldsymbol{\theta}$ has been marginalized out. These distributions are used in Gibbs sampling \citep{gelfand1990sampling} to obtain samples from the joint posterior distribution of all model parameters.

After integrating out $\boldsymbol{\theta}$ from the model in~\eqref{eq:model}, the distribution of $\boldsymbol{Y}$ is given as \begin{equation} \boldsymbol{Y} \mid \beta_0, \beta_1, \sigma^2, \tau^2_{sp}, \rho \sim \mathcal{N}_{nm}\left( \beta_0 \boldsymbol{1}_{nm} + \beta_1 \textbf{x}, \, \Omega \right), \quad \Omega = \tau^2_{sp} \text{Z}Q(\rho)^{-1}\text{Z}^\top + \sigma^2 I_{nm}, \label{eq:marginal} \end{equation} where all terms have been previously described. Based on the data distribution in \eqref{eq:marginal} and given the flat prior distributions on the regression coefficients, the full conditional distribution of $\beta_1$ is given as \begin{equation*} \beta_1 \mid \boldsymbol{Y}, \beta_0, \sigma^2, \tau^2_{sp}, \rho \sim \mathcal{N}\left( \frac{\textbf{x}^\top \Omega^{-1}(\boldsymbol{Y} - \beta_0 \boldsymbol{1}_{nm})}{\textbf{x}^\top \Omega^{-1} \textbf{x}}, \frac{1}{\textbf{x}^\top \Omega^{-1} \textbf{x}}\right),\end{equation*} with full derivation details given in Appendix~\ref{sec:supp_fc_beta}. We note that if $\beta_0$, $\sigma^2$, $\tau^2_{sp}$, and $\rho$ were known, this full conditional represents the posterior distribution of $\beta_1$. Further, the variance of this distribution is only a function of $\sigma^2$, $\tau^2_{sp}$, and $\rho$. Although this expression conditions on unknown parameters, it isolates how $m$ enters the mean and variance formulas. This dependence carries through directly to the marginal posterior, as the full conditional of $\beta_1$ appears as a component of the integrand when marginalizing over the remaining model parameters. Therefore, investigating this full conditional can provide key insights into the marginal posterior of interest. We further validate this connection via simulation in Section~\ref{sec:simulation}.

We compare the means and variances of the full conditional of $\beta_1$ under the spatial and nonspatial models. We denote these quantities as $\mu_{\beta_1}\left(\rho\right)$ and $\sigma^2_{\beta_1}\left(\rho\right)$ under the spatial model in~\eqref{eq:model}, and as $\mu_{\beta_1}\left(0\right)$ and $\sigma^2_{\beta_1}\left(0\right)$ under the nonspatial model with $\rho=0$. Differences in the means and variances directly quantify how the spatial prior affects posterior inference for $\beta_1$ and how this relationship changes as $m$ increases.

To make the roles of $m$ and the covariance parameters in this comparison explicit, we derive reduced-form expressions for $\mu_{\beta_1}\left(\rho\right)$ and $\sigma^2_{\beta_1}\left(\rho\right)$ that avoid matrix calculations involving the quantities of interest. To this end, we begin by simplifying the expression for $Q\left(\rho\right) = \rho \text{L} + \left(1 - \rho\right)I_n$. Because $\text{L}$ is a real and symmetric matrix due to the form of the neighborhood adjacency matrix $\text{W}$, we can apply the spectral decomposition such that $\text{L} = \text{U} \Lambda \text{U}^\top$, where $\text{U} = [\textbf{u}_1, \hdots, \textbf{u}_n]$ is the $n \times n$ matrix of orthonormal eigenvectors such that $\text{U}^\top \text{U} = \text{U} \text{U}^\top= I_n$; and $\Lambda = \text{diag}\left(\lambda_1, \hdots, \lambda_n\right)$ contains the eigenvalues of $\text{L}$, with $0 = \lambda_1 < \lambda_2 \leq \cdots \leq \lambda_n$ since we assume that the spatial map is connected (i.e., every unit can be reached from every other unit via a sequence of neighboring units). Since $\text{L}\boldsymbol{1}_n=\boldsymbol{0}_n$ and $\lambda_1=0$ is a simple eigenvalue, $\textbf{u}_1=\pm\boldsymbol{1}_n/\sqrt{n}$, while orthogonality gives $\textbf{u}_i^\top\boldsymbol{1}_n=0$ for $i\geq2$. Therefore, $Q\left(\rho\right) = \text{U} \, \text{diag}\left(\rho\lambda_1 + 1 - \rho, \hdots, \rho\lambda_n + 1 - \rho \right) \text{U}^\top$ with corresponding inverse $Q(\rho)^{-1} = \sum_{i=1}^n \textbf{u}_i \textbf{u}_i^\top / (\rho\lambda_i + 1 - \rho)$. Substituting the spectral expression for $Q(\rho)^{-1}$ into~\eqref{eq:marginal}, the covariance matrix of the marginal distribution of $\boldsymbol{Y}$ can be written as \begin{equation} \Omega = \tau^2_{sp} \sum_{i=1}^{n} \frac{\text{Z} \textbf{u}_i \textbf{u}_i^\top \text{Z}^\top}{\rho \lambda_i + 1 - \rho} + \sigma^2 I_{nm}.\label{eq:Omega} \end{equation}

Next, we invert $\Omega$ using the Woodbury matrix identity \citep{woodbury1950inverting, harville1998matrix} by noting that $\Omega$ from~\eqref{eq:Omega} can be written more generally as $\sigma^2 I_{nm} + \text{Z}^* C \text{Z}^{*\top}$, where $\text{Z}^* = \text{Z}\text{U}$ and $C=\tau^2_{sp}\operatorname{diag}\{1/(\rho\lambda_1+1-\rho),\ldots,1/(\rho\lambda_n+1-\rho)\}$. Applying the Woodbury matrix identity, \begin{equation*} \Omega^{-1} = \frac{1}{\sigma^2} I_{nm} - \frac{1}{\sigma^4} \text{Z}^* \left( C^{-1} + \frac{\text{Z}^{*\top} \text{Z}^*}{\sigma^2} \right)^{-1} \text{Z}^{*\top}. \label{eq:woodbury} \end{equation*} Because every spatial unit has the same sample size $m$, $\text{Z}^{*\top} \text{Z}^* = \text{U}^\top \left(I_n^\top \otimes \boldsymbol{1}_m^\top\right)\left(I_n \otimes \boldsymbol{1}_m\right) \text{U}= \text{U}^\top \left\{\left(I_n^\top I_n\right) \otimes \left(\boldsymbol{1}_m^\top \boldsymbol{1}_m\right)\right\} \text{U}= m I_n$, such that the inverse simplifies to \begin{equation*} \Omega^{-1} = \frac{1}{\sigma^2} I_{nm} - \frac{\tau^2_{sp}}{\sigma^2} \sum_{i=1}^{n} \frac{\text{Z} \textbf{u}_i \textbf{u}_i^\top \text{Z}^\top}{\sigma^2 \left(\rho \lambda_i + 1 - \rho\right) + m \tau^2_{sp}}.
\end{equation*} Therefore, the precision of the full conditional distribution of $\beta_1$ under the spatial model can be written as \begin{equation}\sigma^{-2}_{\beta_1}\left(\rho\right) = \frac{\textbf{x}^\top \textbf{x}}{\sigma^2} - \frac{\tau^2_{sp}}{\sigma^2} \sum_{i=1}^{n} \frac{(\textbf{x}^\top \text{Z} \textbf{u}_i)^2}{\sigma^2 \left(\rho \lambda_i + 1 - \rho\right) +m\tau^2_{sp}}. \label{eq:precision_step1}\end{equation}

We can further simplify the expression in~\eqref{eq:precision_step1} by expanding $\textbf{x}^\top \text{Z} \textbf{u}_i$. We first note that $\text{Z}^\top\textbf{x} = \left(I_n^\top \otimes \boldsymbol{1}_m^\top\right)\textbf{x} = m \bar{\textbf{x}}$ where $\bar{\textbf{x}}^\top = \left(\bar{\text{x}}_{1\cdot}, \hdots, \bar{\text{x}}_{n\cdot}\right)$ and $\bar{\text{x}}_{i\cdot} = m^{-1} \sum_{j=1}^m \text{x}_{ij}$. Therefore, $\textbf{x}^\top \text{Z} \textbf{u}_i = m \sum_{i'=1}^n \bar{\text{x}}_{i'\cdot} \text{u}_{ii'}$, where $\text{u}_{ii'}$ denotes the $i'^{\text{th}}$ entry of $\textbf{u}_i$. Without loss of generality, we further assume that the covariate has been centered and standardized prior to analysis such that its mean is equal to zero and its population standard deviation (i.e., dividing by $nm$ instead of $nm - 1$) is equal to one such that $\textbf{x}^\top \textbf{x} = nm$. Centering also simplifies the form of $\mu_{\beta_1}\left(\rho\right)$. Since $\textbf{u}_1$ is proportional to $\boldsymbol{1}_n$, centering gives $\textbf{x}^\top\boldsymbol{1}_{nm}=0$ and $\textbf{x}^\top\text{Z}\textbf{u}_1=\textbf{x}^\top\boldsymbol{1}_{nm}/\sqrt{n}=0$. Together with $\text{Z}^\top\boldsymbol{1}_{nm}=m\boldsymbol{1}_n$, these identities imply $\textbf{x}^\top\Omega^{-1}\boldsymbol{1}_{nm}=0$. Thus, the mean of the full conditional distribution does not depend on $\beta_0$ and simplifies to $\mu_{\beta_1}(\rho)=\sigma^2_{\beta_1}(\rho)\textbf{x}^\top\Omega^{-1}\boldsymbol{Y}.$
The same simplification holds under the nonspatial model. Letting $\bar{Y}_{i\cdot}=m^{-1}\sum_{j=1}^{m}Y_{ij}$, the corresponding outcome projection is
$\textbf{u}_i^\top\text{Z}^\top\boldsymbol{Y} = m\sum_{i'=1}^{n}\bar{Y}_{i'\cdot}\text{u}_{ii'}$. Using these projections together with the expression for $\Omega^{-1}$ yields the following precision and mean of the full conditional.

\begin{proposition}[Full conditional distribution means and variances]
\label{prop:conditional_moments}
Under the model specifications and covariate standardization described previously, the precision and mean of the full conditional distribution of $\beta_1$ under the spatial model are
\begin{equation}
\sigma^{-2}_{\beta_1}\left(\rho\right) = \frac{nm}{\sigma^2} - \frac{m^2\tau^2_{sp}}{\sigma^2} \sum_{i=1}^{n} \frac{ \left( \sum_{i'=1}^{n} \bar{\mathrm{x}}_{i'\cdot}\mathrm{u}_{ii'} \right)^2}{ \sigma^2\left(\rho\lambda_i+1-\rho\right)+m\tau^2_{sp}}
\label{eq:precision}
\end{equation}
and
\begin{equation}
\mu_{\beta_1}\left(\rho\right) = \frac{\sigma^2_{\beta_1}\left(\rho\right)}{\sigma^2} \left\{ \mathbf{x}^\top\boldsymbol{Y} - m^2\tau^2_{sp} \sum_{i=1}^{n} \frac{ \left(\sum_{i'=1}^{n}\bar{\mathrm{x}}_{i'\cdot}\mathrm{u}_{ii'}\right) \left(\sum_{i'=1}^{n}\bar{Y}_{i'\cdot}\mathrm{u}_{ii'}\right) }{ \sigma^2\left(\rho\lambda_i+1- \rho\right) + m\tau^2_{sp} } \right\}.
\label{eq:mean}
\end{equation}

Under the nonspatial model, with $\rho=0$ and random-intercept variance $\tau^2_{ns}$, the corresponding expressions are
\begin{equation}
\sigma^{-2}_{\beta_1}\left(0\right) = \frac{nm}{\sigma^2} - \frac{m^2\tau^2_{ns}}{\sigma^2\left(\sigma^2 + m\tau^2_{ns}\right)} \sum_{i=1}^{n} \left( \sum_{i'=1}^{n} \bar{\mathrm{x}}_{i'\cdot}\mathrm{u}_{ii'} \right)^2
\label{eq:precision0}
\end{equation}
and
\begin{equation}
\mu_{\beta_1}\left(0\right) = \frac{\sigma^2_{\beta_1}\left(0\right)}{\sigma^2} \left\{ \mathbf{x}^\top\boldsymbol{Y} - \frac{m^2\tau^2_{ns}}{\sigma^2 + m\tau^2_{ns}} \sum_{i=1}^{n} \left(\sum_{i'=1}^{n}\bar{\mathrm{x}}_{i'\cdot}\mathrm{u}_{ii'}\right) \left(\sum_{i'=1}^{n}\bar{Y}_{i'\cdot}\mathrm{u}_{ii'}\right) \right\}.
\label{eq:mean0}
\end{equation}
\end{proposition}

These expressions are useful in this setting as they make the role of within-area sample size $m$ explicit and can be evaluated using quantities available from the raw data (i.e., sample sizes, covariate and outcome values, and eigenvectors/eigenvalues based on the spatial adjacency matrix) and the covariance parameters. To empirically validate the simplified expressions in Proposition~\ref{prop:conditional_moments}, we compare the precision and mean expressions with the matrix forms, $\textbf{x}^{\top}\Omega^{-1}\textbf{x}$ and $(\textbf{x}^{\top}\Omega^{-1}\textbf{x})^{-1}\textbf{x}^{\top}\Omega^{-1}(\boldsymbol{Y}-\beta_0\boldsymbol{1}_{nm})$, respectively. We do this by randomly sampling different spatial adjacency matrices, covariate values, and parameter values, and computing both forms. Results were in agreement across all simulations. Code for reproducing this validation is available in the \texttt{SpThreshold} repository. 

We focus on balanced replication and a single covariate here because they make the dependence on $m$ and on the covariate's spatial alignment explicit. The same full conditional calculations apply to each coefficient in a model with multiple covariates, conditional on the remaining coefficients, and to unequal within-area sample sizes; Appendices~\ref{sec:multiple_sim} and~\ref{sec:supp_decomp} give the corresponding expressions. 

\subsection{Convergence with increasing within-area sample size}
To determine how the full conditional distribution for $\beta_1$ differs between the spatial and nonspatial models as $m$ increases, we first analyze the absolute relative difference in variances, \begin{equation}\left|\frac{\sigma^2_{\beta_1}\left(\rho\right) - \sigma^2_{\beta_1}\left(0\right)}{\sigma^2_{\beta_1}\left(0\right)}\right| = \left|\frac{\sigma^{-2}_{\beta_1}\left(0\right) - \sigma^{-2}_{\beta_1}\left(\rho\right)}{\sigma^{-2}_{\beta_1}\left(\rho\right)}\right|,\label{eq:abs_diff}\end{equation} using the results from~\eqref{eq:precision} and~\eqref{eq:precision0}. We show that this quantity converges to zero at a rate $O\left(m^{-1}\right)$, provided that the average within-area covariate variance remains bounded away from zero as $m$ increases. 

To simplify notation, from~\eqref{eq:precision} and~\eqref{eq:precision0} we define $\text{d}_i \equiv \left(\sum_{i' = 1}^n \bar{\text{x}}_{i'\cdot} \text{u}_{ii'}\right)^2$ and $\text{d}_{\cdot} \equiv \sum_{i=1}^n \text{d}_i$. The numerator in~\eqref{eq:abs_diff} can then be written as \begin{equation*} \sigma^{-2}_{\beta_1}\left(0\right) - \sigma^{-2}_{\beta_1}\left(\rho\right) = \frac{m^2}{\sigma^2} \left[ \tau^2_{sp} \sum_{i=1}^{n} \frac{\text{d}_i} {\sigma^2\left(\rho\lambda_i+1-\rho\right)+m\tau^2_{sp}} - \frac{\tau^2_{ns}\text{d}_{\cdot}}{\sigma^2+m\tau^2_{ns}} \right].\end{equation*}
To bound this expression, we apply the expansion $\left(1 + g\right)^{-1} = 1 - g + O\left(g^2\right)$ for small $g$ giving \begin{align*}\left\{\sigma^2\left(\rho \lambda_i + 1 - \rho\right) + m\tau^2_{sp}\right\}^{-1} &= \frac{1}{m\tau^2_{sp}} \left\{1 + \frac{\sigma^2\left(\rho \lambda_i + 1 - \rho\right)}{m\tau^2_{sp}}\right\}^{-1}\\ &=  \frac{1}{m\tau^2_{sp}} \left\{1 - \frac{\sigma^2\left(\rho \lambda_i + 1 - \rho\right)}{m\tau^2_{sp}} + O\left(m^{-2}\right)\right\} \\  &=  \frac{1}{m\tau^2_{sp}} \left\{1 - \frac{\sigma^2\left(\rho \lambda_i + 1 - \rho\right)}{m\tau^2_{sp}}\right\} + O\left(m^{-3}\right).\end{align*} Similarly, with $\rho = 0$, $\left(\sigma^2 + m \tau^2_{ns}\right)^{-1} = \frac{1}{m\tau^2_{ns}}\left(1 - \frac{\sigma^2}{m\tau^2_{ns}}\right) + O\left(m^{-3}\right)$. Plugging these results back into~\eqref{eq:precision} and~\eqref{eq:precision0}, we obtain \begin{align*}\sigma^{-2}_{\beta_1}\left(\rho\right) &= \frac{nm}{\sigma^2} - \frac{m^2 \tau^2_{sp}}{\sigma^2} \left[\sum_{i=1}^n \frac{\text{d}_i}{m \tau^2_{sp}} \left\{1 - \frac{\sigma^2\left(\rho \lambda_i + 1 - \rho\right)}{m \tau^2_{sp}}\right\} + O\left(m^{-3}\right)\right]\\ &= \frac{nm}{\sigma^2} - \frac{m \text{d}_{\cdot}}{\sigma^2} + \sum_{i=1}^n \frac{\text{d}_i}{\tau^2_{sp}} \left(\rho \lambda_i + 1 - \rho\right) + O\left(m^{-1}\right)\\ & = \frac{nm}{\sigma^2} - \frac{m \text{d}_{\cdot}}{\sigma^2} + \frac{\text{d}_{\cdot}}{\tau^2_{sp}} + \frac{\rho}{\tau^2_{sp}}\sum_{i=1}^n \text{d}_i \left(\lambda_i - 1\right) + O\left(m^{-1}\right)\end{align*} and $\sigma^{-2}_{\beta_1}\left(0\right) = \frac{nm}{\sigma^2} - \frac{m \text{d}_{\cdot}}{\sigma^2} + \frac{\text{d}_{\cdot}}{\tau^2_{ns}} + O\left(m^{-1}\right).$ Therefore, the numerator in~\eqref{eq:abs_diff} can be further simplified as $\text{d}_{\cdot}
\left(\frac{1}{\tau^2_{ns}}-\frac{1}{\tau^2_{sp}}\right) + \frac{\rho}{\tau^2_{sp}} \sum_{i=1}^{n}\text{d}_i\left(1 - \lambda_i\right) + O\left(m^{-1}\right).$

The denominator of~\eqref{eq:abs_diff} can also be simplified such that $\sigma^{-2}_{\beta_1}\left(\rho\right) = \frac{m\left(n - \text{d}_{\cdot}\right)}{\sigma^2} \left\{1 + O\left(m^{-1}\right)\right\}$, resulting in an absolute relative variance difference of \begin{align} &\left| \frac{ \text{d}_{\cdot} \left(\frac{1}{\tau^2_{ns}}-\frac{1}{\tau^2_{sp}}\right) +\frac{\rho}{\tau^2_{sp}} \sum_{i=1}^{n}\text{d}_i(1-\lambda_i) +O(m^{-1}) }{ \frac{m(n-\text{d}_{\cdot})}{\sigma^2} \left\{1+O(m^{-1})\right\} } \right| \notag\\ &= \left| \left\{ \frac{\sigma^2}{m(n-\text{d}_{\cdot})} \left[ \text{d}_{\cdot} \left(\tfrac{1}{\tau^2_{ns}}-\tfrac{1}{\tau^2_{sp}}\right) +\tfrac{\rho}{\tau^2_{sp}} \sum_{i=1}^{n}\text{d}_i(1-\lambda_i) \right] +O(m^{-2}) \right\} \left\{1+O(m^{-1})\right\} \right|, \label{eq:rel_diff_simp} \end{align} where the previous expansion was again applied to the denominator term.  Expression~\eqref{eq:rel_diff_simp} reduces to $\left\{O\left(m^{-1}\right) + O\left(m^{-2}\right)\right\} \left\{1 + O\left(m^{-1}\right)\right\} = O\left(m^{-1}\right),$ confirming that the absolute relative difference in~\eqref{eq:abs_diff} is $O\left(m^{-1}\right)$. 

We next examine the absolute difference between the full conditional means. Let $w_i=\textbf{x}^{\top}\text{Z}\textbf{u}_i\textbf{u}_i^{\top}\text{Z}^{\top}\boldsymbol{Y}$; subtracting~\eqref{eq:mean0} from~\eqref{eq:mean} and regrouping gives
\begin{align*}
\mu_{\beta_1}(\rho)-\mu_{\beta_1}(0)
&=\frac{\sigma^2_{\beta_1}(\rho)-\sigma^2_{\beta_1}(0)}{\sigma^2}\textbf{x}^{\top}\boldsymbol{Y}\\
&\quad-\frac{\sigma^2_{\beta_1}(0)}{\sigma^2}\sum_{i=1}^{n}w_i\left\{\frac{\tau^2_{sp}}{\sigma^2(\rho\lambda_i+1-\rho)+m\tau^2_{sp}}-\frac{\tau^2_{ns}}{\sigma^2+m\tau^2_{ns}}\right\}\\
&\quad-\frac{\tau^2_{sp}\{\sigma^2_{\beta_1}(\rho)-\sigma^2_{\beta_1}(0)\}}{\sigma^2}\sum_{i=1}^{n}\frac{w_i}{\sigma^2(\rho\lambda_i+1-\rho)+m\tau^2_{sp}}.
\end{align*} To bound these three terms, we first use the preceding variance results. Since $0\leq\text{d}_{\cdot}\leq n$ and $n-\text{d}_{\cdot}$ is bounded away from zero, inverting the nonspatial precision expansion gives
\begin{align*} \sigma^2_{\beta_1}(0) = \left\{\frac{m(n-\text{d}_{\cdot})}{\sigma^2}+\frac{\text{d}_{\cdot}}{\tau^2_{ns}}+O(m^{-1})\right\}^{-1} = \frac{\sigma^2}{m(n-\text{d}_{\cdot})}\{1+O(m^{-1})\} = O(m^{-1}). \end{align*}
Together with~\eqref{eq:rel_diff_simp}, this yields
\begin{align*} \sigma^2_{\beta_1}(\rho)-\sigma^2_{\beta_1}(0) = \sigma^2_{\beta_1}(0)\left\{\frac{\sigma^2_{\beta_1}(\rho)}{\sigma^2_{\beta_1}(0)}-1\right\} = O(m^{-1})O(m^{-1}) = O(m^{-2}). \end{align*}
For each $i$, the coefficient difference in the second term satisfies
\begin{align*} \frac{\tau^2_{sp}}{\sigma^2(\rho\lambda_i+1-\rho)+m\tau^2_{sp}}-\frac{\tau^2_{ns}}{\sigma^2+m\tau^2_{ns}} = \frac{\sigma^2\{\tau^2_{sp}-\tau^2_{ns}(\rho\lambda_i+1-\rho)\}}{\{\sigma^2(\rho\lambda_i+1-\rho)+m\tau^2_{sp}\}(\sigma^2+m\tau^2_{ns})}
=O(m^{-2}),
\end{align*} while $\{\sigma^2(\rho\lambda_i+1-\rho)+m\tau^2_{sp}\}^{-1}=O(m^{-1})$.

Covariate standardization gives $\sum_{i=1}^{n}\bar{\text{x}}_{i\cdot}^{\,2}\leq n$. Under model~\eqref{eq:model}, $\bar{Y}_{i\cdot}=\beta_0+\beta_1\bar{\text{x}}_{i\cdot}+\theta_i+O_p(m^{-1/2})=O_p(1)$. Also, $\textbf{x}^{\top}\boldsymbol{\varepsilon}\sim\mathcal{N}(0,\sigma^2nm)$, so
\begin{equation*}
\textbf{x}^{\top}\boldsymbol{Y}=nm\beta_1+m\sum_{i=1}^{n}\bar{\text{x}}_{i\cdot}\theta_i+\textbf{x}^{\top}\boldsymbol{\varepsilon}=O(m)+O_p(m)+O_p(m^{1/2})=O_p(m).
\end{equation*}
Similarly,
\begin{equation*} w_i = (\textbf{x}^\top\text{Z}\textbf{u}_i)(\textbf{u}_i^\top\text{Z}^\top\boldsymbol{Y}) = m^2\left(\sum_{i'=1}^{n}\bar{\text{x}}_{i'\cdot}\text{u}_{ii'}\right)\left(\sum_{i'=1}^{n}\bar{Y}_{i'\cdot}\text{u}_{ii'}\right)=m^2O(1)O_p(1)=O_p(m^2). \end{equation*}
Since $n$ is fixed, substituting these bounds into the three-term decomposition gives
\begin{align*}
\mu_{\beta_1}(\rho)-\mu_{\beta_1}(0)
&=O(m^{-2})O_p(m)+O(m^{-1})O_p(m^2)O(m^{-2}) + O(m^{-2})O_p(m^2)O(m^{-1})\notag\\
&=O_p(m^{-1})+O_p(m^{-1})+O_p(m^{-1})=O_p(m^{-1}).
\end{align*}
\begin{theorem}[Convergence of full conditional means and variances]
\label{thm:conditional_agreement}
Suppose the data follow model~\eqref{eq:model}, conditional on the centered and standardized covariates described previously. Hold $n$, the spatial graph, and the model parameters fixed, with $\sigma^2,\tau^2_{ns},\tau^2_{sp}>0$ and $0\leq\rho<1$. If the average within-area covariate variance remains bounded away from zero as $m\to\infty$, then
\begin{align*}
\left|\frac{\sigma^2_{\beta_1}(\rho)-\sigma^2_{\beta_1}(0)}{\sigma^2_{\beta_1}(0)}\right|&=O(m^{-1}),\\
\left|\mu_{\beta_1}(\rho)-\mu_{\beta_1}(0)\right|&=O_p(m^{-1}).
\end{align*}
\end{theorem} Thus, the variance ratio of the full conditional distribution converges to one, and the difference between the means of the full conditional distributions becomes negligible relative to posterior uncertainty. Under unequal replication, the relative variance difference also vanishes as the within-area covariate sum of squares grows (Appendix~\ref{sec:unequal_sim}).

\subsection{Within-area sample size threshold}
\label{sec:threshold}

We derive a closed-form threshold $m^*_{\mathrm{approx}}$ that keeps the absolute relative difference in full conditional variances within a desired tolerance $\gamma$, and compute an exact threshold $m^*_{\mathrm{exact}}$ numerically from the variance expressions in Proposition~\ref{prop:conditional_moments}. We base the threshold on the variance difference for two reasons: it quantifies agreement in posterior precision, and it depends only on the covariate structure and covariance parameters, not on the observed outcomes. The marginal posteriors of interest differ from the full conditionals in that they additionally integrate over uncertainty in the covariance parameters. Whether agreement in the full conditionals implies agreement in the marginal posteriors is therefore not immediate, and we assess this empirically in Section~\ref{sec:simulation} by comparing the exact full conditional mean and variance differences with their marginal posterior counterparts estimated from MCMC samples.

Using the form in~\eqref{eq:rel_diff_simp}, the leading-order term of the absolute relative variance difference is $\frac{\sigma^2}{m(n-\text{d}_{\cdot})}\left|\text{d}_{\cdot}\left(\frac{1}{\tau^2_{ns}}-\frac{1}{\tau^2_{sp}}\right)+\frac{\rho}{\tau^2_{sp}}\sum_{i=1}^{n}\text{d}_i(1-\lambda_i)\right|$. By the spectral decomposition of $Q(\rho)$, $\operatorname{tr}\left\{Q(\rho)^{-1}\right\} = \sum_{i=1}^{n}\left(\rho\lambda_i+1-\rho\right)^{-1}$. Solving the variance-matching relation~\eqref{eq:spatial_variance_scale} for $\tau_{sp}^2$ therefore gives \begin{equation}
\tau_{sp}^2 = \frac{n\tau_{ns}^2}{\sum_{i=1}^{n}\left(\rho\lambda_i+1-\rho\right)^{-1}}. \label{eq:matching_spectral}\end{equation}
The leading-order term then becomes \begin{align*} &\frac{\sigma^2}{m(n-\text{d}_{\cdot})}\left|\text{d}_{\cdot}\left(\frac{1}{\tau^2_{ns}}-\frac{1}{\tau^2_{sp}}\right)+\frac{\rho}{\tau^2_{sp}}\sum_{i=1}^{n}\text{d}_i(1-\lambda_i)\right| \\ &\quad=\frac{\sigma^2}{m(n-\text{d}_{\cdot})}\left|\sum_{i=1}^{n}\text{d}_i\left[\frac{1}{\tau^2_{ns}}-\frac{\rho\lambda_i+1-\rho}{\tau^2_{sp}}\right]\right| \\ &\quad=\frac{1}{m\kappa(n-\text{d}_{\cdot})}\left|\sum_{i=1}^{n}\text{d}_i\left[1-\frac{\rho\lambda_i+1-\rho}{n}\sum_{k=1}^{n}(\rho\lambda_k+1-\rho)^{-1}\right]\right|, \end{align*} where $\kappa=\tau^2_{ns}/\sigma^2$.  
\begin{corollary}[Approximate replication threshold]
\label{cor:replication_threshold}
Under the assumptions of Theorem~\ref{thm:conditional_agreement} and the variance-matching relation in~\eqref{eq:spatial_variance_scale}, fix the covariate projections $\text{d}_i$ and let $\gamma>0$. The smallest integer $m\geq2$ for which the leading-order absolute relative variance difference is no larger than $\gamma$ is
\begin{equation}
m^*_{\mathrm{approx}}=\max\left\{2,\,
\left\lceil
\frac{\left|\sum_{i=1}^{n}\text{d}_i\left[1-\frac{\rho\lambda_i+1-\rho}{n}\sum_{k=1}^{n}(\rho\lambda_k+1-\rho)^{-1}\right]\right|}{\gamma\kappa(n-\text{d}_{\cdot})}
\right\rceil\right\},
\label{eq:threshold}
\end{equation}
where $\kappa=\tau^2_{ns}/\sigma^2$.
\end{corollary}

We set a lower bound of $2$ given that it is the smallest possible sample size such that we are still working with multilevel data. Since $\text{U}$ is orthonormal, $\text{d}_{\cdot}=\|\text{U}^{\top}\bar{\textbf{x}}\|^2=\|\bar{\textbf{x}}\|^2=\sum_{i=1}^{n}\bar{\text{x}}_{i\cdot}^2$  The threshold in~\eqref{eq:threshold} is a function of three interpretable quantities: the spatial correlation parameter $\rho$, the variance ratio $\kappa = \tau^2_{ns} / \sigma^2$ (which enters the numerator as $1/\kappa$), and the covariate structure through the projections $\text{d}_i$. A smaller variance ratio can require more replication before spatial smoothing becomes negligible, as it makes the spatial prior more influential. Stronger spatial correlation can have a similar effect, depending on the spatial network and covariate structure.

The projections $\text{d}_i = \left(\sum_{i'=1}^{n} \bar{\text{x}}_{i'\cdot}\, \text{u}_{ii'}\right)^2$ decompose the between-location covariate variation across the eigenvectors of the graph Laplacian, which represent spatial patterns at progressively finer scales: $\textbf{u}_1$ corresponds to a constant (global mean), $\textbf{u}_2$ to the smoothest nontrivial contrast, and successive eigenvectors to increasingly rough spatial variation. When covariate means vary little across locations, the $\text{d}_i$ are uniformly small and the threshold is low. When they exhibit smooth spatial trends aligned with the leading eigenvectors, the $\text{d}_i$ concentrate on the components where the two models differ most, and a larger $m$ is needed. With several covariates, the threshold applies to each coefficient conditional on the others, with $\text{d}_i$ computed from the corresponding covariate's area means. The required replication can therefore differ between coefficients in the same model (Appendix~\ref{sec:multiple_sim}).

We also calculate an exact threshold, $m^*_{\mathrm{exact}}$, numerically from~\eqref{eq:precision} and~\eqref{eq:precision0} for the same fixed inputs, as the smallest integer $m\geq2$ for which the absolute relative variance difference is at most $\gamma$ and remains so at every larger sample size. Unlike $m^*_{\mathrm{approx}}$, it has no closed form and is found by evaluating the exact difference at successive values of $m$. The finite-sample bound in Appendix~\ref{sec:supp_threshold_search} provides an explicit upper limit for this search.

\section{Simulation study}\label{sec:simulation}

\subsection{Study design}
We design a simulation study to evaluate the theoretical results of Section~\ref{sec:spatial}. We have three primary objectives: first, to assess how well the full conditional distribution comparisons of Section~\ref{sec:spatial} describe the marginal posterior distributions of interest and how the differences in the marginal posterior variance and mean of $\beta_1$ between the spatial and nonspatial models change as within-area sample size $m$ increases; second, to assess how well the approximate and exact replication thresholds identify when the marginal posterior variances agree within a specified relative tolerance; and third, to examine how these results vary across levels of spatial correlation, signal-to-noise ratio, and covariate structure.

We simulate data from \eqref{eq:model} with $n \in \left\{25, 100, 400\right\}$ spatial units. For a selected $n$, we first simulate a random connected spatial map with $n$ units to define the adjacency matrix $\text{W}$. Summary statistics of the resulting graphs and example realizations are provided in Appendix~\ref{sec:supp_graph}. Next, we vary spatial correlation $\rho \in \{0.05, 0.50, 0.95\}$ and variance ratio $\kappa = \tau^2_{ns}/\sigma^2 \in \left\{\tfrac{0.05}{0.95},\; \tfrac{0.50}{0.50},\; \tfrac{0.95}{0.05}\right\}$, where $\tau^2_{ns} + \sigma^2 = 1$ so that the total marginal variance is held constant. For each spatial map and value of $\rho$, we calculate $\tau^2_{sp}=n\tau^2_{ns}/\sum_{i=1}^{n}(\rho\lambda_i+1-\rho)^{-1}$ from~\eqref{eq:matching_spectral} and generate spatial random intercepts $\boldsymbol{\theta}\mid\tau^2_{sp},\rho\sim\mathcal{N}_n(\boldsymbol{0}_n,\tau^2_{sp}Q(\rho)^{-1})$. This calibration holds the average marginal variance of $\theta_i+\varepsilon_{ij}$ across spatial units at one. Small values of $\kappa$ correspond to settings where spatial random effects contribute little relative to observation-level noise, while large $\kappa$ indicates that most variability is attributable to location-level differences. Finally, we sample $\beta_0, \beta_1 \overset{\text{iid}}{\sim} \mathcal{N}\left(0, 1\right)$.  
\begin{table}[ht]
    \centering
    \caption{Summary of simulation study settings.}
    \label{tab:sim_settings}
    \begin{tabular}{ll}
        \toprule
        Parameter & Values \\
        \midrule
Datasets per setting & 100 \\
Spatial units ($n$) & 25, 100, 400 \\
Spatial correlation ($\rho$) & 0.05, 0.50, 0.95 \\
Variance ratio ($\kappa = \tau^2_{ns}/\sigma^2$) & 0.05/0.95, 0.50/0.50, 0.95/0.05 \\
Within-area replication ($m$) & 2, 5, 10, 20, 50, 80, 100, 200 \\
Covariate structure & C1: $\text{x}_{ij} \overset{\text{iid}}{\sim} \mathcal{N}\left(0, 1\right)$, \\
                     & C2: $\text{x}_{ij} \mid \mu_i \overset{\text{ind}}{\sim} \mathcal{N}\left(\mu_i, 1\right)$; $\mu_i \overset{\text{iid}}{\sim} \mathcal{N}\left(0, 1\right)$, \\
                     & C3: $\text{x}_{ij} = \mu_i$; $\mu_i \overset{\text{iid}}{\sim} \mathcal{N}\left(0, 1\right)$ \\
   \bottomrule
    \end{tabular}
\end{table}

We also consider three different covariate structure settings (C1-C3). In C1, $\text{x}_{ij} \overset{\text{iid}}{\sim} \mathcal{N}\left(0, 1\right)$ for all $i$ and $j$, so covariate variation is entirely within locations. The $\bar{\text{x}}_{i\cdot}$ are approximately zero resulting in small $\text{d}_i$ and $m^*$, so we expect spatial and nonspatial inference to be similar even at small $m$. In C2, $\text{x}_{ij} \mid \mu_i \overset{\text{ind}}{\sim} \mathcal{N}\left(\mu_i, 1\right)$ with location-specific means $\mu_i \overset{\text{iid}}{\sim} \mathcal{N}\left(0, 1\right)$, introducing between-location variation. In this case, $m^*$ may be larger as the $\bar{\text{x}}_{i\cdot}$ are not all approximately equal to zero. In C3, $\text{x}_{ij} = \mu_i$ with $\mu_i \overset{\text{iid}}{\sim} \mathcal{N}\left(0, 1\right)$, so the covariate is entirely a location-level attribute. Under such covariate structure, $\text{x}_{ij} = \bar{\text{x}}_{i\cdot}$, and $\sum_{j=1}^m \text{x}_{ij}^2 = m \bar{\text{x}}_{i\cdot}^2$ for every $i$. Combined with the population standardization that gives $\sum_{i=1}^n\sum_{j=1}^m \text{x}_{ij}^2 = nm$, this suggests $\sum_{i=1}^n \bar{\text{x}}_{i\cdot}^2 = n$ and therefore, the denominator in~\eqref{eq:threshold} is exactly zero. Hence, increasing within-area replication alone need not bring the two models into agreement.

Within-area replication sample sizes range over $m \in \{2, 5, 10, 20, 50, 80, 100, 200\}$. For each combination of $\left(n, m, \rho, \kappa\right)$ and covariate structure setting (in total $648$ unique settings), we generate $100$ independent datasets for analysis. A summary of all simulation settings is provided in Table~\ref{tab:sim_settings}. Increasing $\rho$ produces smoother spatial patterns, while increasing $\kappa$ amplifies the magnitude of the spatial random effects (Figure~\ref{fig:sim_theta}).

For each dataset, we fit both models using the MCMC algorithm that directly includes $\boldsymbol{\theta}$ as part of the parameter space. This algorithm is detailed in Appendix~\ref{sec:supp_fc}. 
Priors are given in Section~\ref{sec:model}, with inverse-gamma hyperparameters $a = b = 0.01$ to induce weakly informative priors on $\sigma^2$ and the random-intercept variance parameter ($\tau^2_{sp}$ under the spatial model and $\tau^2_{ns}$ under the nonspatial model). Without loss of generality, we center and standardize the covariate to have mean zero and population variance one prior to analysis. 

When updating $\rho$ using the Metropolis algorithm, the proposal distribution standard deviation is adapted during the burn-in period using the Robbins-Monro scheme of \citet{roberts2009examples}, targeting an acceptance rate of $0.234$ \citep{gelman1997weak} and held fixed thereafter. The nonspatial model is fit using a Gibbs sampler with all updates drawn directly from closed-form full conditional distributions. For both models we collect $75,000$ iterations, discard the first $15,000$ as burn-in, and thin by a factor of $5$, yielding $12,000$ posterior draws. Convergence was assessed by visual inspection of trace plots across a subset of analyses. We compare the models using the absolute difference in marginal posterior means and the absolute relative difference in marginal posterior variances of $\beta_1$. The corresponding conditional differences are calculated using Proposition~\ref{prop:conditional_moments}, evaluated at the generating covariance parameters and observed data. We average these differences, and the approximate and exact thresholds under C2, across the $100$ datasets within each setting. For each model, we also evaluate posterior mean bias and empirical coverage of the $95\%$ credible interval.

\subsection{Results}

\begin{figure}[ht!]
\centering
\includegraphics[scale = 0.5]{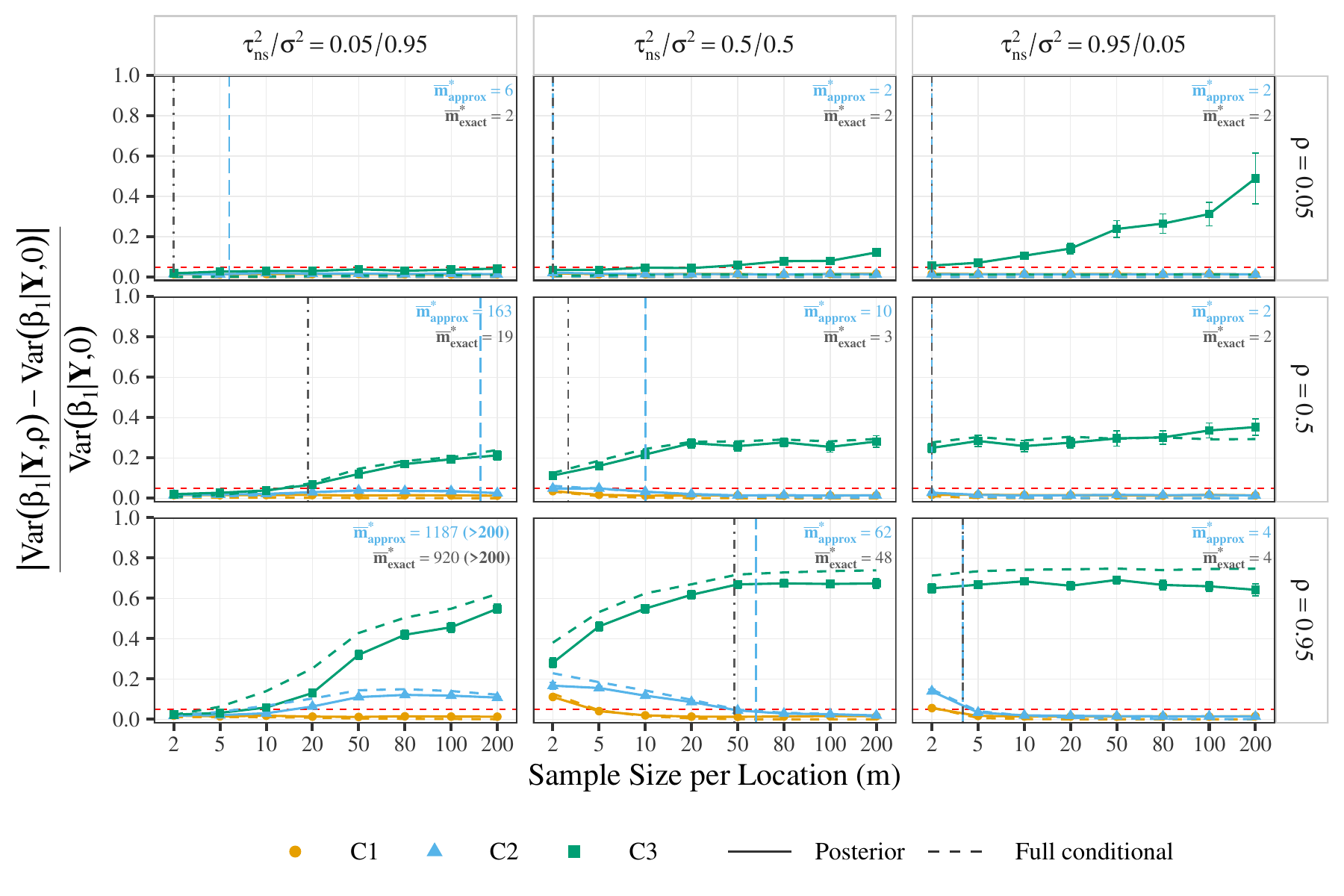}
\caption{Absolute relative difference in the marginal posterior variance of $\beta_1$ between the spatial (i.e., $\text{Var}\left(\beta_1 | \boldsymbol{Y}, \rho\right)$) and nonspatial (i.e., $\text{Var}\left(\beta_1 | \boldsymbol{Y}, 0\right)$) models, averaged over $100$ datasets, for $n = 100$ spatial units. Columns correspond to the variance ratio $\kappa = \tau^2_{ns}/\sigma^2$ and rows to the spatial correlation $\rho$. The three colors represent the three covariate structures, C1-C3, described in Table~\ref{tab:sim_settings}. Solid and dashed curves show results based on the marginal posterior inference and exact full conditional comparisons, respectively. The dashed red horizontal line marks the tolerance $\gamma = 0.05$; blue and grey vertical lines mark the approximate and exact C2 thresholds, respectively, each averaged over $100$ datasets and rounded up to an integer. Error bars are pointwise $95\%$ Monte Carlo confidence intervals for the marginal posterior averages across datasets.}
\label{fig:rel_var}
\end{figure}

Figure~\ref{fig:rel_var} displays the absolute relative difference in the marginal posterior variance of $\beta_1$ between the spatial (i.e., $\text{Var}\left(\beta_1 | \boldsymbol{Y}, \rho\right)$) and nonspatial (i.e., $\text{Var}\left(\beta_1 | \boldsymbol{Y}, 0\right)$) models, averaged over $100$ simulations, as a function of the within-area replication $m$, for $n=100$ spatial units. The corresponding full conditional relative variance differences, averaged over the same datasets, are shown for comparison. Results for $n=25$ and $n=400$ are shown in Figures~\ref{fig:plot1_J25}--\ref{fig:plot2_J400}. 

In C1, where all covariate variation is within locations, the absolute relative difference in variances is near or below $\gamma$ across all settings with $m\geq5$, consistent with the theoretical prediction that small $\text{d}_i$ values yield a small threshold. When between-location variation is introduced (C2), the difference is larger at small $m$ but generally decreases steadily as $m$ grows. The full conditional and marginal posterior curves substantially overlap under C2, although the conditional differences tend to be larger at small $m$ when $\rho$ is large. The two vertical reference lines show that the averaged approximate thresholds are at least as large as the averaged exact thresholds across the nine settings. The thresholds are most demanding when $\rho$ is large and $\kappa$ is small (bottom-left panel, rounded means $m^*_{\mathrm{approx}}=1187$ and $m^*_{\mathrm{exact}}=920$); conversely, when $\kappa$ is large (right column), the within-area likelihood dominates even at small $m$ and the thresholds reduce to ranging from $2$ to $4$. Under C3, substantial differences persist at large $m$, particularly when spatial correlation is moderate or strong, indicating that a spatial model remains necessary when the covariate is entirely a location-level attribute.

\begin{figure}[ht!]
\centering
\includegraphics[scale = 0.5]{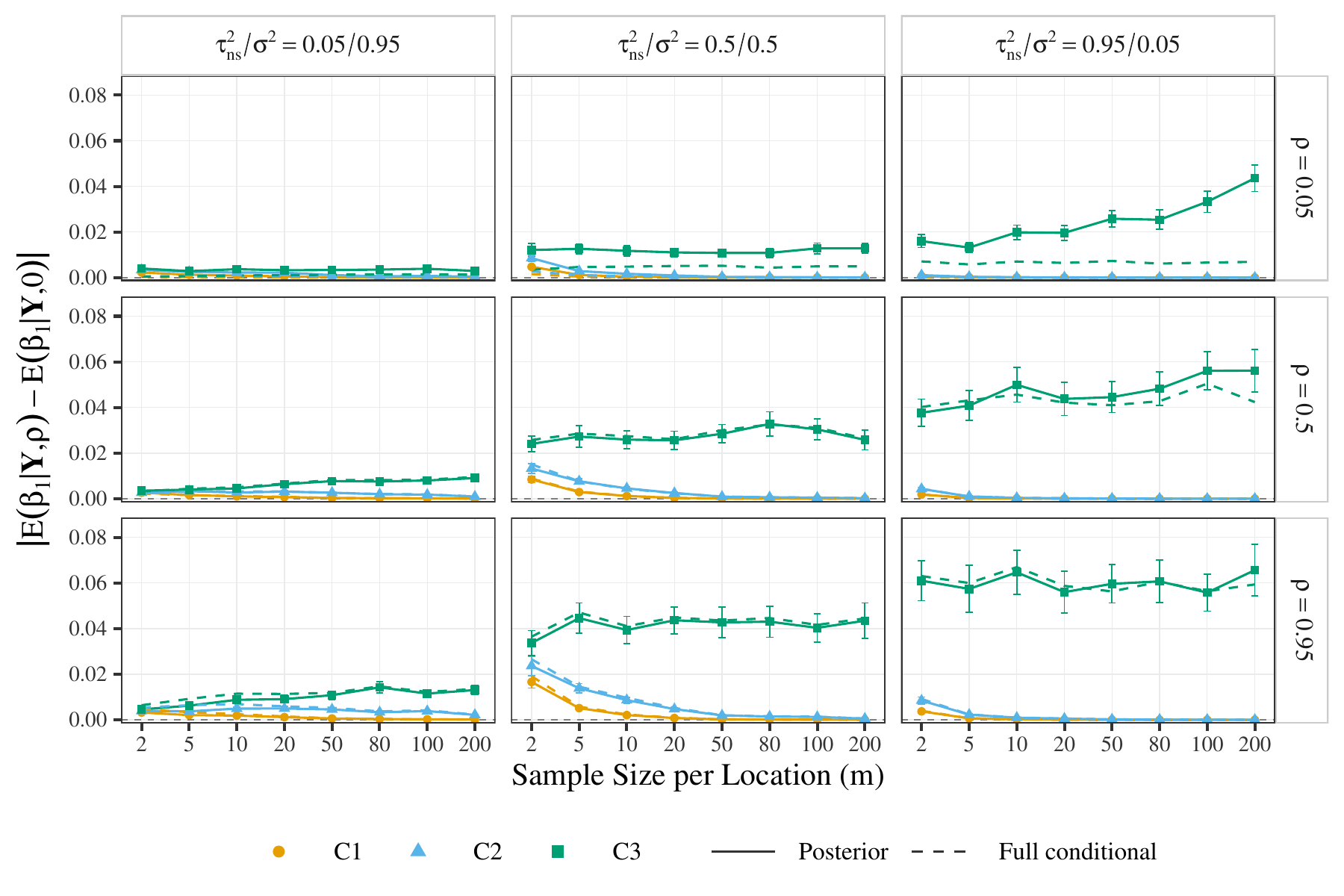}
\caption{Absolute difference in the marginal posterior mean of $\beta_1$ between the spatial (i.e., $\text{E}\left(\beta_1 | \boldsymbol{Y}, \rho\right)$) and nonspatial (i.e., $\text{E}\left(\beta_1 | \boldsymbol{Y}, 0\right)$) models, averaged over $100$ datasets, for $n = 100$ spatial units. Columns correspond to the variance ratio $\kappa = \tau^2_{ns}/\sigma^2$ and rows to the spatial correlation $\rho$. The three colors represent the three covariate structure settings, C1-C3, described in Table~\ref{tab:sim_settings}. Solid and dashed curves show results based on the marginal posterior inference and exact full conditional comparisons, respectively. Error bars are pointwise $95\%$ Monte Carlo confidence intervals for the marginal posterior averages across datasets.}
\label{fig:abs_mean}
\end{figure}

Under C2, the bottom-left panel shows an initial rise where $\rho$ is large and $\kappa$ is small. With other inputs fixed, the full conditional relative variance difference depends on $m$ through $m\kappa$. For sufficiently small $m\kappa$, both precisions have the same leading term $nm/\sigma^2$, giving a small relative difference. Increasing $m$ reduces the observation-level variance $\sigma^2/m$ in the location means, making the models' different between-location contributions more apparent. These contributions remain bounded, while the shared within-location precision $m(n-\text{d}_{\cdot})/\sigma^2$ grows linearly, giving the eventual $O(m^{-1})$ decline (Figure~\ref{fig:nonmonotone_fixed_map}).

Figure~\ref{fig:abs_mean} shows the absolute difference in marginal posterior means of $\beta_1$ between the two models, averaged over $100$ simulations. The corresponding full conditional mean differences, calculated using~\eqref{eq:mean} and~\eqref{eq:mean0} at the generating covariance parameters and averaged over the same datasets, are shown for comparison. Under C1 and C2, the average difference is close to zero across all settings. In C3, the discrepancy is larger and does not diminish as $m$ increases, particularly at high $\rho$ and $\kappa$. When the covariate has no within-location variation, estimation of $\beta_1$ relies entirely on between-location comparisons, making the assumed covariance structure of the random intercepts consequential for the posterior mean. The close agreement between the full conditional and marginal posterior comparisons in Figures~\ref{fig:rel_var} and~\ref{fig:abs_mean} shows that the theoretical patterns remain evident after accounting for uncertainty in the covariance parameters.

Both models showed estimated bias close to zero for $\beta_1$ across the settings considered; bias and empirical coverage of the $95\%$ credible intervals are reported in Appendix~\ref{sec:bias_coverage}. Additional Gaussian simulations with unequal within-area sample sizes showed broadly similar patterns to those under balanced replication. When simulating with two covariates, small posterior variance differences for a C1 or C2 coefficient could coexist with substantial differences for a C3 coefficient, indicating that agreement can differ between coefficients within the same model. Results for Bernoulli and negative binomial outcomes are qualitatively consistent with the Gaussian findings: differences in marginal posterior means and variances generally decrease with replication under C1 and C2, while substantial differences persist under C3 (Appendix~\ref{sec:supp_additional_settings}).

\section{Application with unknown covariance parameters}
\label{sec:practice}

\subsection{Two-stage estimation}
\label{sec:practice_estimation}
Using the replication findings of Section~\ref{sec:spatial} requires $\sigma^2$, $\tau^2_{ns}$, and $\rho$, which are unknown in practice. We therefore propose a two-stage procedure that estimates them without fitting the full spatial model or requiring a Bayesian analysis at the preliminary stage. This is important because fitting the spatial model to obtain these parameters would defeat the purpose of the threshold, which is to determine whether spatial modeling is necessary in the first place.

To do so, the estimated random intercepts from a non-Bayesian nonspatial fit are treated as noisy realizations of the true, spatially correlated random intercepts. Under the spatial model, $\boldsymbol{\theta}\mid\tau^2_{sp},\rho\sim\mathcal{N}_n(\boldsymbol{0}_n,\tau^2_{sp}Q(\rho)^{-1})$, and if $\boldsymbol{\theta}$ were observed, we could estimate $\rho$ by maximizing this density, with $\widehat{\tau}^2_{sp}$ determined for each $\rho$ by $\widehat{\tau}_{ns}^2$ through~\eqref{eq:spatial_variance_scale}. We instead use the estimates from the nonspatial fit, $\widehat{\boldsymbol{\theta}}$, to carry out the estimation. With the first-stage variance parameter estimates held fixed, these estimates are linear in $\boldsymbol{Y}$, giving a Gaussian working distribution whose covariance accounts for shrinkage, observation noise, and regression-estimation error, and depends on $\rho$ through $Q\left(\rho\right)$. Maximizing the corresponding likelihood gives the estimator $\widehat{\rho}$.

Specifically, in the first stage, we fit the nonspatial model of Section~\ref{sec:model}, $\boldsymbol{Y}=\beta_0\boldsymbol{1}_{\sum_i m_i}+\beta_1\textbf{x}+\text{Z}\boldsymbol{\theta}+\boldsymbol{\varepsilon}$ with $\boldsymbol{\theta}\mid\tau^2_{ns}\sim\mathcal{N}_n(\boldsymbol{0}_n,\tau^2_{ns}I_n)$, by restricted maximum likelihood using \texttt{lmer} in the \texttt{lme4} R package \citep{bates2015fitting}. We allow the within-area sample sizes $m_i$ to differ here. This gives $\widehat{\sigma}^2$, $\widehat{\tau}_{ns}^2$, $\widehat{\beta}_0$, $\widehat{\beta}_1$, and $\widehat{\theta}_i$ for all $i$.

In the second stage, $\rho$ is the only remaining covariance parameter to estimate, since $\widehat{\tau}_{sp}^2(\rho)$ is determined by $\widehat{\tau}_{ns}^2$ through~\eqref{eq:spatial_variance_scale}. To estimate $\rho$, we use the joint working distribution of the estimated random intercepts under the spatial model, \begin{equation*} \widehat{\boldsymbol{\theta}} \sim \mathcal{N}_n\left( \boldsymbol{0}_n,\widehat{\Sigma}_{\theta}(\rho) \right), \end{equation*} with covariance \begin{equation} \begin{aligned} \widehat{\Sigma}_{\theta}(\rho) &=(\widehat{\tau}^2_{ns})^2\,\text{Z}^{\top}\left\{\widehat{\Omega}_{ns}^{-1}-\widehat{\Omega}_{ns}^{-1}\text{X}(\text{X}^{\top}\widehat{\Omega}_{ns}^{-1}\text{X})^{-1}\text{X}^{\top}\widehat{\Omega}_{ns}^{-1}\right\}
\left\{\widehat{\tau}^2_{sp}(\rho)\text{Z}Q(\rho)^{-1}\text{Z}^{\top}+\widehat{\sigma}^2I\right\}\\
&\qquad\times\left\{\widehat{\Omega}_{ns}^{-1}-\widehat{\Omega}_{ns}^{-1}\text{X}(\text{X}^{\top}\widehat{\Omega}_{ns}^{-1}\text{X})^{-1}\text{X}^{\top}\widehat{\Omega}_{ns}^{-1}\right\}\text{Z},\end{aligned}\label{eq:two_stage_covariance}\end{equation}
where $\text{X}=[\boldsymbol{1}_{\sum_i m_i},\textbf{x}]$ and $\widehat{\Omega}_{ns}=\widehat{\tau}^2_{ns}\text{Z}\text{Z}^{\top}+\widehat{\sigma}^2I$ is the fitted nonspatial observation covariance. We obtain $\widehat{\rho}$ by maximizing the corresponding Gaussian working likelihood over $0\leq\rho<1$.

\begin{figure}[h!]
\centering
\includegraphics[width=\linewidth]{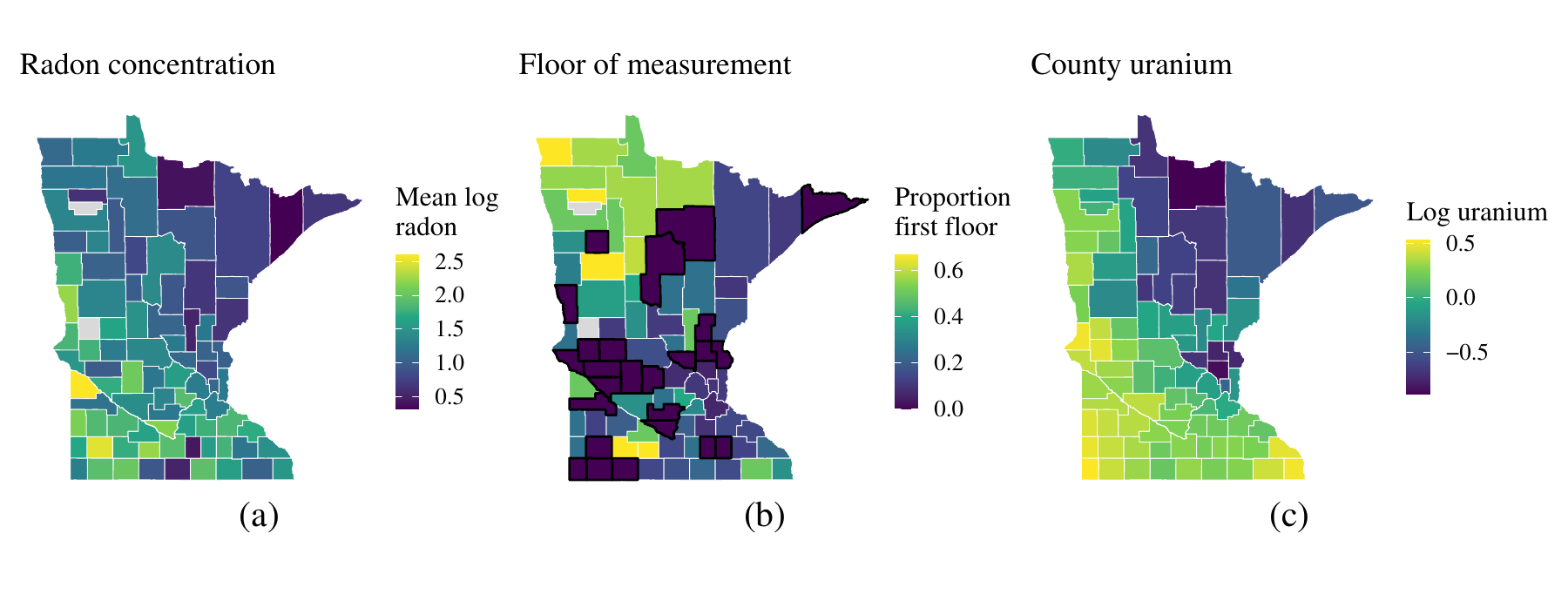}
\caption{Minnesota radon data. (a) Mean log radon concentration of the measured homes in each county. (b) Proportion of measurements taken on the first floor in each county. (c) County-level log uranium. Grey counties in (a) and (b) have no measurements.}
\label{fig:radon}
\end{figure}

With data in hand, we plug the resulting estimates into the full conditional distribution variance of $\beta_1$ under each model at the observed design and compare the absolute relative difference~\eqref{eq:abs_diff} with $\gamma$. This calculation accommodates unequal within-area sample sizes, as illustrated for the radon data in Section~\ref{sec:radon}. For study planning under balanced replication, the same estimates can be used as inputs to calculate the thresholds of Section~\ref{sec:threshold}. Appendix~\ref{sec:supp_practice} derives~\eqref{eq:two_stage_covariance} and gives the computational details, and evaluates the accuracy of the resulting thresholds by comparison with those computed from the generating parameters. The procedure yields accurate threshold estimates on average across most settings, with reduced accuracy mainly when the preliminary within-area sample size and $\kappa$ are both small.

\subsection{Case study illustration}
\label{sec:radon}

We illustrate the procedure using the Minnesota radon dataset \citep{price1996bayesian, gelman2007data}, which contains indoor radon measurements from $919$ homes in $85$ counties, with between $1$ and $116$ homes per county. We model log radon concentration as $Y_{ij}=\beta_0+\beta_1\text{x}_{1ij}+\beta_2\text{x}_{2i}+\theta_i+\varepsilon_{ij}$, where $\text{x}_{1ij}$ indicates whether the measurement was taken on the first floor rather than in the basement, and $\text{x}_{2i}$ is county-level log uranium. Both predictors are centered across homes. The two covariates fall on opposite sides of the distinction in Section~\ref{sec:spatial}. Floor is recorded for each home, and $83\%$ of its variation lies within counties. Uranium is constant within counties (C3) and varies smoothly across the state (Figure~\ref{fig:radon}). We exclude three zero radon readings before taking logarithms, leaving $916$ homes in the analysis.

\begin{table}[ht]
    \centering
    \caption{Covariance parameter estimates from the two-stage procedure and posterior medians ($95\%$ credible intervals) from the fully Bayesian model fits.}
    \label{tab:radon_covariance}
    \begin{tabular}{lccc}
        \toprule
Parameter & Two-stage & Nonspatial & Spatial \\
$\sigma^2$    & 0.54 & 0.54 (0.49, 0.60) & 0.54 (0.49, 0.59) \\
$\tau_{ns}^2$ & 0.03 & 0.03 (0.01, 0.07) &  \\
$\tau_{sp}^2$ & 0.05 &  & 0.07 (0.02, 0.18) \\
$\rho$        & 0.87 &  & 0.71 (0.10, 0.99) \\
\bottomrule
    \end{tabular}
\end{table}

Using the two-stage covariance estimates, we evaluate the full conditional variance of each coefficient under each model at the observed design and calculate the absolute relative differences in~\eqref{eq:abs_diff}. These are $2.31\%$ for floor and $49.57\%$ for uranium. We then fit the spatial and nonspatial models by MCMC as in Section~\ref{sec:simulation} and compare their marginal posterior variances, obtaining differences of $2.90\%$ and $93.27\%$, respectively. Both comparisons therefore indicate variance agreement for floor at tolerance $\gamma=0.05$ and a substantial difference for uranium. Table~\ref{tab:radon_covariance} compares the two-stage estimates with posterior summaries from the full fits. The residual variance estimates agree closely, and all two-stage estimates lie within the corresponding $95\%$ credible intervals. 

\begin{figure}[ht!]
\centering
\includegraphics[width=0.9\linewidth]{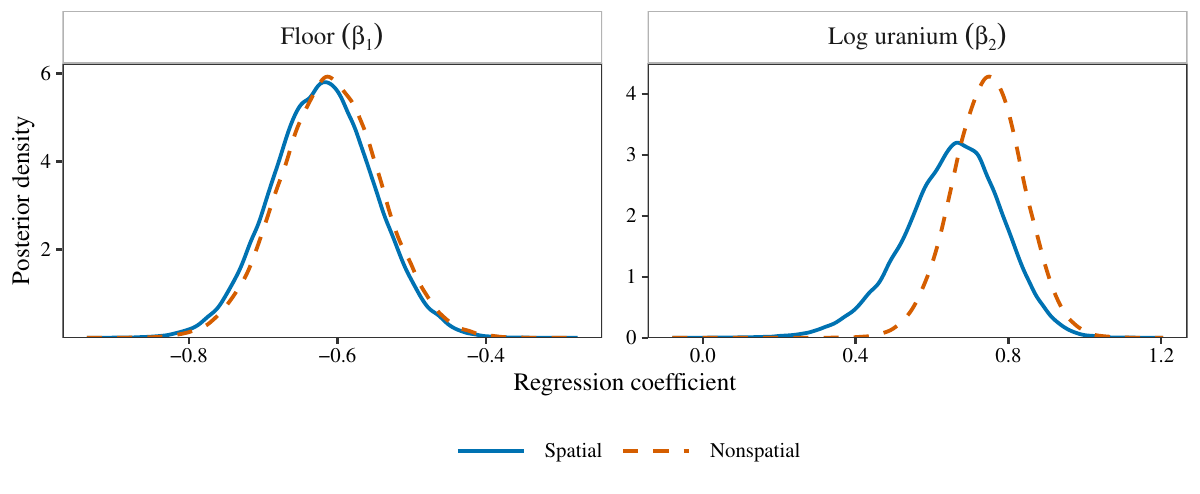}
\caption{Marginal posterior densities for the floor coefficient $\beta_1$ and log uranium coefficient $\beta_2$ in the Minnesota radon analysis. Solid blue and dashed orange curves show the spatial and nonspatial models, respectively.}
\label{fig:radon_posterior}
\end{figure}

Figure~\ref{fig:radon_posterior} shows nearly identical posterior densities for floor but clear differences for uranium. Using only the nonspatial fit, the two-stage criterion correctly anticipated this contrast in posterior variance, with uranium illustrating the persistent differences expected for a covariate with no within-county variation (C3).

\section{Discussion} \label{sec:discussion}
The criterion developed here can inform both study design and analysis. At the design stage, investigators can examine the required replication using pilot data or existing knowledge. At the analysis stage, the two-stage procedure uses a preliminary nonspatial fit to assess model agreement at the observed design. The estimates introduce uncertainty into the assessment, particularly when the preliminary sample size and variance ratio are small. In such settings, we recommend evaluating the criterion over plausible covariance parameter values to assess whether the conclusion is sensitive to the preliminary estimates. The criterion concerns regression inference; if the goal is spatial prediction (i.e., inferring outcomes at unobserved or sparsely observed locations, where information must be borrowed from neighbors rather than drawn from within-area replication), spatial structure remains important regardless of replication.

Developing an area-specific threshold that accounts for heterogeneous sample sizes would be useful for applications where some areas are data-rich and others sparse. This raises further questions about how to quantify increasing replication and when restrictions on the degree of imbalance are needed.

Deriving analogous results for count outcomes modeled using Poisson, negative binomial, or binomial likelihoods is a natural next step. While auxiliary variable approaches exist for some of these likelihoods that yield conjugacy under carefully chosen priors \citep{holmes2006bayesian,polson2013bayesian}, the auxiliary variables themselves are often conditioned on in the derivations, making a useful closed-form solution difficult to obtain. Spatio-temporal extensions, where replication arises across both space and time and the model parameters may vary temporally, are another important direction given the increasing availability of longitudinal spatially referenced datasets.

By making the role of within-area replication explicit, our derivations also provide a starting point for investigating whether replication can mitigate bias from unmeasured area-level covariates. Understanding when this may occur, and its implications for interval coverage under spatial confounding, is an important direction for future work.

{
\bibliographystyle{informs2014}
\bibliography{reference}
}

\newpage
\setcounter{page}{1}
\begin{APPENDICES}
\renewcommand{\theHsection}{appendix.\Alph{section}}
\renewcommand{\theHsubsection}{appendix.\Alph{section}.\arabic{subsection}}
\renewcommand{\thefigure}{\thesection.\arabic{figure}}
\renewcommand{\thetable}{\thesection.\arabic{table}}
\renewcommand{\theequation}{\thesection.\arabic{equation}}
\makeatletter
\@addtoreset{figure}{section}
\@addtoreset{table}{section}
\@addtoreset{equation}{section}
\makeatother
    \section{Full conditional distributions}\label{sec:supp_fc}
In this section, we provide derivation details for the full conditional distributions used to fit the model from the main text via Markov chain Monte Carlo (MCMC) sampling \citep{metropolis1953equation, geman1984stochastic, gelfand1990sampling}. Derivations are based on the conditional likelihood (i.e., treating $\boldsymbol{\theta}$ as part of the parameter space), which is the version used to fit both the spatial and nonspatial models in the simulation study in Section~\ref{sec:simulation}. We additionally derive the full conditional distribution for $\boldsymbol{\beta}$ based on the marginal likelihood (i.e., after integrating $\boldsymbol{\theta}$ out), which is the starting point for the derivations in Section~\ref{sec:spatial}.

\subsection{Regression coefficients \texorpdfstring{$\protect\boldsymbol{\beta}$}{beta}}
\label{sec:supp_fc_beta}
Under the conditional likelihood, let $\textbf{r} = \boldsymbol{Y} - \text{Z}\boldsymbol{\theta}$. Write $\text{X}=[\boldsymbol{1}_{nm},\textbf{x}]$ and $\boldsymbol{\beta}=(\beta_0,\beta_1)^\top$. Because $f\left(\boldsymbol{\beta}\right) \propto 1$, the full conditional distribution of $\boldsymbol{\beta}$ is proportional to the likelihood contribution from the observed data,
\begin{align*}
f\left(\boldsymbol{\beta} \mid \text{rest}\right)
&\propto \exp\left\{-\frac{1}{2\sigma^2}\left(\textbf{r} - \text{X}\boldsymbol{\beta}\right)^\top\left(\textbf{r} - \text{X}\boldsymbol{\beta}\right)\right\}\\
&\propto \exp\left\{-\frac{1}{2\sigma^2}\left(\boldsymbol{\beta}^\top \text{X}^\top \text{X} \boldsymbol{\beta} - 2\boldsymbol{\beta}^\top \text{X}^\top \textbf{r}\right)\right\}\\
&\propto \exp\left\{-\frac{1}{2}\left(\boldsymbol{\beta} - \hat{\boldsymbol{\beta}}\right)^\top \frac{\text{X}^\top \text{X}}{\sigma^2} \left(\boldsymbol{\beta} - \hat{\boldsymbol{\beta}}\right)\right\},
\end{align*}
where $\hat{\boldsymbol{\beta}} = \left(\text{X}^\top \text{X}\right)^{-1}\text{X}^\top \textbf{r}$. Therefore,
\begin{equation*}
\boldsymbol{\beta} \mid \text{rest} \sim \mathcal{N}_2\left(\left(\text{X}^\top \text{X}\right)^{-1} \text{X}^\top \left(\boldsymbol{Y} - \text{Z}\boldsymbol{\theta}\right),\; \sigma^2 \left(\text{X}^\top \text{X}\right)^{-1}\right).
\end{equation*}

After integrating $\boldsymbol{\theta}$ out of the model, the marginal likelihood is $\boldsymbol{Y} \mid \boldsymbol{\beta}, \sigma^2, \tau^2_{sp}, \rho \sim \mathcal{N}_{nm}\left(\text{X}\boldsymbol{\beta}, \Omega\right)$, where $\Omega = \tau^2_{sp} \text{Z} Q\left(\rho\right)^{-1} \text{Z}^\top + \sigma^2 I_{nm}$. The full conditional distribution of $\boldsymbol{\beta}$ satisfies
\begin{align*}
f\left(\boldsymbol{\beta} \mid \text{rest}\right)
&\propto \exp\left\{-\frac{1}{2}\left(\boldsymbol{Y} - \text{X}\boldsymbol{\beta}\right)^\top \Omega^{-1} \left(\boldsymbol{Y} - \text{X}\boldsymbol{\beta}\right)\right\}\\
&\propto \exp\left\{-\frac{1}{2}\left(\boldsymbol{\beta}^\top \text{X}^\top \Omega^{-1} \text{X} \boldsymbol{\beta} - 2\boldsymbol{\beta}^\top \text{X}^\top \Omega^{-1} \boldsymbol{Y}\right)\right\}\\
&\propto \exp\left\{-\frac{1}{2}\left(\boldsymbol{\beta} - \tilde{\boldsymbol{\beta}}\right)^\top \text{X}^\top \Omega^{-1} \text{X} \left(\boldsymbol{\beta} - \tilde{\boldsymbol{\beta}}\right)\right\},
\end{align*}
where $\tilde{\boldsymbol{\beta}} = \left(\text{X}^\top \Omega^{-1} \text{X}\right)^{-1}\text{X}^\top \Omega^{-1} \boldsymbol{Y}$. Therefore,
\begin{equation*}
\boldsymbol{\beta} \mid \text{rest} \sim \mathcal{N}_2\left(\left(\text{X}^\top \Omega^{-1} \text{X}\right)^{-1}\text{X}^\top \Omega^{-1} \boldsymbol{Y},\; \left(\text{X}^\top \Omega^{-1} \text{X}\right)^{-1}\right).
\end{equation*}
In particular, the full conditional of $\beta_1$ conditional on $\beta_0$ is
\begin{equation*}
\beta_1 \mid \boldsymbol{Y}, \beta_0, \sigma^2, \tau^2_{sp}, \rho \sim \mathcal{N}\left(\mu_{\beta_1},\; \sigma^2_{\beta_1}\right),
\end{equation*}
where
\begin{equation*}
\sigma^2_{\beta_1} = \frac{1}{\textbf{x}^\top \Omega^{-1} \textbf{x}}, \quad \mu_{\beta_1} = \sigma^2_{\beta_1}\, \textbf{x}^\top \Omega^{-1}\left(\boldsymbol{Y} - \beta_0 \boldsymbol{1}_{nm}\right).
\end{equation*}

\subsection{Spatial random effects \texorpdfstring{$\protect\boldsymbol{\theta}$}{theta}}
Under the Leroux CAR prior, the full conditional density of $\theta_i$ satisfies
\begin{equation*}
f\left(\theta_i \mid \text{rest}\right) \propto f\left(\boldsymbol{Y} \mid \boldsymbol{\beta}, \boldsymbol{\theta}, \sigma^2\right) f\left(\theta_i \mid \boldsymbol{\theta}_{-i}, \tau^2_{sp}, \rho\right),
\end{equation*}
for $i = 1, \ldots, n$. Define $\bar{\text{r}}_{i\cdot} = m^{-1}\sum_{j=1}^m \left(Y_{ij} - \beta_0 - \beta_1 \text{x}_{ij}\right)$. The observations within spatial unit $i$ contribute
\begin{align*}
L_i\left(\theta_i\right)
&\propto \exp\left\{-\frac{1}{2\sigma^2}\sum_{j=1}^m \left(Y_{ij} - \beta_0 - \beta_1 \text{x}_{ij} - \theta_i\right)^2\right\}\\
&= \exp\left\{-\frac{1}{2\sigma^2}\left[\sum_{j=1}^m \left(Y_{ij} - \beta_0 - \beta_1 \text{x}_{ij} - \bar{\text{r}}_{i\cdot}\right)^2 + m\left(\bar{\text{r}}_{i\cdot} - \theta_i\right)^2\right]\right\}\\
&\propto \exp\left\{-\frac{m}{2\sigma^2}\left(\theta_i - \bar{\text{r}}_{i\cdot}\right)^2\right\},
\end{align*}
where the cross term vanishes because $\sum_{j=1}^m \left(Y_{ij} - \beta_0 - \beta_1 \text{x}_{ij} - \bar{\text{r}}_{i\cdot}\right) = 0$ by definition of $\bar{\text{r}}_{i\cdot}$. Therefore,
\begin{equation*}
\theta_i \mid \text{rest} \sim \mathcal{N}\left(\mu_i^*,\; \sigma_i^{2*}\right),
\end{equation*}
where
\begin{equation*}
\sigma_i^{2*} = \left(\frac{m}{\sigma^2} + \frac{\left(1-\rho\right) + \rho \text{w}_{i\cdot}}{\tau^2_{sp}}\right)^{-1}, \quad \mu_i^* = \sigma_i^{2*}\left(\frac{m}{\sigma^2}\bar{\text{r}}_{i.} + \frac{\rho \sum_{i'=1}^n \text{w}_{ii'}\theta_{i'}}{\tau^2_{sp}}\right).
\end{equation*}
Under the nonspatial model ($\rho = 0$), the prior reduces to $\theta_i \mid \tau^2_{ns} \overset{\text{iid}}{\sim} \mathcal{N}\left(0, \tau^2_{ns}\right)$, and the full conditional simplifies to
\begin{equation*}
\theta_i \mid \text{rest} \sim \mathcal{N}\left(\sigma_{i,0}^{2*}\cdot \frac{m\bar{\text{r}}_{i\cdot}}{\sigma^2},\; \sigma_{i,0}^{2*}\right), \quad \sigma_{i,0}^{2*} = \left(\frac{m}{\sigma^2} + \frac{1}{\tau^2_{ns}}\right)^{-1}.
\end{equation*}

\subsection{Error variance \texorpdfstring{$\sigma^2$}{sigma squared}}
\label{sec:supp_fc_sigma}
Under the prior $\sigma^2 \sim \text{IG}\left(a, b\right)$, the full conditional density of $\sigma^2$ satisfies 
\begin{align*} 
f\left(\sigma^2 \mid \text{rest}\right) 
&\propto f\left(\boldsymbol{Y} \mid \boldsymbol{\beta}, \boldsymbol{\theta}, \sigma^2\right) f\left(\sigma^2\right) \\ 
&\propto \left(\sigma^2\right)^{-nm/2} \exp\left\{-\frac{1}{2\sigma^2}\sum_{i=1}^n\sum_{j=1}^m \left(Y_{ij} - \beta_0 - \beta_1 \text{x}_{ij} - \theta_i\right)^2\right\} \left(\sigma^2\right)^{-\left(a+1\right)} \exp\left\{-\frac{b}{\sigma^2}\right\} \\ 
&= \left(\sigma^2\right)^{-\left(a + nm/2 + 1\right)} \exp\left\{-\frac{1}{\sigma^2}\left[b + \frac{1}{2}\sum_{i=1}^n\sum_{j=1}^m \left(Y_{ij} - \beta_0 - \beta_1 \text{x}_{ij} - \theta_i\right)^2\right]\right\}. \end{align*} 
Therefore, 
\begin{equation*} 
\sigma^2 \mid \text{rest} \sim \text{IG}\left(a + \frac{nm}{2},\; b + \frac{1}{2}\sum_{i=1}^n\sum_{j=1}^m \left(Y_{ij} - \beta_0 - \beta_1 \text{x}_{ij} - \theta_i\right)^2\right).
\end{equation*}
With $a = b = 0.01$, we have 
\begin{equation*} 
\sigma^2 \mid \text{rest} \sim \text{IG}\left(0.01 + \frac{nm}{2},\; 0.01 + \frac{1}{2}\sum_{i=1}^n\sum_{j=1}^m \left(Y_{ij} - \beta_0 - \beta_1 \text{x}_{ij} - \theta_i\right)^2\right). 
\end{equation*}

\subsection{Spatial variance \texorpdfstring{$\tau^2_{sp}$}{tau sp squared}}
\label{sec:supp_fc_tau}
Because $|Q\left(\rho\right)|$ does not depend on $\tau^2_{sp}$, under the prior $\tau^2_{sp} \sim \text{IG}\left(a, b\right)$ the full conditional density of $\tau^2_{sp}$ satisfies
\begin{align*}
f\left(\tau^2_{sp} \mid \text{rest}\right)
&\propto f\left(\boldsymbol{\theta} \mid \tau^2_{sp}, \rho\right) f\left(\tau^2_{sp}\right)\\
&\propto \left(\tau^2_{sp}\right)^{-n/2} \exp\left\{-\frac{\boldsymbol{\theta}^\top Q\left(\rho\right) \boldsymbol{\theta}}{2\tau^2_{sp}}\right\} \left(\tau^2_{sp}\right)^{-\left(a+1\right)} \exp\left\{-\frac{b}{\tau^2_{sp}}\right\}\\
&= \left(\tau^2_{sp}\right)^{-\left(a + n/2 + 1\right)} \exp\left\{-\frac{1}{\tau^2_{sp}}\left(b + \frac{\boldsymbol{\theta}^\top Q\left(\rho\right) \boldsymbol{\theta}}{2}\right)\right\}.
\end{align*}
Therefore,
\begin{equation*}
\tau^2_{sp} \mid \text{rest} \sim \text{IG}\left(a + \frac{n}{2},\; b + \frac{1}{2}\boldsymbol{\theta}^\top Q\left(\rho\right) \boldsymbol{\theta}\right).
\end{equation*}
With $a = b = 0.01$, we have
\begin{equation*}
\tau^2_{sp} \mid \text{rest} \sim \text{IG}\left(0.01 + \frac{n}{2},\; 0.01 + \frac{1}{2}\boldsymbol{\theta}^\top Q\left(\rho\right) \boldsymbol{\theta}\right).
\end{equation*}
Under the nonspatial model, the corresponding full conditional distribution is
\begin{equation*}
\tau^2_{ns} \mid \text{rest} \sim \text{IG}\left(a + \frac{n}{2},\; b + \frac{1}{2}\boldsymbol{\theta}^\top\boldsymbol{\theta}\right).
\end{equation*}
With $a = b = 0.01$, we have
\begin{equation*}
\tau^2_{ns} \mid \text{rest} \sim \text{IG}\left(0.01 + \frac{n}{2},\; 0.01 + \frac{1}{2}\boldsymbol{\theta}^\top\boldsymbol{\theta}\right).
\end{equation*}

\subsection{Spatial correlation parameter \texorpdfstring{$\rho$}{rho}}
Under the prior $\rho \sim \text{Uniform}\left(0, 1\right)$, the full conditional density of $\rho$ satisfies
\begin{align*}
f\left(\rho \mid \text{rest}\right)
&\propto f\left(\boldsymbol{\theta} \mid \tau^2_{sp}, \rho\right)\\
&\propto |Q\left(\rho\right)|^{1/2} \exp\left\{-\frac{1}{2\tau^2_{sp}}\boldsymbol{\theta}^\top Q\left(\rho\right) \boldsymbol{\theta}\right\}.
\end{align*}
Using the spectral decomposition $\text{L} = \text{U} \Lambda \text{U}^\top$ described in Section~\ref{sec:conditional}, and letting $\boldsymbol{\alpha} = \text{U}^\top \boldsymbol{\theta}$,
\begin{equation*}
|Q\left(\rho\right)| = \prod_{i=1}^n \left(\rho\lambda_i + 1 - \rho\right), \quad \boldsymbol{\theta}^\top Q\left(\rho\right) \boldsymbol{\theta} = \sum_{i=1}^n \left(\rho \lambda_i + 1 - \rho\right)\alpha_i^2,
\end{equation*}
such that
\begin{equation*}
\log f\left(\rho \mid \text{rest}\right) = \frac{1}{2}\sum_{i=1}^n \log\left(\rho\lambda_i + 1 - \rho\right) - \frac{1}{2\tau^2_{sp}}\sum_{i=1}^n \left(\rho\lambda_i + 1 - \rho\right)\alpha_i^2 + \text{const}.
\end{equation*}
This does not correspond to a standard distribution, so $\rho$ is updated via a Metropolis step with a Gaussian random walk proposal on the logit scale \citep{gamerman2006markov}. Letting $\psi = \log\left\{\rho/\left(1-\rho\right)\right\}$, a proposal at iteration $t$ is generated as $\psi^* = \psi^{\left(t-1\right)} + \eta$ with $\eta \sim \mathcal{N}\left(0,s^2\right)$, where $s^2$ is a tuning parameter calibrated during burn-in, and the proposal on the original scale is recovered as $\rho^* = 1/\left(1+e^{-\psi^*}\right)$. The proposal is accepted with probability $\min\left(1,R\right)$, where
\begin{equation*}
R = \frac{f\left(\rho^* \mid \text{rest}\right)}{f\left(\rho^{\left(t-1\right)} \mid \text{rest}\right)} \cdot \frac{q\left(\psi^{\left(t-1\right)} \mid \psi^*\right)}{q\left(\psi^* \mid \psi^{\left(t-1\right)}\right)} \cdot \frac{\rho^*\left(1-\rho^*\right)}{\rho^{\left(t-1\right)}\left(1-\rho^{\left(t-1\right)}\right)},
\end{equation*}
and the proposal density ratio $q\left(\psi^{\left(t-1\right)} \mid \psi^*\right)/q\left(\psi^* \mid \psi^{\left(t-1\right)}\right)=1$ by the symmetry of $\eta$.

\clearpage
\section{Numerical calculation of the exact replication threshold with a finite-sample bound on the search limit}
\label{sec:supp_threshold_search}

The exact threshold $m^*_{\mathrm{exact}}$ is computed numerically by evaluating the absolute relative variance difference from the exact precisions at successive integers $m$. To determine when the search can stop, we derive an upper limit beyond which the difference is guaranteed to remain within tolerance. We then evaluate the exact variance difference at every integer from $2$ through this limit and set $m^*_{\mathrm{exact}}$ to one greater than the last integer at which the tolerance is exceeded, or to $2$ if there is no exceedance.

We hold the spatial graph, covariance parameters and covariate projections $\text{d}_i$ fixed as $m$ varies. Subtracting the spatial precision in~\eqref{eq:precision} from the nonspatial precision in~\eqref{eq:precision0} gives
\begin{align*}
\sigma^{-2}_{\beta_1}(0)-\sigma^{-2}_{\beta_1}(\rho)
&=\left\{\frac{nm}{\sigma^2}
-\frac{m^2\tau_{ns}^2\text{d}_{\cdot}}
{\sigma^2(\sigma^2+m\tau_{ns}^2)}\right\}
-\left\{\frac{nm}{\sigma^2}
-\frac{m^2\tau_{sp}^2}{\sigma^2}
\sum_{i=1}^n
\frac{\text{d}_i}
{\sigma^2(\rho\lambda_i+1-\rho)+m\tau_{sp}^2}\right\}\\
&=\frac{m^2}{\sigma^2}\sum_{i=1}^n\text{d}_i
\left\{
\frac{\tau_{sp}^2}
{\sigma^2(\rho\lambda_i+1-\rho)+m\tau_{sp}^2}
-\frac{\tau_{ns}^2}{\sigma^2+m\tau_{ns}^2}
\right\}\\
&=\frac{m^2}{\sigma^2}\sum_{i=1}^n
\frac{\text{d}_i
\left[
\tau_{sp}^2(\sigma^2+m\tau_{ns}^2)
-\tau_{ns}^2\{\sigma^2(\rho\lambda_i+1-\rho)+m\tau_{sp}^2\}
\right]}
{(\sigma^2+m\tau_{ns}^2)
\{\sigma^2(\rho\lambda_i+1-\rho)+m\tau_{sp}^2\}}\\
&=\sum_{i=1}^n
\frac{m^2\text{d}_i
\{\tau_{sp}^2-\tau_{ns}^2(\rho\lambda_i+1-\rho)\}}
{(\sigma^2+m\tau_{ns}^2)
\{\sigma^2(\rho\lambda_i+1-\rho)+m\tau_{sp}^2\}}\\
&=\sum_{i=1}^n
\frac{m^2\text{d}_i
\{\tau_{sp}^2-\tau_{ns}^2(\rho\lambda_i+1-\rho)\}}
{m^2\tau_{ns}^2\tau_{sp}^2
\left(1+\frac{\sigma^2}{m\tau_{ns}^2}\right)
\left\{1+\frac{\sigma^2(\rho\lambda_i+1-\rho)}{m\tau_{sp}^2}\right\}}\\
&=\sum_{i=1}^n
\frac{\text{d}_i
\left\{\frac{1}{\tau_{ns}^2}
-\frac{\rho\lambda_i+1-\rho}{\tau_{sp}^2}\right\}}
{\left(1+\frac{\sigma^2}{m\tau_{ns}^2}\right)
\left\{1+\frac{\sigma^2(\rho\lambda_i+1-\rho)}{m\tau_{sp}^2}\right\}}.
\end{align*}
Both factors in the denominator are at least one, and $\sigma^{-2}_{\beta_1}(\rho)\geq m(n-\text{d}_{\cdot})/\sigma^2$. Therefore, when $n-\text{d}_{\cdot}>0$, the absolute relative variance difference satisfies
\begin{equation}
\frac{\left|\sigma^{-2}_{\beta_1}(0)-\sigma^{-2}_{\beta_1}(\rho)\right|}{\sigma^{-2}_{\beta_1}(\rho)}
\leq\frac{\sigma^2}{m(n-\text{d}_{\cdot})}\sum_{i=1}^{n}\text{d}_i\left|\frac{1}{\tau^2_{ns}}-\frac{\rho\lambda_i+1-\rho}{\tau^2_{sp}}\right|.
\label{eq:supp_variance_bound}
\end{equation}
For a tolerance $\gamma>0$, define
\begin{equation*}
M_{\gamma}=\max\left\{2,\,\left\lceil\frac{\sigma^2}{\gamma(n-\text{d}_{\cdot})}\sum_{i=1}^{n}\text{d}_i\left|\frac{1}{\tau^2_{ns}}-\frac{\rho\lambda_i+1-\rho}{\tau^2_{sp}}\right|\right\rceil\right\}.
\end{equation*}
By~\eqref{eq:supp_variance_bound}, the tolerance holds for every integer $m\geq M_{\gamma}$. We therefore evaluate the exact relative variance difference at each integer $m=2,\ldots,M_{\gamma}$. If none exceeds $\gamma$, then $m^*_{\mathrm{exact}}=2$; otherwise, $m^*_{\mathrm{exact}}$ is one greater than the largest integer at which the difference exceeds $\gamma$. This search is implemented in the R package \texttt{SpThreshold} and yielded finite exact thresholds for all C2 datasets underlying the threshold reference lines in Figure~\ref{fig:rel_var}.

\clearpage
\section{Additional simulation results}
\label{sec:supp_gaussian}

This section provides additional simulation results for $n = 25$ and $n = 400$ spatial units. The main text presents results for $n = 100$; the figures below show that the findings are consistent across different numbers of spatial units. We additionally examine the full conditional relative variance difference over a wider range of $m$ for a fixed spatial map and covariate structure, and report estimated bias and empirical coverage of the $95\%$ credible intervals.

\clearpage 
\subsection{Results for \texorpdfstring{$n = 25$}{n = 25}}

\begin{figure}[htbp]
\centering
\includegraphics[width=\linewidth]{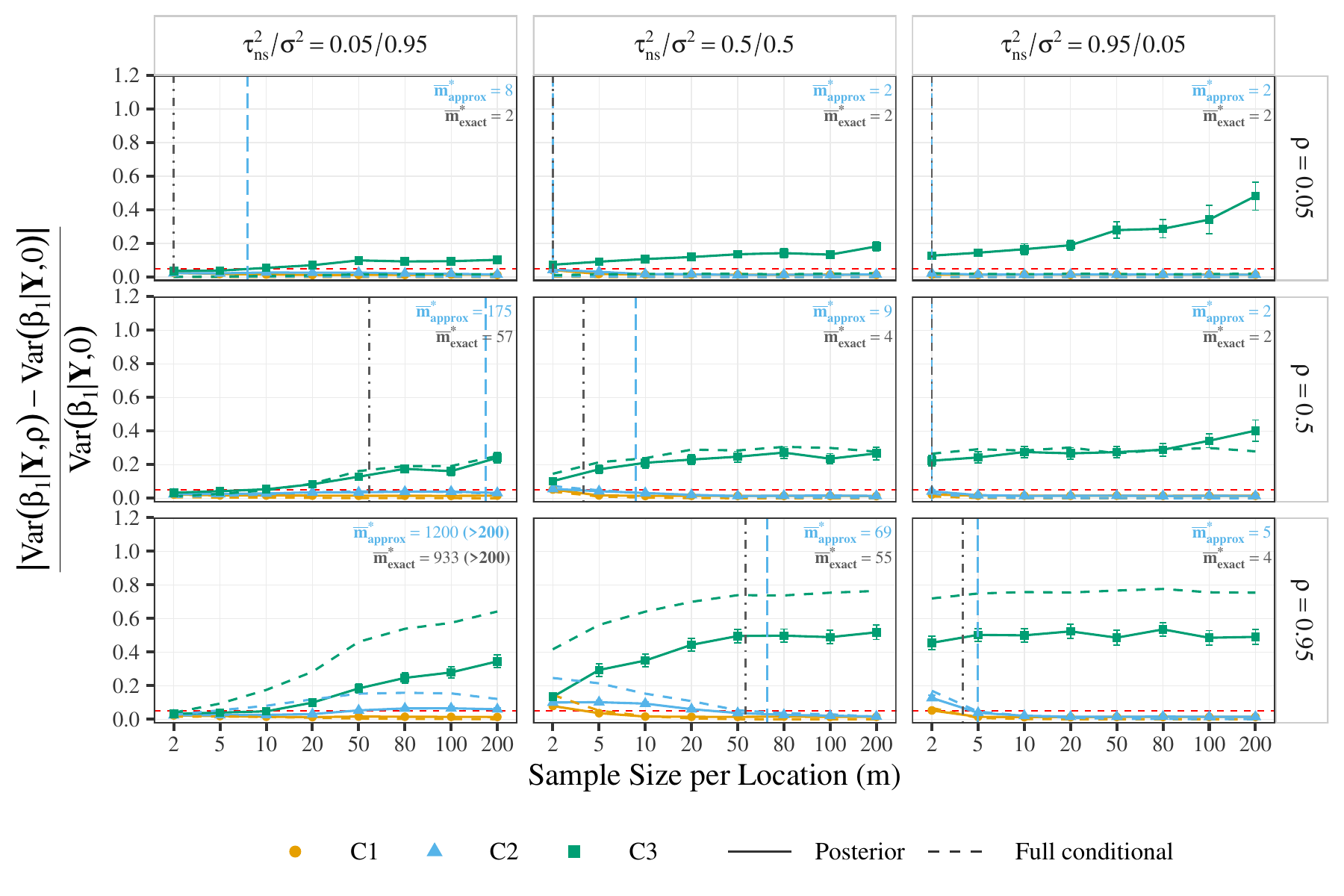}
\caption{Absolute relative difference in the marginal posterior variance of $\beta_1$ between the spatial (i.e., $\text{Var}\left(\beta_1 | \boldsymbol{Y}, \rho\right)$) and nonspatial (i.e., $\text{Var}\left(\beta_1 | \boldsymbol{Y}, 0\right)$) models, averaged over $100$ datasets, for $n = 25$ spatial units. Columns correspond to the variance ratio $\kappa = \tau^2_{ns}/\sigma^2$ and rows to the spatial correlation $\rho$. The three colors represent the three covariate structures, C1-C3, described in Table~\ref{tab:sim_settings}. Solid and dashed curves show marginal posterior inference and exact full conditional comparisons, respectively. The dashed red horizontal line marks the tolerance $\gamma = 0.05$; blue and grey vertical lines mark the approximate and exact C2 thresholds, respectively, each averaged over $100$ datasets and rounded up to an integer. Error bars are pointwise $95\%$ Monte Carlo confidence intervals for the marginal averages across datasets.}
\label{fig:plot1_J25}
\end{figure}

\clearpage 
\begin{figure}[htbp]
\centering
\includegraphics[width=\linewidth]{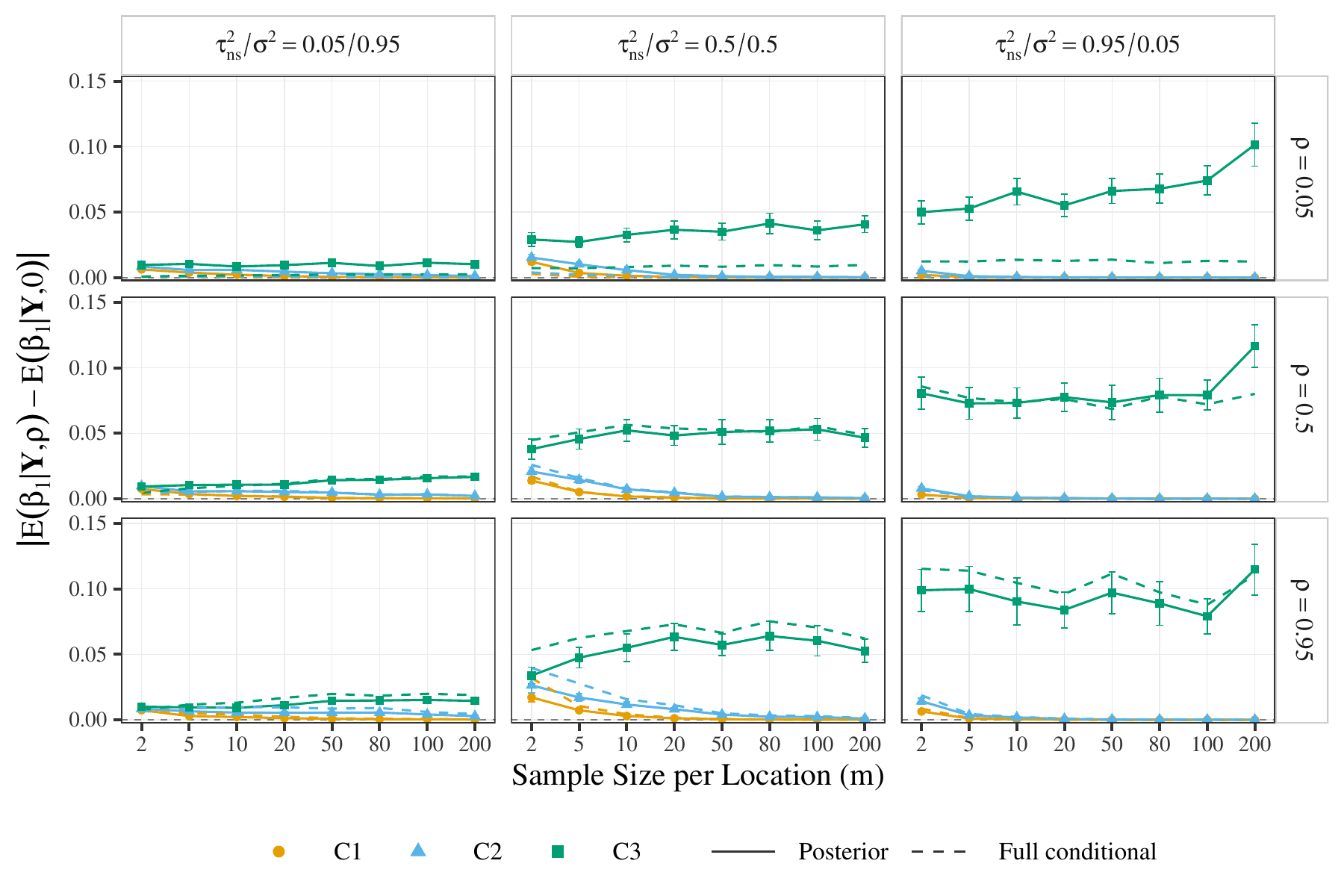}
\caption{Absolute difference in the marginal posterior mean of $\beta_1$ between the spatial (i.e., $\text{E}\left(\beta_1 | \boldsymbol{Y}, \rho\right)$) and nonspatial (i.e., $\text{E}\left(\beta_1 | \boldsymbol{Y}, 0\right)$) models, averaged over $100$ datasets, for $n = 25$ spatial units. Columns correspond to the variance ratio $\kappa = \tau^2_{ns}/\sigma^2$ and rows to the spatial correlation $\rho$. The three colors represent the three covariate structure settings, C1-C3, described in Table~\ref{tab:sim_settings}. Solid and dashed curves show marginal posterior inference and exact full conditional comparisons, respectively. Error bars are pointwise $95\%$ Monte Carlo confidence intervals across datasets.}
\label{fig:plot2_J25}
\end{figure}

\clearpage 
\subsection{Results for \texorpdfstring{$n = 400$}{n = 400}}

\begin{figure}[htbp]
\centering
\includegraphics[width=\linewidth]{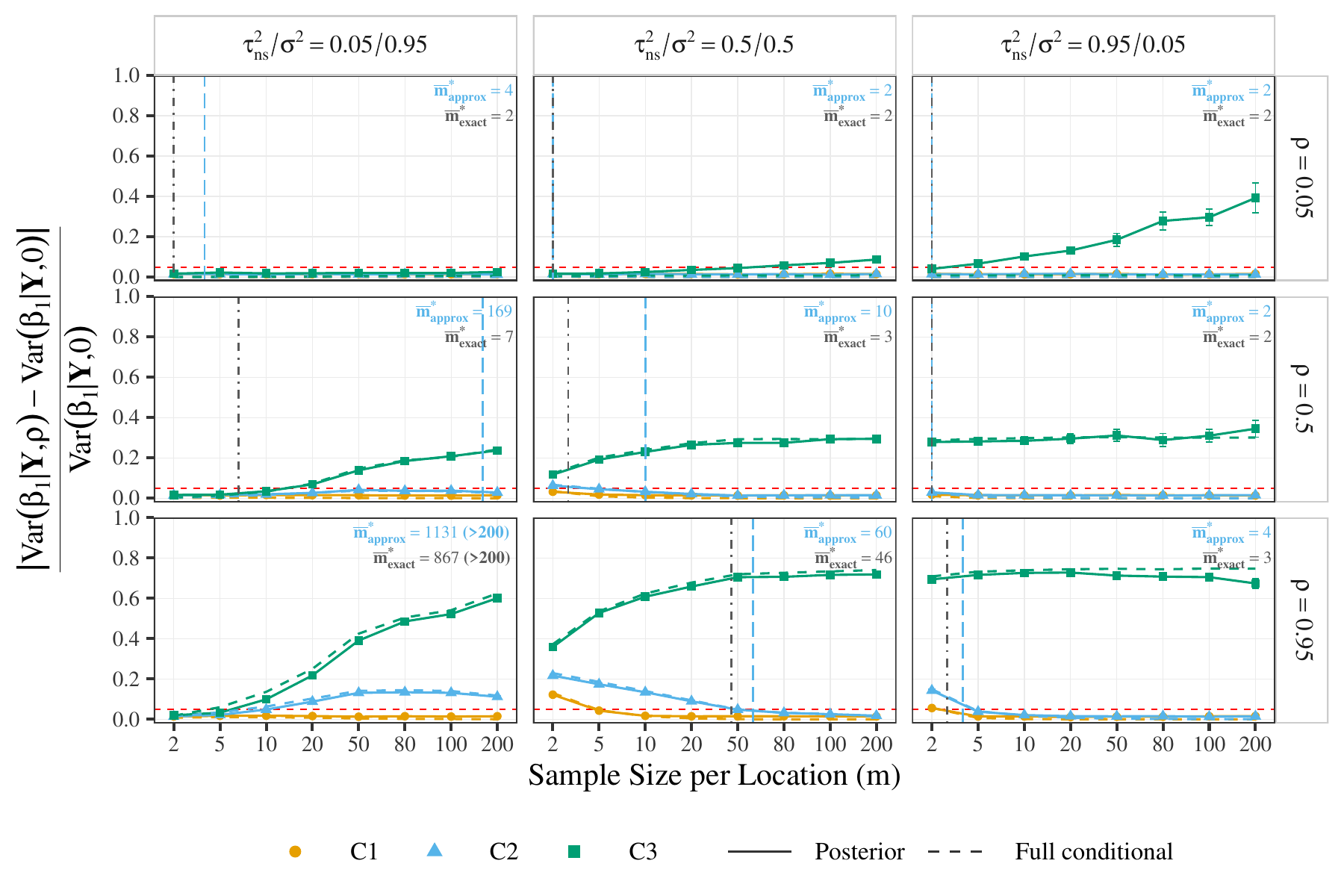}
\caption{Absolute relative difference in the marginal posterior variance of $\beta_1$ between the spatial (i.e., $\text{Var}\left(\beta_1 | \boldsymbol{Y}, \rho\right)$) and nonspatial (i.e., $\text{Var}\left(\beta_1 | \boldsymbol{Y}, 0\right)$) models, averaged over $100$ datasets, for $n = 400$ spatial units. Columns correspond to the variance ratio $\kappa = \tau^2_{ns}/\sigma^2$ and rows to the spatial correlation $\rho$. The three colors represent the three covariate structures, C1-C3, described in Table~\ref{tab:sim_settings}. Solid and dashed curves show marginal posterior inference and exact full conditional comparisons, respectively. The dashed red horizontal line marks the tolerance $\gamma = 0.05$; blue and grey vertical lines mark the approximate and exact C2 thresholds, respectively, each averaged over $100$ datasets and rounded up to an integer. Error bars are pointwise $95\%$ Monte Carlo confidence intervals for the marginal averages across datasets.}
\label{fig:plot1_J400}
\end{figure}
\clearpage

\begin{figure}[htbp]
\centering
\includegraphics[width=\linewidth]{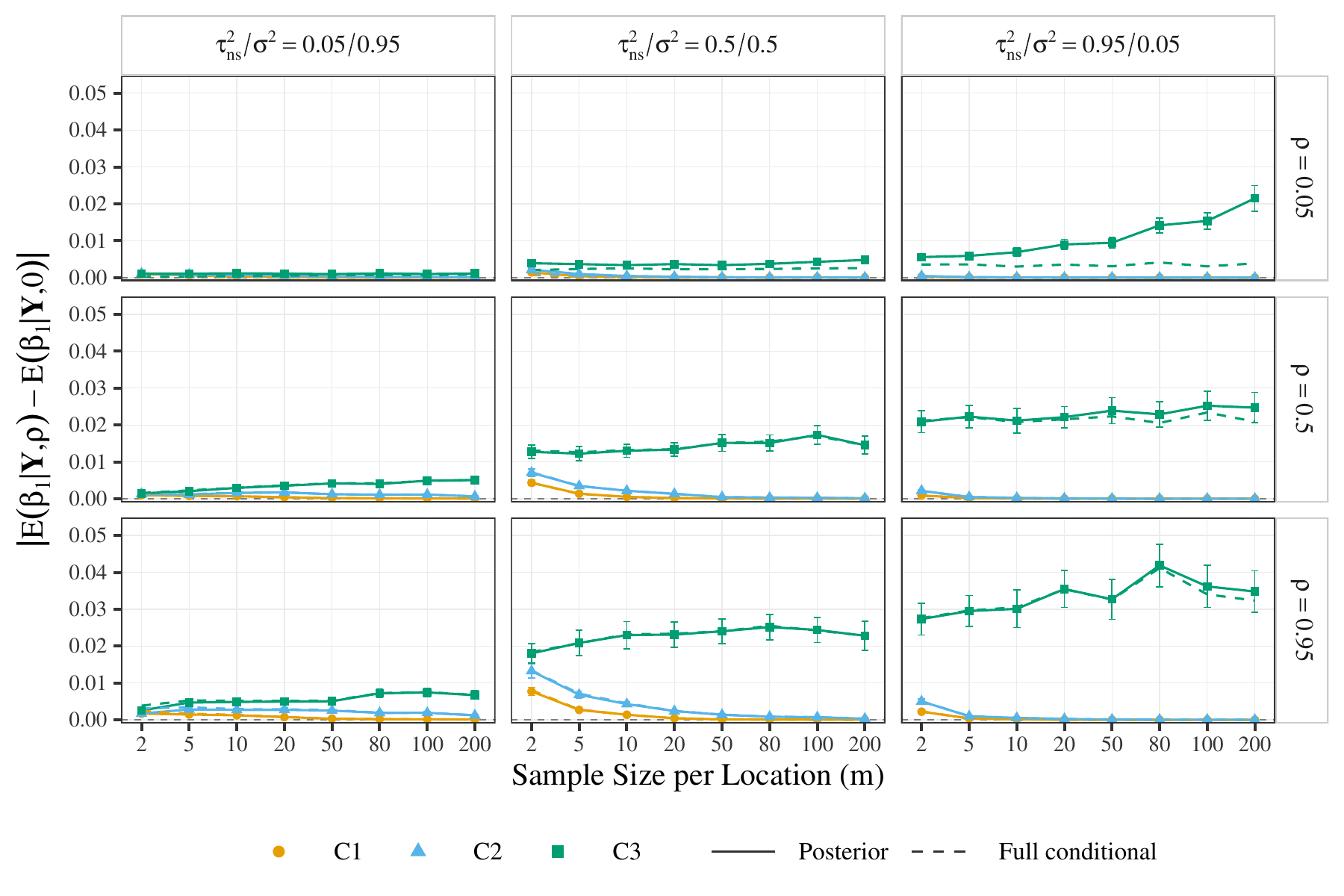}
\caption{Absolute difference in the marginal posterior mean of $\beta_1$ between the spatial (i.e., $\text{E}\left(\beta_1 | \boldsymbol{Y}, \rho\right)$) and nonspatial (i.e., $\text{E}\left(\beta_1 | \boldsymbol{Y}, 0\right)$) models, averaged over $100$ datasets, for $n = 400$ spatial units. Columns correspond to the variance ratio $\kappa = \tau^2_{ns}/\sigma^2$ and rows to the spatial correlation $\rho$. The three colors represent the three covariate structure settings, C1-C3, described in Table~\ref{tab:sim_settings}. Solid and dashed curves show marginal posterior inference and exact full conditional comparisons, respectively. Error bars are pointwise $95\%$ Monte Carlo confidence intervals across datasets.}
\label{fig:plot2_J400}
\end{figure}

\clearpage
\subsection{Results for a fixed spatial map}
\label{sec:supp_nonmonotone}

\begin{figure}[htbp]
\centering
\includegraphics[width=\linewidth]{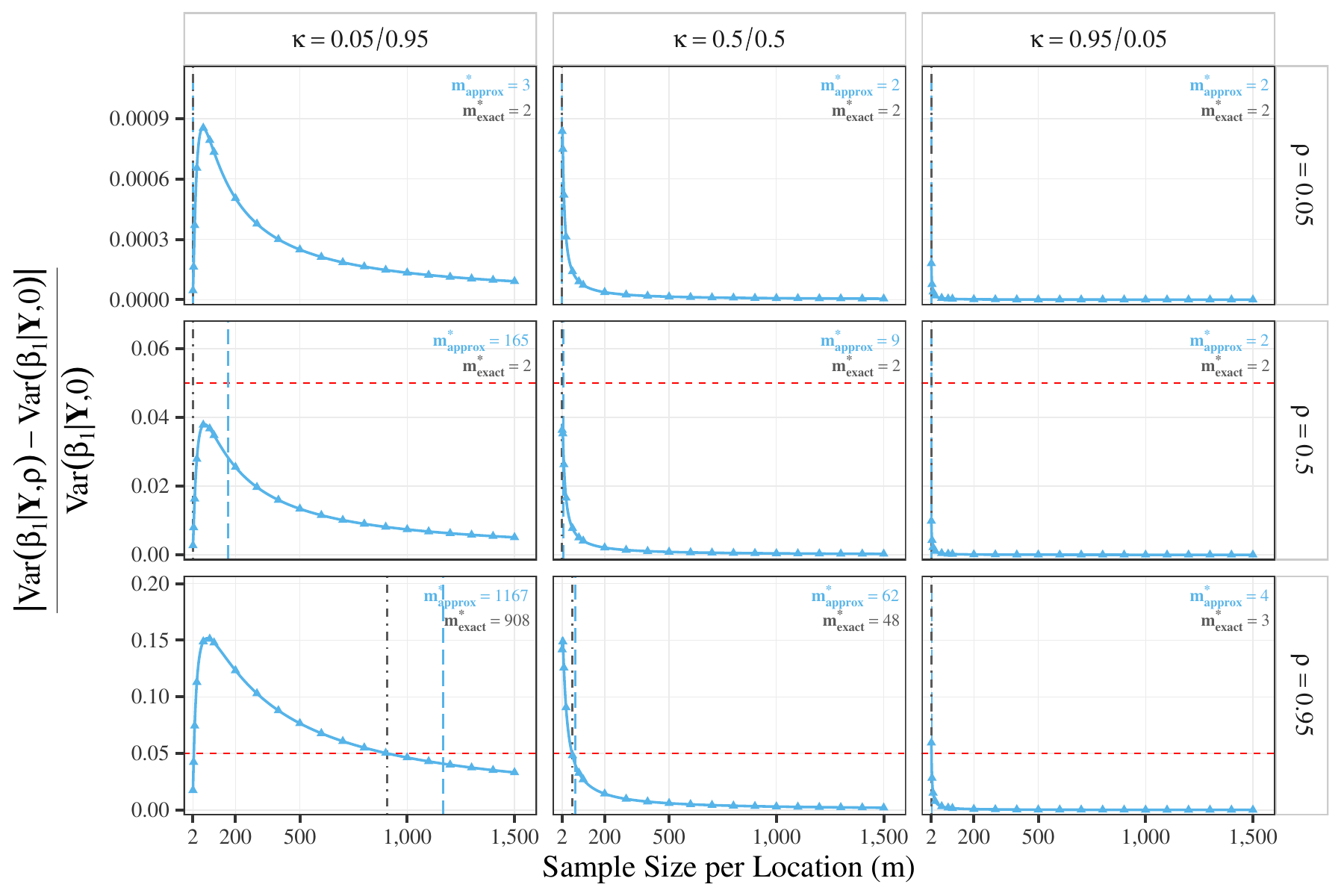}
\caption{Absolute relative variance difference of the full conditional distributions for $\beta_1$ over an extended range of within-area sample sizes $m$, using one fixed spatial map with $n=100$ units and a fixed C2 covariate structure. The dashed red horizontal line marks the tolerance $\gamma=0.05$, and the two vertical reference lines mark the approximate and exact thresholds for this design. All curves are calculated directly from the full conditional variance expressions.}
\label{fig:nonmonotone_fixed_map}
\end{figure}

\clearpage
\subsection{Bias and coverage}
\label{sec:bias_coverage}

Figures~\ref{fig:supp_bias} and~\ref{fig:supp_coverage} show estimated bias and empirical coverage, respectively, for both models.

\begin{figure}[htbp]
\centering
\includegraphics[width=\linewidth]{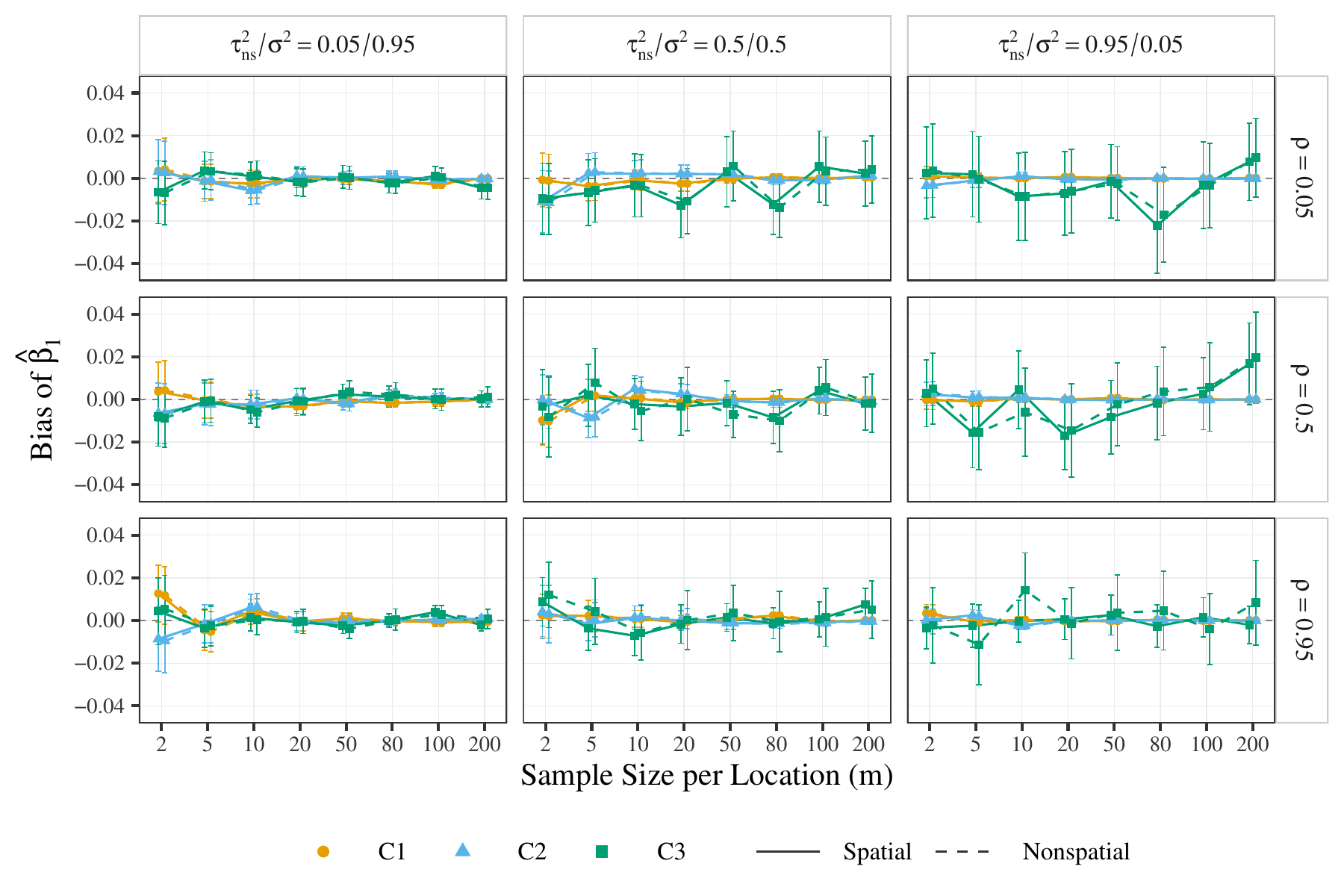}
\caption{Estimated bias of the marginal posterior mean of $\beta_1$ under the spatial and nonspatial models for covariate structures C1--C3. Bias is calculated as the average difference between the posterior mean and the generating coefficient over $100$ datasets per setting. Error bars show pointwise $95\%$ Monte Carlo confidence intervals. The horizontal reference line marks zero bias.}
\label{fig:supp_bias}
\end{figure}

\clearpage
\begin{figure}[htbp]
\centering
\includegraphics[width=\linewidth]{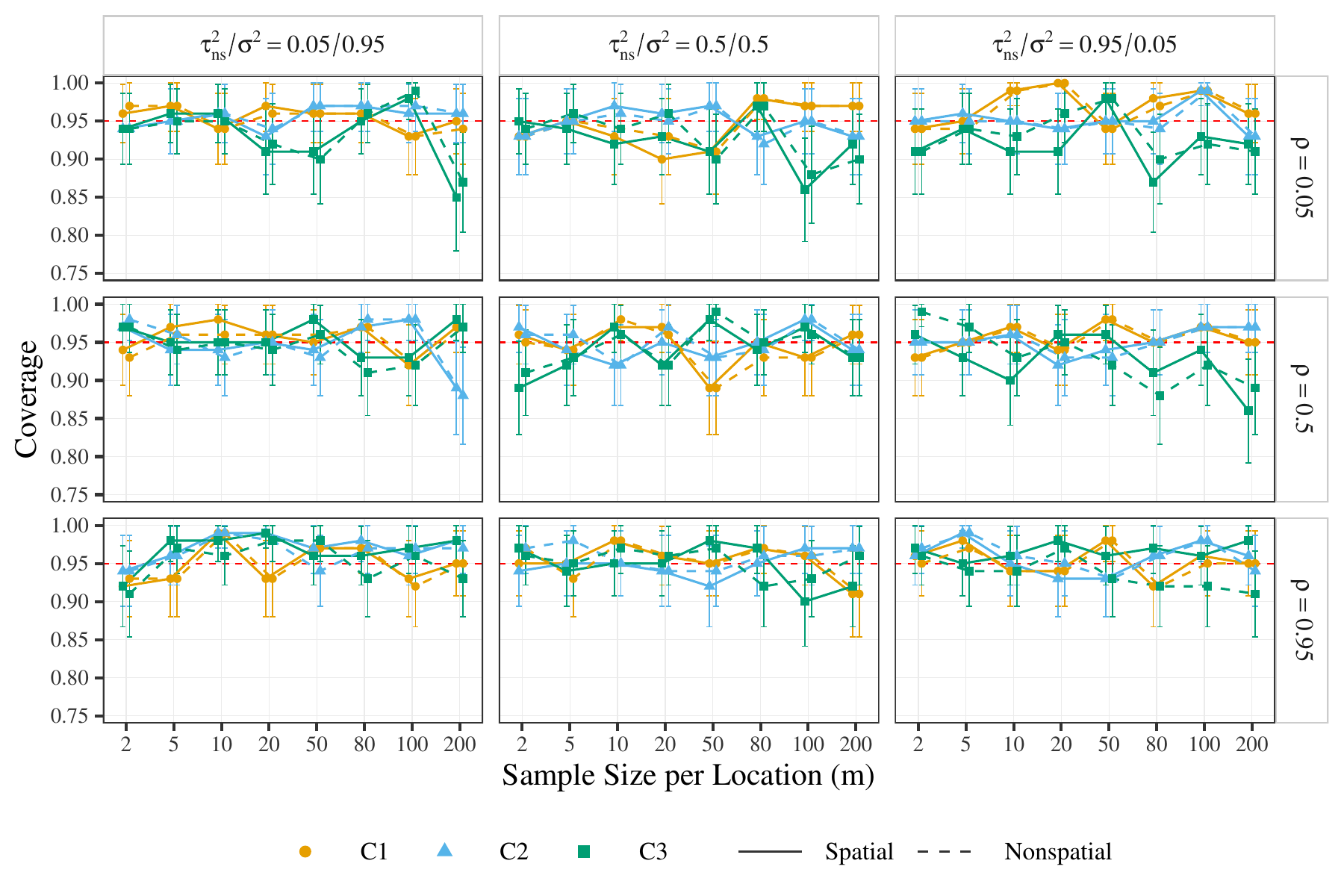}
\caption{Empirical coverage of the $95\%$ quantile-based marginal posterior credible intervals for $\beta_1$ under the spatial and nonspatial models for covariate structures C1--C3. Coverage is calculated as the proportion of intervals containing the generating coefficient over $100$ datasets per setting. Error bars show pointwise $95\%$ Monte Carlo confidence intervals. The horizontal reference line marks the nominal coverage of $0.95$.}
\label{fig:supp_coverage}
\end{figure}

\clearpage
\section{Extensions to other designs and outcomes}
\label{sec:supp_additional_settings}
The main text focuses on balanced replication to make the dependence on a common within-area sample size and the covariate's alignment with spatial eigenvectors explicit. Here, we complement that analysis by deriving the full conditional distribution precisions for arbitrary within-area sample sizes and show that the relative difference in full conditional distribution variances vanishes as within-area covariate variation accumulates, however the observations are allocated among areas. We then report simulations with unequal replication, with two covariates, and with Bernoulli or negative binomial outcomes.

\subsection{Within-area and between-area contributions to precision}
\label{sec:supp_decomp}

As in Section~\ref{sec:supp_fc_beta}, after integrating out the random intercepts, the full conditional distribution precision of $\beta_1$, given $\beta_0$ and the covariance parameters, is
\[\sigma_{\beta_1}^{-2}(\rho)=\textbf{x}^{\top}\Omega^{-1}\textbf{x}.\]
In Section~\ref{sec:spatial}, the spectral decomposition of $Q(\rho)$ and the identity $\text{Z}^{\top}\text{Z}=mI_n$ reduce the matrix inversion in the Woodbury formula to inversions of diagonal entries. Here, we allow arbitrary within-area sample sizes $m_i\geq1$, for which the observation covariance remains \[ \Omega=\tau_{sp}^2\text{Z}Q(\rho)^{-1}\text{Z}^{\top}+\sigma^2I_{\sum_i m_i}, \qquad \text{Z}^{\top}\text{Z}=\operatorname{diag}(m_1,\ldots,m_n). \]
To evaluate $\textbf{x}^{\top}\Omega^{-1}\textbf{x}$, we separate the covariate into area means and within-area deviations. This follows the structure of the model, in which each random intercept is shared by all observations within its area. Using the area-mean vector $\bar{\textbf{x}}$ from Section~\ref{sec:conditional}, with $\bar{\text{x}}_{i\cdot}=m_i^{-1}\sum_{j=1}^{m_i}\text{x}_{ij}$, we write \[ \text{x}_{ij} =(\text{x}_{ij}-\bar{\text{x}}_{i\cdot})+\bar{\text{x}}_{i\cdot}, \qquad \textbf{x} =(\textbf{x}-\text{Z}\bar{\textbf{x}})+\text{Z}\bar{\textbf{x}}. \]
By the definition of $\bar{\text{x}}_{i\cdot}$, subtracting the area mean gives deviations that sum to zero within each area, so
\begin{equation}
\text{Z}^{\top}(\textbf{x}-\text{Z}\bar{\textbf{x}})
=\text{Z}^{\top}\textbf{x}
-\operatorname{diag}(m_1,\ldots,m_n)\bar{\textbf{x}}
=\begin{pmatrix}
\sum_{j=1}^{m_1}\text{x}_{1j}-m_1\bar{\text{x}}_{1\cdot}\\
\vdots\\
\sum_{j=1}^{m_n}\text{x}_{nj}-m_n\bar{\text{x}}_{n\cdot}
\end{pmatrix}
=\boldsymbol{0}_n.
\label{eq:supp_within_zero_sum}
\end{equation}
Using~\eqref{eq:supp_within_zero_sum},
\begin{align*}
\Omega(\textbf{x}-\text{Z}\bar{\textbf{x}})
&=\left\{\tau_{sp}^2\text{Z}Q(\rho)^{-1}\text{Z}^{\top}
+\sigma^2I_{\sum_i m_i}\right\}
(\textbf{x}-\text{Z}\bar{\textbf{x}})\\
&=\tau_{sp}^2\text{Z}Q(\rho)^{-1}
\left\{\text{Z}^{\top}(\textbf{x}-\text{Z}\bar{\textbf{x}})\right\}
+\sigma^2(\textbf{x}-\text{Z}\bar{\textbf{x}})\\
&=\sigma^2(\textbf{x}-\text{Z}\bar{\textbf{x}}),
\end{align*}
so $\Omega^{-1}(\textbf{x}-\text{Z}\bar{\textbf{x}})=\sigma^{-2}(\textbf{x}-\text{Z}\bar{\textbf{x}})$. Since $\Omega^{-1}$ is symmetric, we also have
\begin{align*}
(\textbf{x}-\text{Z}\bar{\textbf{x}})^{\top}
\Omega^{-1}\text{Z}\bar{\textbf{x}}
&=\left\{\Omega^{-1}(\textbf{x}-\text{Z}\bar{\textbf{x}})\right\}^{\top}
\text{Z}\bar{\textbf{x}}\\
&=\frac{1}{\sigma^2}
(\textbf{x}-\text{Z}\bar{\textbf{x}})^{\top}
\text{Z}\bar{\textbf{x}}\\
&=\frac{1}{\sigma^2}
\left\{\text{Z}^{\top}(\textbf{x}-\text{Z}\bar{\textbf{x}})\right\}^{\top}
\bar{\textbf{x}}
=0.
\end{align*}
Substituting $\textbf{x}=(\textbf{x}-\text{Z}\bar{\textbf{x}})+\text{Z}\bar{\textbf{x}}$ into the precision gives
\begin{align}
\textbf{x}^{\top}\Omega^{-1}\textbf{x}
&=\left\{(\textbf{x}-\text{Z}\bar{\textbf{x}})+\text{Z}\bar{\textbf{x}}\right\}^{\top}
\Omega^{-1}
\left\{(\textbf{x}-\text{Z}\bar{\textbf{x}})+\text{Z}\bar{\textbf{x}}\right\}\notag\\
&=(\textbf{x}-\text{Z}\bar{\textbf{x}})^{\top}
\Omega^{-1}(\textbf{x}-\text{Z}\bar{\textbf{x}})
+2(\textbf{x}-\text{Z}\bar{\textbf{x}})^{\top}
\Omega^{-1}\text{Z}\bar{\textbf{x}}
+\bar{\textbf{x}}^{\top}\text{Z}^{\top}\Omega^{-1}\text{Z}\bar{\textbf{x}}\notag\\
&=\frac{1}{\sigma^2}
(\textbf{x}-\text{Z}\bar{\textbf{x}})^{\top}
(\textbf{x}-\text{Z}\bar{\textbf{x}})
+2\cdot 0
+\bar{\textbf{x}}^{\top}\text{Z}^{\top}\Omega^{-1}\text{Z}\bar{\textbf{x}}\notag\\
&=\frac{1}{\sigma^2}\sum_{i=1}^n\sum_{j=1}^{m_i}
(\text{x}_{ij}-\bar{\text{x}}_{i\cdot})^2
+\bar{\textbf{x}}^{\top}\text{Z}^{\top}\Omega^{-1}\text{Z}\bar{\textbf{x}}.
\label{eq:supp_precision_split}
\end{align}
To evaluate the remaining matrix $\text{Z}^{\top}\Omega^{-1}\text{Z}$ in~\eqref{eq:supp_precision_split}, we first note that
\begin{align*}
\Omega\text{Z}
&=\left\{\tau_{sp}^2\text{Z}Q(\rho)^{-1}\text{Z}^{\top}
+\sigma^2I_{\sum_i m_i}\right\}\text{Z}\\
&=\tau_{sp}^2\text{Z}Q(\rho)^{-1}\text{Z}^{\top}\text{Z}
+\sigma^2\text{Z}\\
&=\text{Z}\left\{\tau_{sp}^2Q(\rho)^{-1}
\text{Z}^{\top}\text{Z}+\sigma^2I_n\right\}.
\end{align*}
Thus,
\begin{align*}
\Omega^{-1}\Omega\text{Z}
&=\text{Z}
=\Omega^{-1}\text{Z}
\left\{\tau_{sp}^2Q(\rho)^{-1}\text{Z}^{\top}\text{Z}
+\sigma^2I_n\right\},
\end{align*}
and therefore
\begin{align*}
\Omega^{-1}\text{Z}
&=\Omega^{-1}\text{Z}
\left\{\tau_{sp}^2Q(\rho)^{-1}\text{Z}^{\top}\text{Z}
+\sigma^2I_n\right\}
\left\{\tau_{sp}^2Q(\rho)^{-1}\text{Z}^{\top}\text{Z}
+\sigma^2I_n\right\}^{-1}\\
&=\text{Z}
\left\{\tau_{sp}^2Q(\rho)^{-1}\text{Z}^{\top}\text{Z}
+\sigma^2I_n\right\}^{-1}.
\end{align*}
Since
\[
\tau_{sp}^2Q(\rho)^{-1}\text{Z}^{\top}\text{Z}+\sigma^2I_n
=\left\{\tau_{sp}^2Q(\rho)^{-1}
+\sigma^2(\text{Z}^{\top}\text{Z})^{-1}\right\}
\text{Z}^{\top}\text{Z},
\]
we obtain
\begin{align}
\text{Z}^{\top}\Omega^{-1}\text{Z}
&=\text{Z}^{\top}\text{Z}
\left\{\tau_{sp}^2Q(\rho)^{-1}\text{Z}^{\top}\text{Z}
+\sigma^2I_n\right\}^{-1}\notag\\
&=\text{Z}^{\top}\text{Z}
\left[
\left\{\tau_{sp}^2Q(\rho)^{-1}
+\sigma^2(\text{Z}^{\top}\text{Z})^{-1}\right\}
\text{Z}^{\top}\text{Z}
\right]^{-1}\notag\\
&=\text{Z}^{\top}\text{Z}(\text{Z}^{\top}\text{Z})^{-1}
\left\{\tau_{sp}^2Q(\rho)^{-1}
+\sigma^2(\text{Z}^{\top}\text{Z})^{-1}\right\}^{-1}\notag\\
&=\left\{\tau_{sp}^2Q(\rho)^{-1}
+\sigma^2(\text{Z}^{\top}\text{Z})^{-1}\right\}^{-1}\notag\\
&=\left\{\tau_{sp}^2Q(\rho)^{-1}
+\operatorname{diag}\left(
\frac{\sigma^2}{m_1},\ldots,\frac{\sigma^2}{m_n}
\right)\right\}^{-1}.
\label{eq:supp_area_mean_precision}
\end{align}

Substituting~\eqref{eq:supp_area_mean_precision} into~\eqref{eq:supp_precision_split} gives
\begin{equation}
\begin{aligned}
\sigma_{\beta_1}^{-2}(\rho)
&=\frac{1}{\sigma^2}\sum_{i=1}^n\sum_{j=1}^{m_i}
(\text{x}_{ij}-\bar{\text{x}}_{i\cdot})^2
+\bar{\textbf{x}}^{\top}\text{Z}^{\top}\Omega^{-1}\text{Z}\bar{\textbf{x}}\\
&=\frac{1}{\sigma^2}\sum_{i=1}^n\sum_{j=1}^{m_i}
(\text{x}_{ij}-\bar{\text{x}}_{i\cdot})^2
+\bar{\textbf{x}}^{\top}
\left\{\tau_{sp}^2Q(\rho)^{-1}
+\operatorname{diag}\left(
\frac{\sigma^2}{m_1},\ldots,\frac{\sigma^2}{m_n}
\right)\right\}^{-1}
\bar{\textbf{x}}.
\end{aligned}
\label{eq:supp_precision_sp_general}
\end{equation}
Under the nonspatial model of Section~\ref{sec:model}, the random-intercept covariance $\tau_{sp}^2Q(\rho)^{-1}$ is replaced by $\tau_{ns}^2I_n$, yielding
\begin{equation}
\begin{aligned}
\sigma_{\beta_1}^{-2}(0)
&=\frac{1}{\sigma^2}\sum_{i=1}^n\sum_{j=1}^{m_i}
(\text{x}_{ij}-\bar{\text{x}}_{i\cdot})^2
+\bar{\textbf{x}}^{\top}
\left\{\tau_{ns}^2I_n
+\operatorname{diag}\left(
\frac{\sigma^2}{m_1},\ldots,\frac{\sigma^2}{m_n}
\right)\right\}^{-1}\bar{\textbf{x}}\\
&=\frac{1}{\sigma^2}\sum_{i=1}^n\sum_{j=1}^{m_i}
(\text{x}_{ij}-\bar{\text{x}}_{i\cdot})^2
+\bar{\textbf{x}}^{\top}
\operatorname{diag}\left(
\tau_{ns}^2+\frac{\sigma^2}{m_1},\ldots,
\tau_{ns}^2+\frac{\sigma^2}{m_n}
\right)^{-1}\bar{\textbf{x}}\\
&=\frac{1}{\sigma^2}\sum_{i=1}^n\sum_{j=1}^{m_i}
(\text{x}_{ij}-\bar{\text{x}}_{i\cdot})^2
+\sum_{i=1}^n
\frac{\bar{\text{x}}_{i\cdot}^2}{\tau_{ns}^2+\sigma^2/m_i}.
\end{aligned}
\label{eq:supp_precision_ns_general}
\end{equation}

Under balanced replication, the observation-error covariance of the area means is $\sigma^2I_n/m$, so the spatial area-mean covariance has the same eigenvectors as $Q(\rho)$. Using the covariate standardization and projections from Section~\ref{sec:spatial}, $\text{d}_{\cdot}=\sum_i\bar{\text{x}}_{i\cdot}^2$ and
\[
\sum_{i=1}^n\sum_{j=1}^m
(\text{x}_{ij}-\bar{\text{x}}_{i\cdot})^2
=\textbf{x}^{\top}\textbf{x}
-m\sum_{i=1}^n\bar{\text{x}}_{i\cdot}^2
=m(n-\text{d}_{\cdot}).
\]
Equations~\eqref{eq:supp_precision_sp_general} and~\eqref{eq:supp_precision_ns_general} therefore become
\begin{equation}
\begin{aligned}
\sigma_{\beta_1}^{-2}(\rho)
&=\frac{m(n-\text{d}_{\cdot})}{\sigma^2}
+\sum_{i=1}^n
\frac{\text{d}_i}
{\tau_{sp}^2/(\rho\lambda_i+1-\rho)+\sigma^2/m},\\
\sigma_{\beta_1}^{-2}(0)
&=\frac{m(n-\text{d}_{\cdot})}{\sigma^2}
+\frac{\text{d}_{\cdot}}{\tau_{ns}^2+\sigma^2/m}.
\end{aligned}
\label{eq:supp_precision_balanced}
\end{equation}
These are equivalent to expressions~\eqref{eq:precision} and~\eqref{eq:precision0}.

Equations~\eqref{eq:supp_precision_sp_general} and~\eqref{eq:supp_precision_ns_general} show that the two precisions differ only in their area-mean contributions. This decomposition complements the spectral approach in Section~\ref{sec:spatial}, which quantifies how those differences depend on the covariate's alignment with the spatial eigenvectors. Under balanced replication, the area-mean covariance shares the eigenvectors of $Q(\rho)$, and the area-mean contribution splits into one term per eigenvector, each depending on the corresponding Laplacian eigenvalue $\lambda_i$ and covariate projection $\text{d}_i$. This split identifies the spatial patterns in a covariate for which the choice between spatial and nonspatial models materially affects the precision of its regression coefficient. Two covariates with the same $\text{d}_{\cdot}$ have the same within-area and between-area variation under the standardization used in the main text, yet the models can yield similar precisions for one and substantially different precisions for the other, depending on which eigenvectors their area means align with. The simulations exploit the same split by varying spatial correlation, the variance ratio and covariate structure.

\subsection{Variance agreement under unequal replication}
\label{sec:unequal_sim}

Subtracting~\eqref{eq:supp_precision_sp_general} from~\eqref{eq:supp_precision_ns_general} gives
\[
\sigma_{\beta_1}^{-2}(0)-\sigma_{\beta_1}^{-2}(\rho)
=
\sum_{i=1}^n
\frac{\bar{\text{x}}_{i\cdot}^2}{\tau_{ns}^2+\sigma^2/m_i}
-
\bar{\textbf{x}}^{\top}
\left\{\tau_{sp}^2Q(\rho)^{-1}
+\operatorname{diag}\left(
\frac{\sigma^2}{m_1},\ldots,\frac{\sigma^2}{m_n}
\right)\right\}^{-1}
\bar{\textbf{x}}.
\]
Hold $n$, the graph, the covariance parameters and $\bar{\textbf{x}}$ fixed:
\begin{align*}
0
&\leq
\sum_{i=1}^n
\frac{\bar{\text{x}}_{i\cdot}^2}{\tau_{ns}^2+\sigma^2/m_i}
\leq
\frac{\bar{\textbf{x}}^{\top}\bar{\textbf{x}}}{\tau_{ns}^2}, \quad \text{and}\quad
0\leq
\bar{\textbf{x}}^{\top}
\left\{\tau_{sp}^2Q(\rho)^{-1}
+\operatorname{diag}\left(
\frac{\sigma^2}{m_1},\ldots,\frac{\sigma^2}{m_n}
\right)\right\}^{-1}
\bar{\textbf{x}}
\leq
\frac{\bar{\textbf{x}}^{\top}Q(\rho)\bar{\textbf{x}}}{\tau_{sp}^2}.
\end{align*}
Consequently, for $\sum_{i=1}^n\sum_{j=1}^{m_i}(\text{x}_{ij}-\bar{\text{x}}_{i\cdot})^2>0$,
\begin{equation}
\begin{aligned}
\left|
\frac{\sigma_{\beta_1}^2(\rho)}
{\sigma_{\beta_1}^2(0)}-1
\right|
&=
\frac{
\left|\sigma_{\beta_1}^{-2}(0)-\sigma_{\beta_1}^{-2}(\rho)\right|
}{
\sigma_{\beta_1}^{-2}(\rho)
}\\
&\leq
\frac{
\max\left\{
\bar{\textbf{x}}^{\top}\bar{\textbf{x}}/\tau_{ns}^2,\,
\bar{\textbf{x}}^{\top}Q(\rho)\bar{\textbf{x}}/\tau_{sp}^2
\right\}
}{
\displaystyle
\frac{1}{\sigma^2}
\sum_{i=1}^n\sum_{j=1}^{m_i}
(\text{x}_{ij}-\bar{\text{x}}_{i\cdot})^2
+
\bar{\textbf{x}}^{\top}
\left\{\tau_{sp}^2Q(\rho)^{-1}
+\operatorname{diag}\left(
\frac{\sigma^2}{m_1},\ldots,\frac{\sigma^2}{m_n}
\right)\right\}^{-1}
\bar{\textbf{x}}
}\\
&\leq
\frac{
\max\left\{
\bar{\textbf{x}}^{\top}\bar{\textbf{x}}/\tau_{ns}^2,\,
\bar{\textbf{x}}^{\top}Q(\rho)\bar{\textbf{x}}/\tau_{sp}^2
\right\}
}{
\displaystyle
\frac{1}{\sigma^2}
\sum_{i=1}^n\sum_{j=1}^{m_i}
(\text{x}_{ij}-\bar{\text{x}}_{i\cdot})^2
}\\
&=
\frac{
\sigma^2\max\left\{
\bar{\textbf{x}}^{\top}\bar{\textbf{x}}/\tau_{ns}^2,\,
\bar{\textbf{x}}^{\top}Q(\rho)\bar{\textbf{x}}/\tau_{sp}^2
\right\}
}{
\displaystyle\sum_{i=1}^n\sum_{j=1}^{m_i}
(\text{x}_{ij}-\bar{\text{x}}_{i\cdot})^2
}.
\end{aligned}
\label{eq:supp_unequal_relative_difference}
\end{equation}
The numerator is fixed, so as long as the within-area sum of squares grows without bound (e.g., under C1 and C2), the relative variance difference vanishes at rate $O\{[\sum_{i=1}^n\sum_{j=1}^{m_i}(\text{x}_{ij}-\bar{\text{x}}_{i\cdot})^2]^{-1}\}$. Thus, agreement is governed by the accumulated within-area covariate variation.

More generally, let the Gaussian random intercepts have a fixed
positive-definite covariance matrix $\Sigma$. Replacing
$\tau_{sp}^2Q(\rho)^{-1}$ by $\Sigma$ in the precision decomposition
leaves the within-area contribution unchanged. The contribution
from the area means satisfies
\[
0\leq
\bar{\textbf{x}}^\top
\left\{
\Sigma+\operatorname{diag}\left(
\frac{\sigma^2}{m_1},\ldots,\frac{\sigma^2}{m_n}
\right)
\right\}^{-1}
\bar{\textbf{x}}
\leq
\bar{\textbf{x}}^\top\Sigma^{-1}\bar{\textbf{x}}.
\]
Thus, for bounded covariate area means, the covariance-dependent
contribution remains bounded as replication increases. Together
with the corresponding nonspatial bound, this yields
\eqref{eq:supp_unequal_relative_difference} with
$\bar{\textbf{x}}^\top Q(\rho)\bar{\textbf{x}}/\tau_{sp}^2$
replaced by $\bar{\textbf{x}}^\top\Sigma^{-1}\bar{\textbf{x}}$.
The relative variance difference therefore vanishes as the
within-area covariate sum of squares grows.

\subsection{Simulations with unequal replication}
\label{sec:unequal_simulation}

We examine unequal replication using $n=100$ spatial units, $\tau^2_{ns}=\sigma^2=0.5$, $\rho\in\{0.50,0.95\}$, and $m\in\{2,5,10,20,50,80,100,200\}$. We compare balanced sample sizes $m_i=m$ with unequal sample sizes generated as $m_i-1\sim\operatorname{Poisson}(m-1)$, independently of the covariates and random intercepts. Thus, each area contributes at least one observation and $E(m_i)=m$, giving the two allocations the same expected total sample size. For each setting and covariate structure C1--C3, we generate $100$ datasets and fit both models.

Figures~\ref{fig:unequal_sim} and~\ref{fig:unequal_mean} compare marginal posterior variances and means under balanced and unequal replication. The variance curves closely overlap under the two allocations. At larger $m$, variance differences are small under C1 and C2 but remain substantial under C3. Table~\ref{tab:unequal_sizes} summarizes the variation in within-area sample sizes under the unequal allocation.

\begin{table}[htbp]
\centering
\caption{Theoretical standard deviations and percentiles of the within-area sample size $m_i$ under $m_i-1\sim\operatorname{Poisson}(m-1)$}
\begin{tabular}{rrc}
Expected sample size $m$ & $\operatorname{SD}(m_i)$ & 2.5th--97.5th percentiles \\
2   & 1.00  & 1--4     \\
5   & 2.00  & 2--9     \\
10  & 3.00  & 5--16    \\
20  & 4.36  & 12--29   \\
50  & 7.00  & 37--64   \\
80  & 8.89  & 63--98   \\
100 & 9.95  & 81--120  \\
200 & 14.11 & 173--228 \\
\end{tabular}
\label{tab:unequal_sizes}
\end{table}

\begin{figure}[htbp]
\centering
\includegraphics[width=\linewidth]{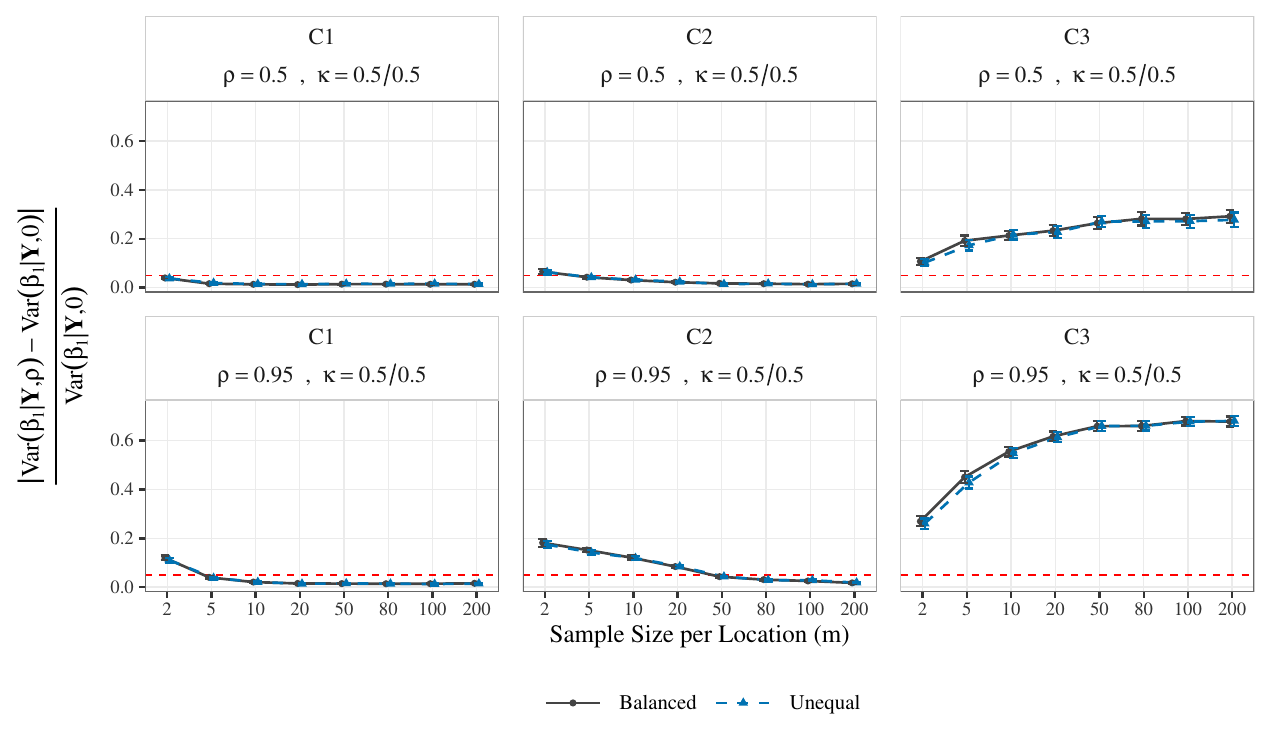}
\caption{Absolute relative difference in the marginal posterior variance of $\beta_1$ between the spatial and nonspatial models under balanced and unequal within-area sample sizes, averaged over $100$ datasets per setting. Results are shown for $n=100$, $\tau^2_{ns}=\sigma^2=0.5$, $\rho\in\{0.50,0.95\}$, and covariate structures C1--C3. Under balanced replication, $m_i=m$; under unequal replication, $m_i-1\sim\operatorname{Poisson}(m-1)$, so $m$ denotes the expected within-area sample size. The horizontal reference line marks the tolerance $\gamma=0.05$. Error bars are pointwise $95\%$ Monte Carlo confidence intervals across datasets.}
\label{fig:unequal_sim}
\end{figure}

\begin{figure}[htbp]
\centering
\includegraphics[width=\linewidth]{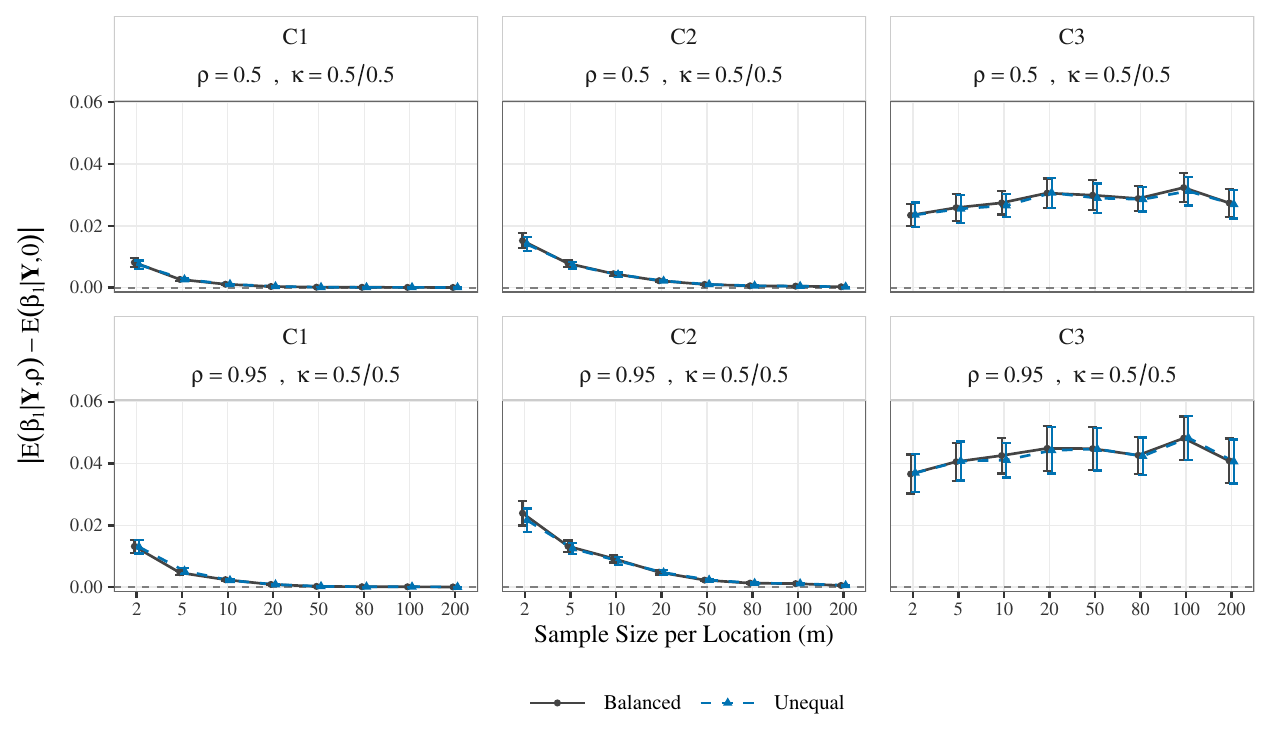}
\caption{Absolute difference in the marginal posterior mean of $\beta_1$ between the spatial and nonspatial models under balanced and unequal within-area sample sizes, averaged over $100$ datasets per setting. Settings are as in Figure~\ref{fig:unequal_sim}, with $m$ denoting the expected within-area sample size under unequal replication. Error bars are pointwise $95\%$ Monte Carlo confidence intervals across datasets.}
\label{fig:unequal_mean}
\end{figure}

\clearpage
\subsection{Multiple covariates}
\label{sec:multiple_sim}

We extend the Gaussian observation model to include $p$ covariates,
\[
Y_{ij}=\beta_0+\sum_{k=1}^{p}\beta_k\text{x}_{kij}+\theta_i+\varepsilon_{ij},
\qquad
\varepsilon_{ij}\mid\sigma^2
\overset{\mathrm{iid}}{\sim}\mathcal{N}(0,\sigma^2).
\]
For coefficient $\beta_k$, define the adjusted response
\[
\textbf{r}_{-k}
=\boldsymbol{Y}-\beta_0\boldsymbol{1}
-\sum_{\ell\ne k}\beta_\ell\textbf{x}_\ell.
\]
After integrating out the random intercepts, a flat coefficient prior gives
\[
\beta_k\mid\text{rest}\sim\mathcal{N}\left(
\frac{\textbf{x}_k^\top\Omega^{-1}\textbf{r}_{-k}}
{\textbf{x}_k^\top\Omega^{-1}\textbf{x}_k},
\frac{1}{\textbf{x}_k^\top\Omega^{-1}\textbf{x}_k}
\right),
\]
where conditioning is on the remaining coefficients and covariance parameters, and $\Omega$ is the observation covariance under the relevant model.

These are the single-covariate expressions applied to $\textbf{x}_k$ and the adjusted response $\textbf{r}_{-k}$. The precision decomposition and variance bound apply separately to each coefficient under both balanced and unequal replication, with comparisons depending on the corresponding covariate's within-area variation and spatial alignment. Under balanced replication, the projections $\text{d}_i$ are computed from that covariate's area means, and the convergence rates hold under the same assumptions. The simulations below examine the corresponding marginal posterior comparisons.

We consider six covariate pairings, C1--C1, C1--C2, C1--C3, C2--C2, C2--C3, and C3--C3, generating each predictor separately according to its assigned structure. We use balanced replication with $n=100$, $\tau^2_{ns}=\sigma^2=0.5$, $\rho\in\{0.50,0.95\}$, and $m\in\{2,5,10,20,50,80,100,200\}$, with $100$ datasets per setting.

Figures~\ref{fig:multiple_sim} and~\ref{fig:multiple_mean} show the absolute relative variance differences and absolute mean differences, respectively, for the two coefficients. Both differences diminish with replication for C1 and C2 coefficients, including when the other predictor is C3. Differences persist for C3 coefficients, and for both coefficients under C3--C3. Thus, agreement between spatial and nonspatial inference can differ between coefficients within the same model.

\clearpage
\begin{figure}[htbp]
\centering
\includegraphics[width=\linewidth,height=0.72\textheight,keepaspectratio]{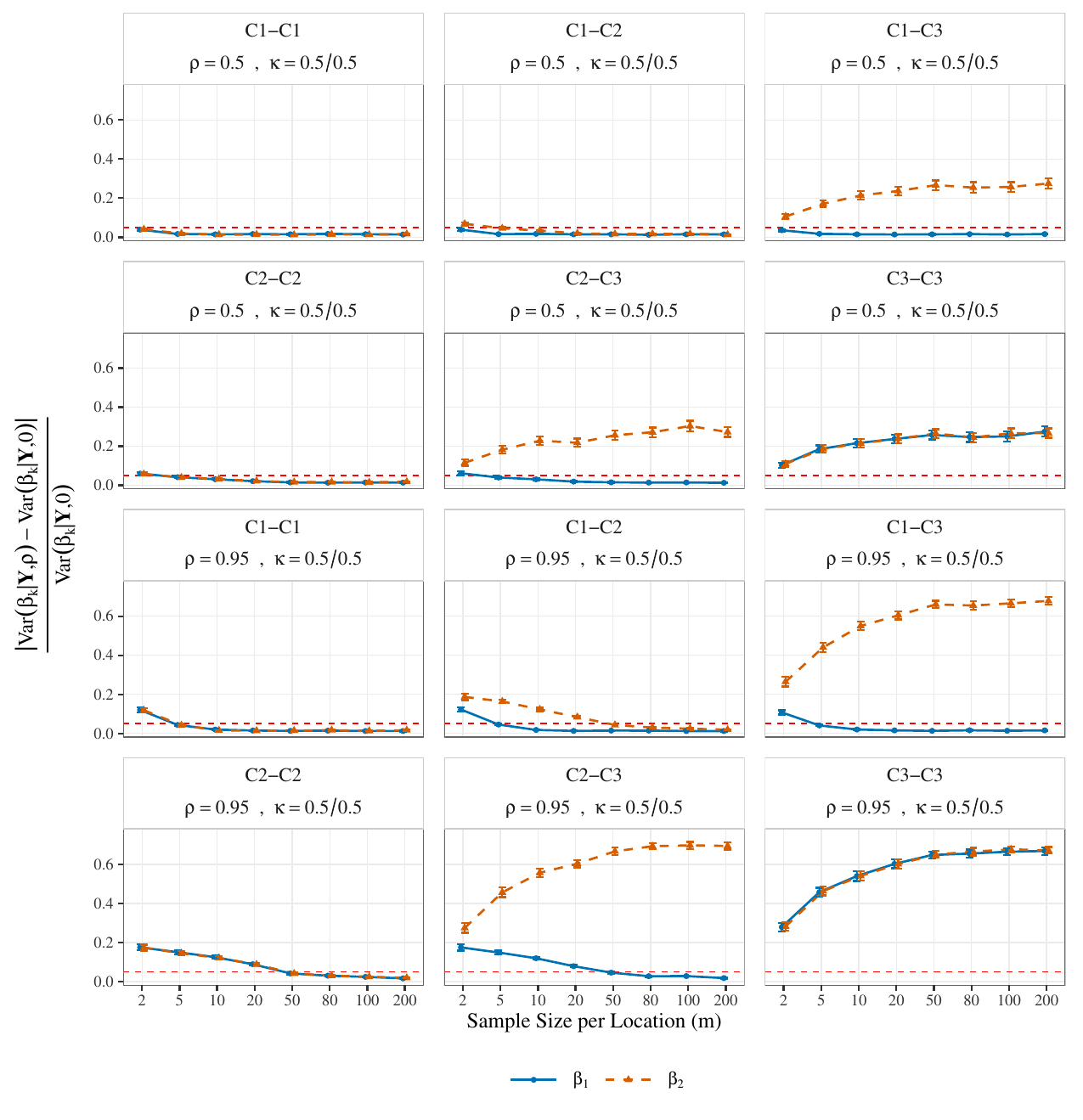}
\caption{Absolute relative differences in the marginal posterior variances of $\beta_1$ and $\beta_2$ between the spatial and nonspatial models, averaged over $100$ datasets per setting. Results are shown for the six covariate pairings with $n=100$, $\tau^2_{ns}=\sigma^2=0.5$, and $\rho\in\{0.50,0.95\}$. Each pairing lists the structure of $\text{x}_1$ followed by that of $\text{x}_2$. Blue solid curves represent $\beta_1$ and orange dashed curves represent $\beta_2$. The horizontal reference line marks the tolerance $\gamma=0.05$. Error bars are pointwise $95\%$ Monte Carlo confidence intervals across datasets.}
\label{fig:multiple_sim}
\end{figure}

\clearpage
\begin{figure}[htbp]
\centering
\includegraphics[width=\linewidth,height=0.72\textheight,keepaspectratio]{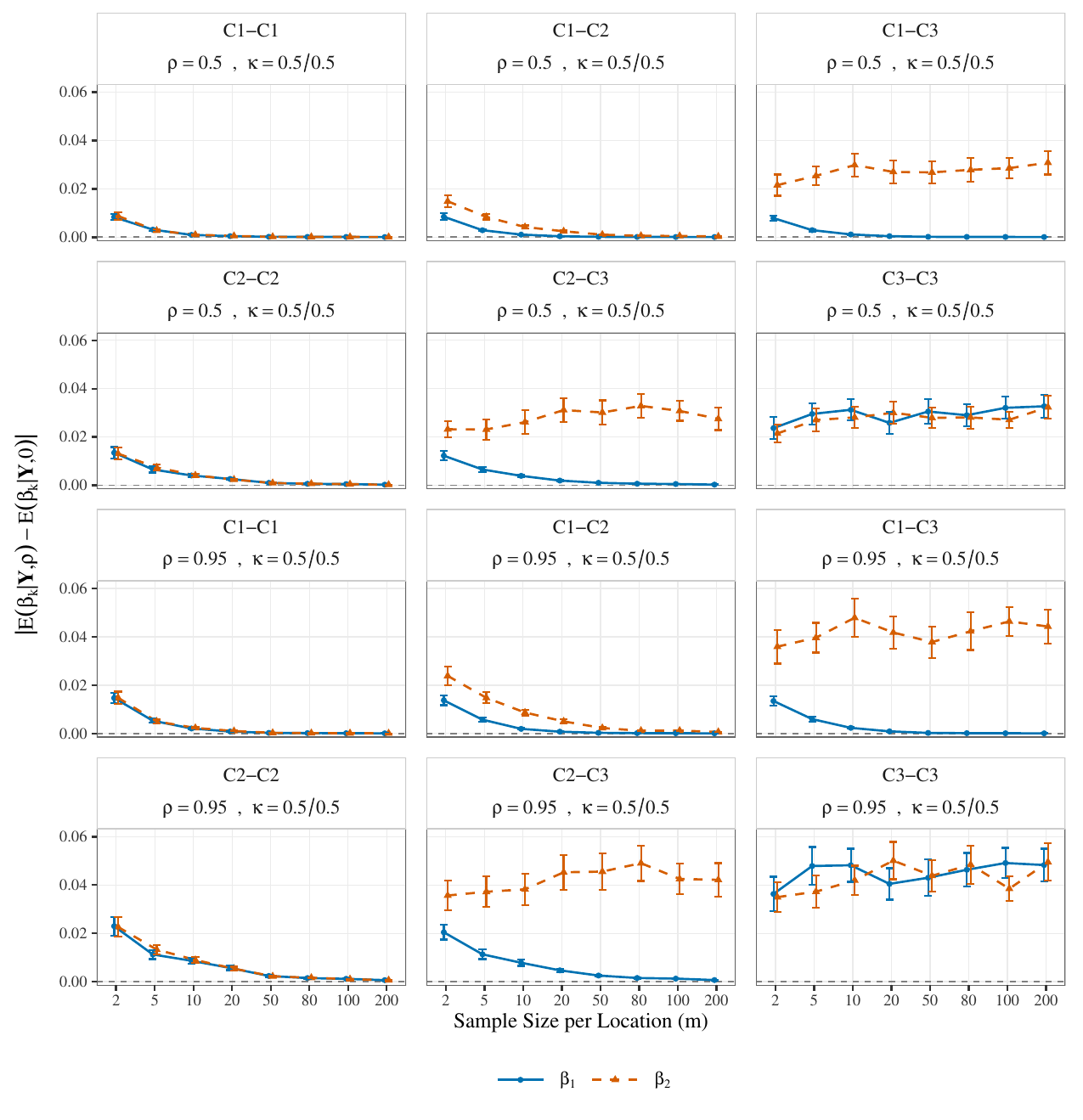}
\caption{Absolute differences in the marginal posterior means of $\beta_1$ and $\beta_2$ between the spatial and nonspatial models, averaged over $100$ datasets per setting. Settings and covariate pairings are as in Figure~\ref{fig:multiple_sim}. Blue solid curves represent $\beta_1$ and orange dashed curves represent $\beta_2$. Error bars are pointwise $95\%$ Monte Carlo confidence intervals across datasets.}
\label{fig:multiple_mean}
\end{figure}

\clearpage
\subsection{Bernoulli outcomes}
\label{sec:supp_extensions}

We examine whether the replication patterns observed under the Gaussian model extend to Bernoulli and negative binomial outcomes. Both extensions retain the spatial and nonspatial random-intercept distributions described in the main text, with $\tau^2_{sp}$ calibrated to $\tau^2_{ns}$ through the same variance-matching relation.

P\'olya--Gamma augmentation provides a connection between these models and the Gaussian calculations \citep{polson2013bayesian}. Write $\eta_{ij}=\beta_0+\beta_1\text{x}_{ij}+\theta_i$. Conditional on the augmentation variables $\omega_{ij}$, the augmented likelihood has the Gaussian kernel
\begin{equation*}
L_{\mathrm{aug}}(\boldsymbol{\beta},\boldsymbol{\theta}\mid\boldsymbol{Y},\boldsymbol{\omega}) \propto \exp\left\{-\frac{1}{2}\sum_{i=1}^{n}\sum_{j=1}^{m}\omega_{ij}\left(\widetilde{Y}_{ij}-\eta_{ij}\right)^2\right\},
\end{equation*}
where $\widetilde{Y}_{ij}=(Y_{ij}-1/2)/\omega_{ij}$ for Bernoulli outcomes and $\widetilde{Y}_{ij}=\log r+(Y_{ij}-r)/(2\omega_{ij})$ for negative binomial outcomes with conditional mean $\exp(\eta_{ij})$ and fixed size parameter $r>0$. Thus, observation-specific precisions $\omega_{ij}$ replace the common Gaussian precision $\sigma^{-2}$. These latent precisions vary across observations and are integrated out in marginal inference, so the balanced Gaussian threshold formulas do not directly apply. We assess replication through differences in marginal posterior means and variances. For both outcome types, we assign independent $\mathcal{N}(0,100^2)$ priors to $\beta_0$ and $\beta_1$, retaining the covariance-parameter priors specified in the main text. Each model is fitted using $75,000$ MCMC iterations, discarding the first $15,000$ and retaining every fifth subsequent draw, giving $12,000$ posterior draws.

We generate conditionally independent binary outcomes according to
\begin{equation*}
Y_{ij}\mid p_{ij}\sim\operatorname{Bernoulli}(p_{ij}), \qquad \operatorname{logit}(p_{ij})=\beta_0+\beta_1\text{x}_{ij}+\theta_i.
\end{equation*}
We use $n=100$, $m\in\{2,5,10,20,50,80,100,200\}$, $\rho\in\{0.05,0.50,0.95\}$, and $\tau^2_{ns}\in\{0.05,0.50,0.95\}$, with $100$ datasets per setting under each covariate structure C1--C3.

Figures~\ref{fig:bernoulli_sim} and~\ref{fig:bernoulli_mean} show the variance and mean comparisons, respectively. Posterior mean differences generally decrease with replication under C1 and C2 but persist under C3. Variance differences are small under C1 at moderate replication. Under C2, they remain above $\gamma=0.05$ at $m=200$ when $\rho=0.95$ and $\tau^2_{ns}\in\{0.05,0.50\}$. Under C3, variance differences generally increase with replication when spatial correlation is moderate or strong.

\clearpage
\begin{figure}[p]
\centering
\includegraphics[width=\linewidth]{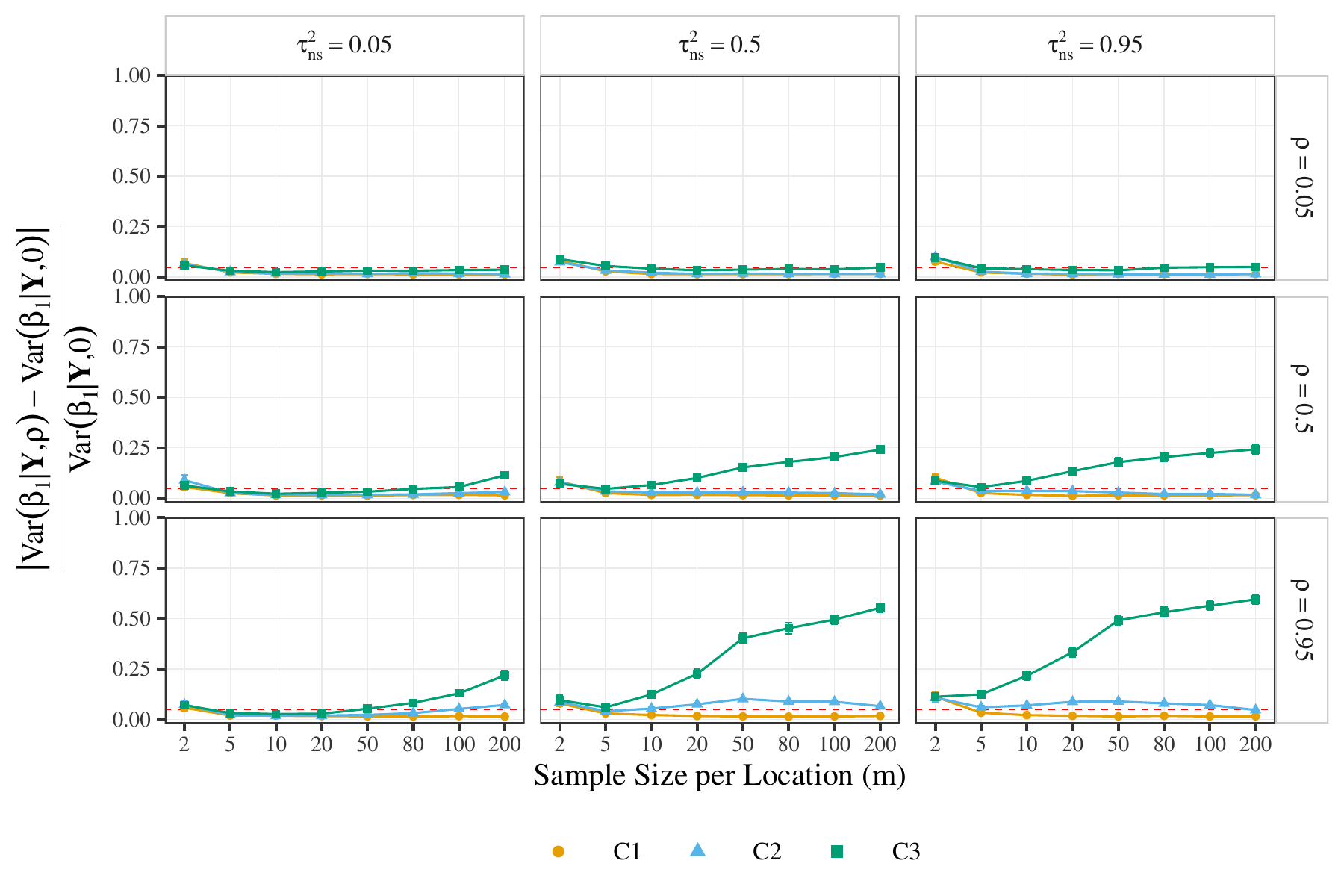}
\caption{Absolute relative difference in the marginal posterior variance of $\beta_1$ between the spatial and nonspatial Bernoulli models, averaged over $100$ datasets per setting, for $n=100$ spatial units. Curves represent covariate structures C1--C3. The dashed red line marks the tolerance $\gamma=0.05$. Error bars are pointwise $95\%$ Monte Carlo confidence intervals across datasets.}
\label{fig:bernoulli_sim}
\end{figure}

\clearpage
\begin{figure}[p]
\centering
\includegraphics[width=\linewidth]{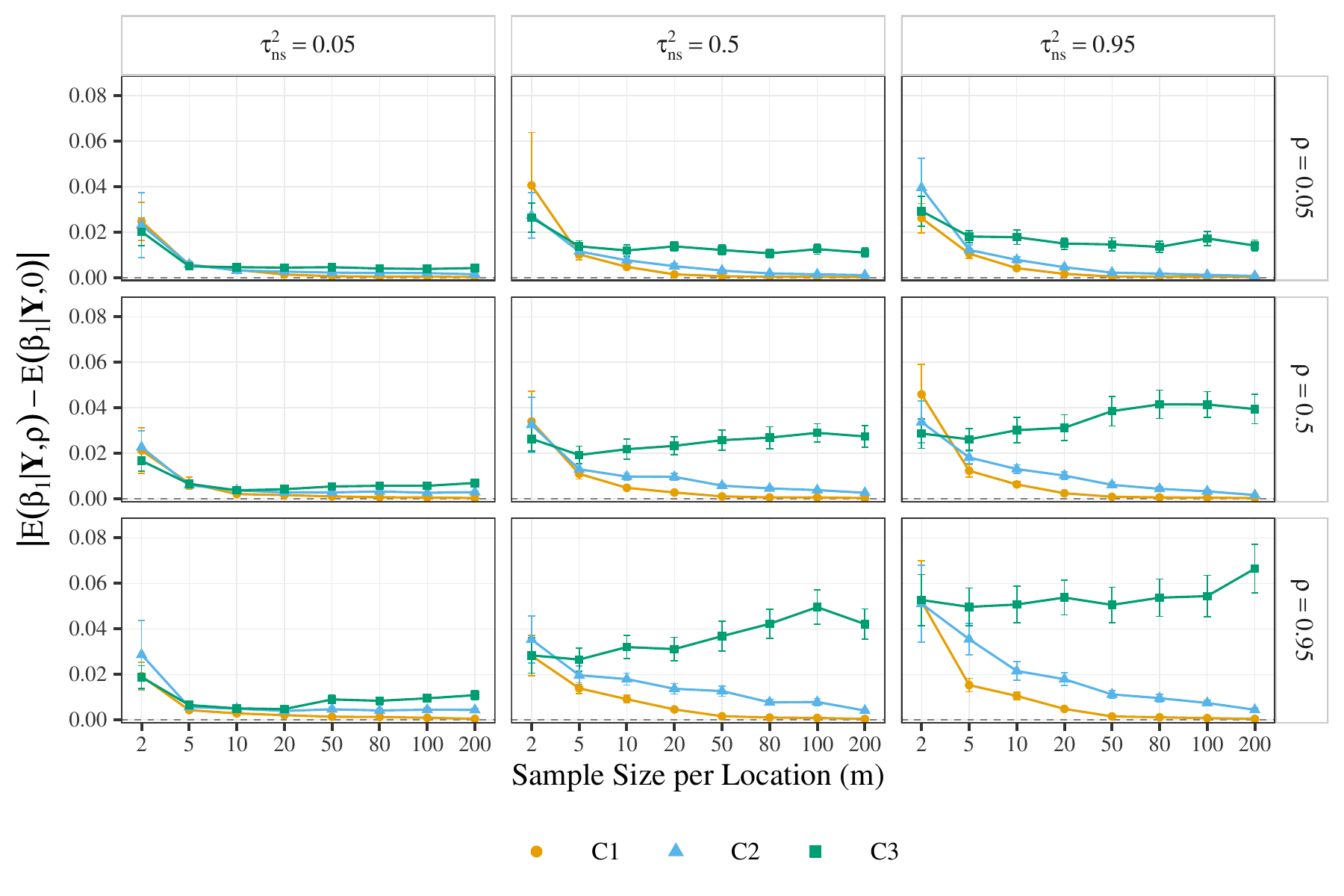}
\caption{Absolute difference in the marginal posterior mean of $\beta_1$ between the spatial and nonspatial Bernoulli models, averaged over $100$ datasets per setting, for $n=100$ spatial units. Curves represent covariate structures C1--C3. Error bars are pointwise $95\%$ Monte Carlo confidence intervals across datasets.}
\label{fig:bernoulli_mean}
\end{figure}

\clearpage
\subsection{Negative binomial outcomes}
\label{sec:nb_sim}

We generate conditionally independent counts according to
\begin{equation*}
Y_{ij}\mid\mu_{ij},r\sim\operatorname{NB}(\mu_{ij},r), \qquad \log\mu_{ij}=\beta_0+\beta_1\text{x}_{ij}+\theta_i,
\end{equation*}
where $\operatorname{E}(Y_{ij}\mid\mu_{ij},r)=\mu_{ij}$ and $\operatorname{Var}(Y_{ij}\mid\mu_{ij},r)=\mu_{ij}+\mu_{ij}^2/r$. We use the same values of $n$, $m$, and $\rho$ and the same covariate structures as in the Bernoulli simulations, with $(\tau^2_{ns},r)\in\{(0.05,1),(0.50,10),(0.95,100)\}$ and $100$ datasets per setting. The size parameter $r$ is fixed at its generating value when fitting both models.

Figures~\ref{fig:nb_sim} and~\ref{fig:nb_mean} show the variance and mean comparisons, respectively. Posterior mean differences generally decrease under C1 and C2, whereas C3 differences persist. Variance differences are small at larger $m$ under C1 and most C2 settings. Substantial variance differences persist under C3 in several settings.

\clearpage
\begin{figure}[p]
\centering
\includegraphics[width=\linewidth]{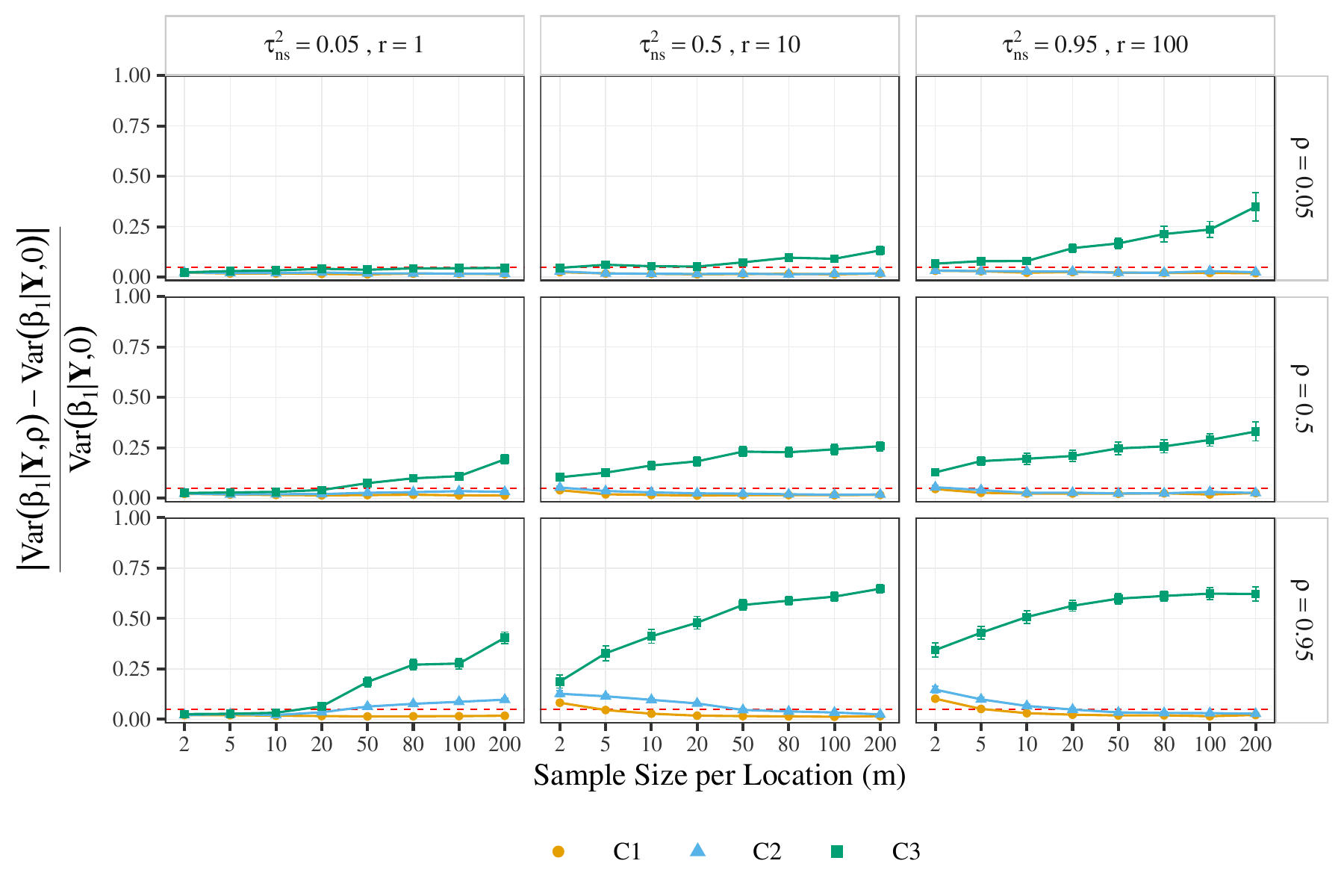}
\caption{Absolute relative difference in the marginal posterior variance of $\beta_1$ between the spatial and nonspatial negative binomial models, averaged over $100$ datasets per setting, for $n=100$ spatial units. Curves represent covariate structures C1--C3. The dashed red line marks the tolerance $\gamma=0.05$. Error bars are pointwise $95\%$ Monte Carlo confidence intervals across datasets.}
\label{fig:nb_sim}
\end{figure}

\clearpage
\begin{figure}[p]
\centering
\includegraphics[width=\linewidth]{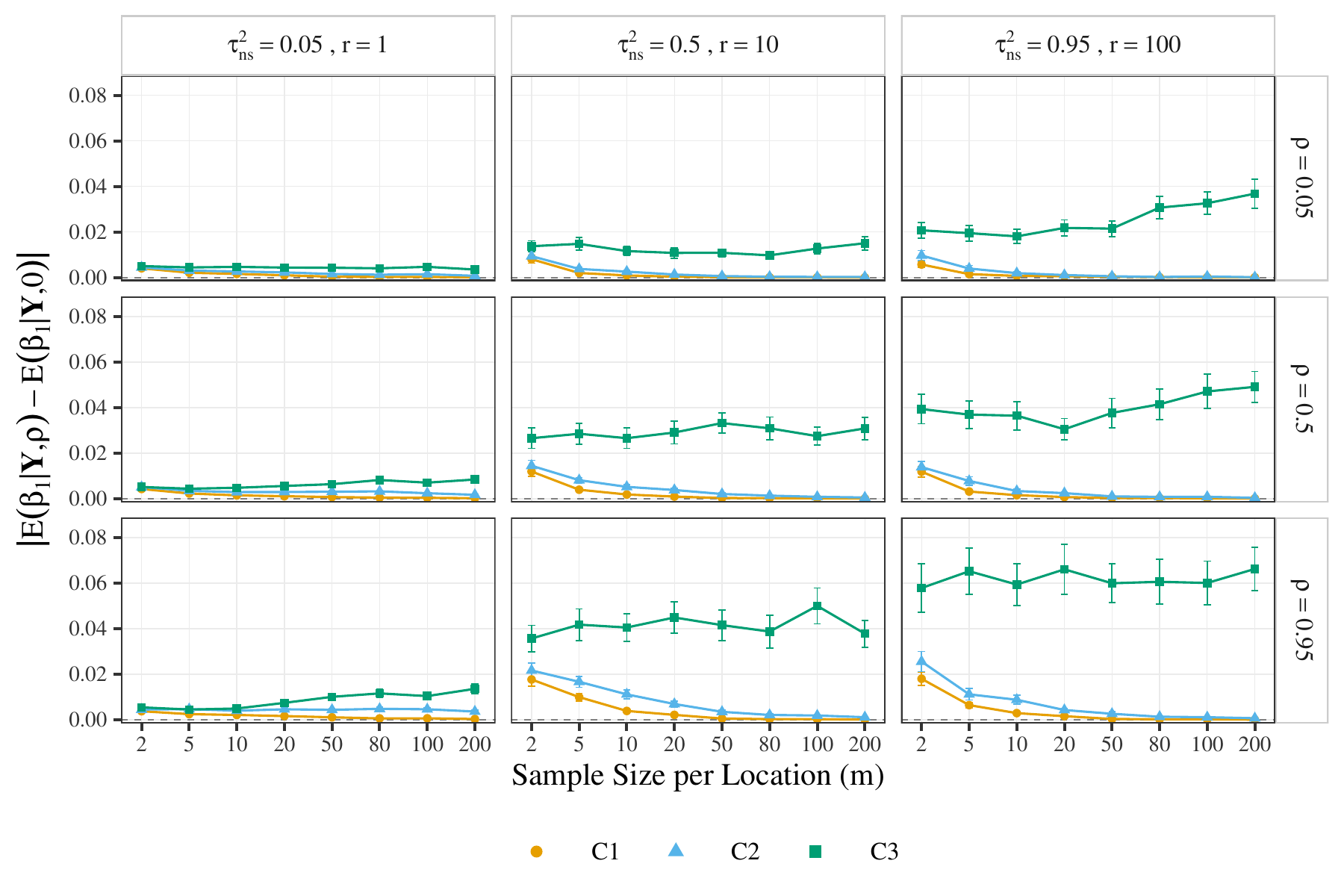}
\caption{Absolute difference in the marginal posterior mean of $\beta_1$ between the spatial and nonspatial negative binomial models, averaged over $100$ datasets per setting, for $n=100$ spatial units. Curves represent covariate structures C1--C3. Error bars are pointwise $95\%$ Monte Carlo confidence intervals across datasets.}
\label{fig:nb_mean}
\end{figure}

\clearpage
\section{Details of the two-stage procedure}
\label{sec:supp_practice}

\subsection{Covariance of the estimated random intercepts}
\label{sec:supp_working_covariance}
We derive the working covariance of the first-stage estimates and the likelihood used to estimate $\rho$ in Section~\ref{sec:practice_estimation}.
Let $N=\sum_{i=1}^{n}m_i$, $\text{X}=[\boldsymbol{1}_N,\textbf{x}]$, and $\boldsymbol{\beta}=(\beta_0,\beta_1)^{\top}$, with $\operatorname{rank}(\text{X})=2$. The incidence matrix satisfies $\text{Z}^{\top}\text{Z}=\operatorname{diag}(m_1,\ldots,m_n)$ and $\text{Z}\boldsymbol{1}_n=\boldsymbol{1}_N$. For this derivation, we treat the positive first-stage variance estimates $\widehat{\tau}_{ns}^2$ and $\widehat{\sigma}^2$ as fixed.

Under the nonspatial model,
\begin{gather*}
\boldsymbol{Y}=\text{X}\boldsymbol{\beta}+\text{Z}\boldsymbol{\theta}
+\boldsymbol{\varepsilon},\qquad
\boldsymbol{\theta}\sim\mathcal{N}_n(\boldsymbol{0}_n,\tau^2_{ns}I_n),\qquad
\boldsymbol{\varepsilon}\sim\mathcal{N}_N(\boldsymbol{0}_N,\sigma^2I_N),
\qquad \boldsymbol{\theta}\perp\boldsymbol{\varepsilon},\\
\operatorname{Cov}(\boldsymbol{\theta},\boldsymbol{Y})
=\tau^2_{ns}\text{Z}^{\top},\qquad
\operatorname{Var}(\boldsymbol{Y})
=\tau^2_{ns}\text{Z}\text{Z}^{\top}+\sigma^2I_N
=\Omega_{ns}.
\end{gather*}
Gaussian conditioning gives
\begin{equation*}
\mathrm{E}(\boldsymbol{\theta}\mid\boldsymbol{Y},
\boldsymbol{\beta},\sigma^2,\tau^2_{ns})
=\tau^2_{ns}\text{Z}^{\top}\Omega_{ns}^{-1}
(\boldsymbol{Y}-\text{X}\boldsymbol{\beta}).
\end{equation*}
Since
\begin{align*}
\text{Z}^{\top}\Omega_{ns}
&=\text{Z}^{\top}
(\tau^2_{ns}\text{Z}\text{Z}^{\top}+\sigma^2I_N)
=(\sigma^2I_n+\tau^2_{ns}\text{Z}^{\top}\text{Z})\text{Z}^{\top},\\
\text{Z}^{\top}\Omega_{ns}^{-1}
&=(\sigma^2I_n+\tau^2_{ns}\text{Z}^{\top}\text{Z})^{-1}
\text{Z}^{\top},
\end{align*}
the conditional mean for area $i$ is
\begin{align*}
\mathrm{E}(\theta_i\mid\boldsymbol{Y},
\boldsymbol{\beta},\sigma^2,\tau^2_{ns})
&=\frac{\tau^2_{ns}}{\sigma^2+m_i\tau^2_{ns}}
\sum_{j=1}^{m_i}(Y_{ij}-\beta_0-\beta_1\text{x}_{ij})\\
&=\frac{\tau^2_{ns}}{\tau^2_{ns}+\sigma^2/m_i}
(\bar{Y}_{i\cdot}-\beta_0-\beta_1\bar{\text{x}}_{i\cdot}).
\end{align*}
Substituting the first-stage estimates gives the random-intercept estimates returned by \texttt{lmer},
\begin{equation}
\widehat{\theta}_i
=\frac{\widehat{\tau}^2_{ns}}
{\widehat{\tau}^2_{ns}+\widehat{\sigma}^2/m_i}
(\bar{Y}_{i\cdot}-\widehat{\beta}_0
-\widehat{\beta}_1\bar{\text{x}}_{i\cdot}).
\label{eq:supp_intercept_prediction}
\end{equation}
To calculate the joint covariance of these estimates, we also need to account for the estimation of the regression coefficients.
Writing $\bar{\varepsilon}_{i\cdot}=m_i^{-1}\sum_{j=1}^{m_i}\varepsilon_{ij}$, the observation model and~\eqref{eq:supp_intercept_prediction} gives
\begin{align*}
\bar{Y}_{i\cdot}
&=\beta_0+\beta_1\bar{\text{x}}_{i\cdot}
+\theta_i+\bar{\varepsilon}_{i\cdot},\\
\widehat{\theta}_i
&=\frac{\widehat{\tau}^2_{ns}}
{\widehat{\tau}^2_{ns}+\widehat{\sigma}^2/m_i}
\left\{
\theta_i+\bar{\varepsilon}_{i\cdot}
-(\widehat{\beta}_0-\beta_0)
-(\widehat{\beta}_1-\beta_1)\bar{\text{x}}_{i\cdot}
\right\}.
\end{align*}
The first-stage generalized least-squares estimator expresses these estimation errors in terms of the same random intercepts and observation errors. With $\widehat{\Omega}_{ns}=\widehat{\tau}_{ns}^2\text{Z}\text{Z}^{\top}+\widehat{\sigma}^2I_N$,
\begin{equation}
\begin{aligned}
\widehat{\boldsymbol{\beta}}
&=(\text{X}^{\top}\widehat{\Omega}_{ns}^{-1}\text{X})^{-1}
\text{X}^{\top}\widehat{\Omega}_{ns}^{-1}\boldsymbol{Y},\\
\widehat{\boldsymbol{\beta}}-\boldsymbol{\beta}
&=(\text{X}^{\top}\widehat{\Omega}_{ns}^{-1}\text{X})^{-1}
\text{X}^{\top}\widehat{\Omega}_{ns}^{-1}
(\text{Z}\boldsymbol{\theta}+\boldsymbol{\varepsilon}).
\label{eq:supp_gls_estimates}
\end{aligned}
\end{equation}

Substituting the generalized least-squares estimator in~\eqref{eq:supp_gls_estimates} into the vector form of~\eqref{eq:supp_intercept_prediction} gives
\begin{equation}
\begin{aligned}
\widehat{\boldsymbol{\theta}}
&=\widehat{\tau}^2_{ns}\text{Z}^{\top}\widehat{\Omega}_{ns}^{-1}
(\boldsymbol{Y}-\text{X}\widehat{\boldsymbol{\beta}})\\
&=\widehat{\tau}^2_{ns}\text{Z}^{\top}
\left\{
\widehat{\Omega}_{ns}^{-1}
-\widehat{\Omega}_{ns}^{-1}\text{X}
(\text{X}^{\top}\widehat{\Omega}_{ns}^{-1}\text{X})^{-1}
\text{X}^{\top}\widehat{\Omega}_{ns}^{-1}
\right\}\boldsymbol{Y}\\
&=\widehat{\tau}^2_{ns}\text{Z}^{\top}
\left\{
\widehat{\Omega}_{ns}^{-1}
-\widehat{\Omega}_{ns}^{-1}\text{X}
(\text{X}^{\top}\widehat{\Omega}_{ns}^{-1}\text{X})^{-1}
\text{X}^{\top}\widehat{\Omega}_{ns}^{-1}
\right\}
(\text{Z}\boldsymbol{\theta}+\boldsymbol{\varepsilon}),
\label{eq:supp_prediction_linear}
\end{aligned}    
\end{equation}

where
\begin{equation*}
\left\{
\widehat{\Omega}_{ns}^{-1}
-\widehat{\Omega}_{ns}^{-1}\text{X}
(\text{X}^{\top}\widehat{\Omega}_{ns}^{-1}\text{X})^{-1}
\text{X}^{\top}\widehat{\Omega}_{ns}^{-1}
\right\}\text{X}
=\widehat{\Omega}_{ns}^{-1}\text{X}
-\widehat{\Omega}_{ns}^{-1}\text{X}=0.
\end{equation*}

We now evaluate the covariance of $\widehat{\boldsymbol{\theta}}$ in~\eqref{eq:supp_prediction_linear} under the spatial model. For each candidate $0\leq\rho<1$, the variance-matching relation determines $\widehat{\tau}_{sp}^2(\rho)$, giving the working model:
\begin{gather*}
\boldsymbol{\theta}\sim
\mathcal{N}_n\left(\boldsymbol{0}_n,
\widehat{\tau}^2_{sp}(\rho)Q(\rho)^{-1}\right),\qquad
\boldsymbol{\varepsilon}\sim
\mathcal{N}_N(\boldsymbol{0}_N,\widehat{\sigma}^2I_N),
\qquad \boldsymbol{\theta}\perp\boldsymbol{\varepsilon},\\
\widehat{\tau}^2_{sp}(\rho)
=\frac{n\widehat{\tau}^2_{ns}}
{\displaystyle\sum_{i=1}^{n}(\rho\lambda_i+1-\rho)^{-1}}.
\end{gather*}
Consequently,
\begin{equation*}
\mathrm{E}_{\mathrm{work}}
(\text{Z}\boldsymbol{\theta}+\boldsymbol{\varepsilon})
=\boldsymbol{0}_N,\qquad
\operatorname{Var}_{\mathrm{work}}
(\text{Z}\boldsymbol{\theta}+\boldsymbol{\varepsilon})
=\widehat{\tau}^2_{sp}(\rho)\text{Z}Q(\rho)^{-1}\text{Z}^{\top}
+\widehat{\sigma}^2I_N.
\end{equation*}
Since~\eqref{eq:supp_prediction_linear} expresses $\widehat{\boldsymbol{\theta}}$ as a linear function of $\text{Z}\boldsymbol{\theta}+\boldsymbol{\varepsilon}$, its Gaussian working distribution is
$\widehat{\boldsymbol{\theta}}\sim
\mathcal{N}_n(\boldsymbol{0}_n,\widehat{\Sigma}_{\theta}(\rho))$, where
\begin{equation}
\begin{aligned}
\widehat{\Sigma}_{\theta}(\rho)
&=(\widehat{\tau}^2_{ns})^2\text{Z}^{\top}
\left\{
\widehat{\Omega}_{ns}^{-1}
-\widehat{\Omega}_{ns}^{-1}\text{X}
(\text{X}^{\top}\widehat{\Omega}_{ns}^{-1}\text{X})^{-1}
\text{X}^{\top}\widehat{\Omega}_{ns}^{-1}
\right\}\\
&\quad\times
\left\{
\widehat{\tau}^2_{sp}(\rho)\text{Z}Q(\rho)^{-1}\text{Z}^{\top}
+\widehat{\sigma}^2I_N
\right\}\\
&\quad\times
\left\{
\widehat{\Omega}_{ns}^{-1}
-\widehat{\Omega}_{ns}^{-1}\text{X}
(\text{X}^{\top}\widehat{\Omega}_{ns}^{-1}\text{X})^{-1}
\text{X}^{\top}\widehat{\Omega}_{ns}^{-1}
\right\}\text{Z}.
\label{eq:supp_working_covariance}
\end{aligned}
\end{equation}
This covariance accounts for the estimation of $\beta_0$ and $\beta_1$ through their contribution to $\widehat{\boldsymbol{\theta}}$.

With the first-stage variance estimates held fixed, $\widehat{\boldsymbol{\beta}}$ minimizes $(\boldsymbol{Y}-\text{X}\boldsymbol{\beta})^{\top}\widehat{\Omega}_{ns}^{-1}(\boldsymbol{Y}-\text{X}\boldsymbol{\beta})$ and therefore satisfies
\[
\text{X}^{\top}\widehat{\Omega}_{ns}^{-1}
(\boldsymbol{Y}-\text{X}\widehat{\boldsymbol{\beta}})
=\boldsymbol{0}.
\]
Since $\boldsymbol{1}_N$ is a column of $\text{X}$ and $\text{Z}\boldsymbol{1}_n=\boldsymbol{1}_N$,
\begin{align*}
\boldsymbol{1}_n^{\top}\widehat{\boldsymbol{\theta}}
&=\widehat{\tau}_{ns}^2\boldsymbol{1}_n^{\top}\text{Z}^{\top}
\widehat{\Omega}_{ns}^{-1}
(\boldsymbol{Y}-\text{X}\widehat{\boldsymbol{\beta}})\\
&=\widehat{\tau}_{ns}^2\boldsymbol{1}_N^{\top}
\widehat{\Omega}_{ns}^{-1}
(\boldsymbol{Y}-\text{X}\widehat{\boldsymbol{\beta}})
=0.
\end{align*}
Thus, the estimates sum to zero and their covariance is singular. The working likelihood must account for this constraint.

If $\textbf{x}=\text{Z}\bar{\textbf{x}}$, the slope equation also gives
\begin{equation*}
\bar{\textbf{x}}^{\top}\widehat{\boldsymbol{\theta}}
=\widehat{\tau}^2_{ns}\textbf{x}^{\top}
\widehat{\Omega}_{ns}^{-1}
(\boldsymbol{Y}-\text{X}\widehat{\boldsymbol{\beta}})
=0.
\end{equation*}

The middle matrix in~\eqref{eq:supp_working_covariance} is positive definite for every $0\leq\rho<1$, while the outer factors are fixed. Hence
\begin{equation*}
\operatorname{null}\{\widehat{\Sigma}_{\theta}(\rho)\}
=\operatorname{null}\{\widehat{\Sigma}_{\theta}(0)\},
\end{equation*}
so the rank and supporting subspace do not depend on $\rho$. The Gaussian likelihood on this fixed subspace is
\begin{equation}
L(\rho;\widehat{\boldsymbol{\theta}})
\propto
\operatorname{pdet}\{\widehat{\Sigma}_{\theta}(\rho)\}^{-1/2}
\exp\left[
-\frac12\widehat{\boldsymbol{\theta}}^{\top}
\{\widehat{\Sigma}_{\theta}(\rho)\}^{+}
\widehat{\boldsymbol{\theta}}
\right].
\label{eq:supp_working_likelihood}
\end{equation}
Here, $\operatorname{pdet}$ is the product of the positive eigenvalues, and the superscript $+$ replaces those eigenvalues by their reciprocals, leaving zero eigenvalues and eigenvectors unchanged. The omitted factor is independent of $\rho$.

Maximizing~\eqref{eq:supp_working_likelihood} is equivalent to minimizing
\begin{equation*}
\log\operatorname{pdet}\{\widehat{\Sigma}_{\theta}(\rho)\}
+\widehat{\boldsymbol{\theta}}^{\top}
\{\widehat{\Sigma}_{\theta}(\rho)\}^{+}
\widehat{\boldsymbol{\theta}}.
\end{equation*}
If $\widehat{\tau}^2_{ns}=0$, then $\widehat{\boldsymbol{\theta}}=\boldsymbol{0}_n$ and $\widehat{\tau}^2_{sp}(\rho)=0$ for every candidate $\rho$, so the estimates provide no information for estimating $\rho$.

\subsection{Numerical implementation and boundary estimates}
\label{sec:supp_boundary}

We estimate $\rho$ by numerically minimizing the negative Gaussian working log likelihood, with the upper limit excluding the singular endpoint $\rho=1$.

In the simulation evaluation, positive first-stage estimates satisfying $\sqrt{\widehat{\tau}^2_{ns}/\widehat{\sigma}^2}<10^{-4}$ were screened for numerical boundary solutions. For each candidate, we compared the profiled restricted maximum likelihood deviance at $\tau^2_{ns}=0$ with its values at the interior stationary points. We set $\widehat{\tau}^2_{ns}=0$ only when this comparison verified the boundary minimum to a deviance tolerance of $10^{-7}$, and recomputed $\widehat{\sigma}^2$ at that boundary fit.

Of the $7{,}200$ datasets, $73$ first-stage fits returned an exactly zero variance estimate, and $58$ additional fits were verified as numerical boundary solutions. For these fits, $\rho$ was recorded as unidentified. Variance matching then gives $\widehat{\tau}^2_{sp}(\rho)=0$, so the two plug-in covariance matrices coincide and the estimated exact threshold is $2$. All $7{,}200$ datasets were retained.

\subsection{Simulation evaluation of the estimated thresholds}
\label{sec:supp_practice_simulation}

We evaluate the procedure using the Gaussian simulations with $n=100$ locations and covariate setting C2. The settings comprise $\rho\in\{0.05,0.50,0.95\}$, $\kappa\in\{0.05/0.95,\,0.50/0.50,\,0.95/0.05\}$, and preliminary within-area sample sizes $m_{\mathrm{fit}}\in\{2,5,10,20,50,80,100,200\}$, with $100$ datasets per setting. For each dataset, we calculate the exact threshold at tolerance $\gamma=0.05$ using either the generating covariance parameters or the two-stage estimates, retaining the same spatial graph and covariate summaries in both calculations.

Figure~\ref{fig:estimated_thresholds} shows the mean thresholds and empirical $2.5$th--$97.5$th percentile ranges across datasets. The means include all estimates, including extreme values. The percentile ranges describe variation across datasets.

Mean estimated thresholds closely follow those calculated using the generating parameters in most settings. For $\kappa=0.95/0.05$, the two means differ by at most $0.23$ across the evaluated settings, including at $m_{\mathrm{fit}}=2$. The largest discrepancies occur at $\kappa=0.05/0.95$ with limited preliminary replication, where a small number of large estimates inflate the means. Here, $m_{\mathrm{fit}}\kappa=\tau^2_{ns}/(\sigma^2/m_{\mathrm{fit}})$ measures the random-intercept variance relative to the observation noise in the area means.

\begin{figure}[htbp]
\centering
\includegraphics[width=\linewidth]{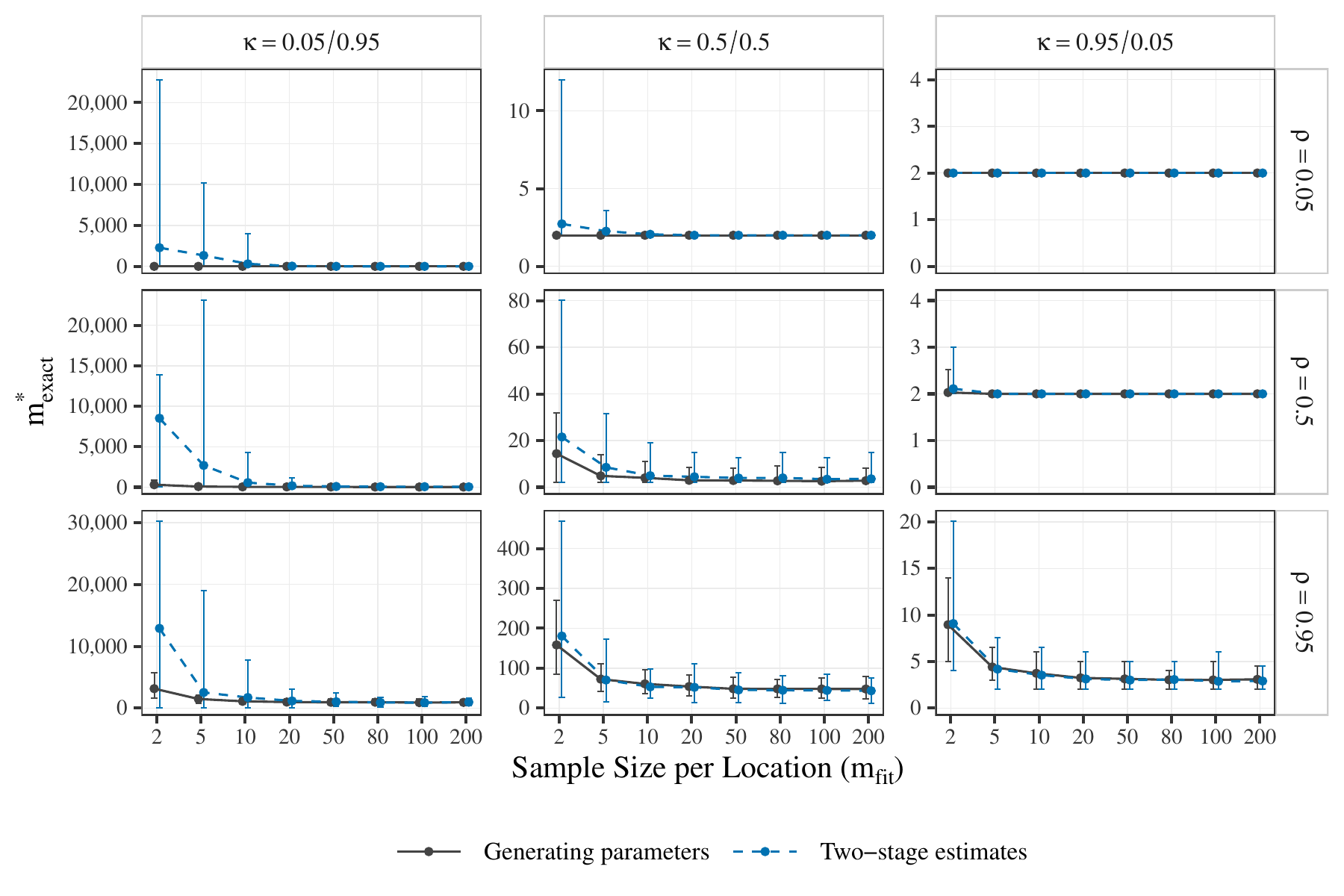}
\caption{Exact replication thresholds calculated using the generating covariance parameters and the two-stage estimates for $n=100$ locations under C2, with tolerance $\gamma=0.05$. The horizontal axis gives the preliminary within-area sample size $m_{\mathrm{fit}}$. Curves show means over $100$ datasets per setting, and bars show the empirical $2.5$th and $97.5$th percentiles. Rows correspond to $\rho\in\{0.05,0.50,0.95\}$ and columns to $\kappa\in\{0.05/0.95,\,0.50/0.50,\,0.95/0.05\}$. All datasets and extreme threshold estimates are retained.}
\label{fig:estimated_thresholds}
\end{figure}

\clearpage 
\section{Spatial graphs and random-effect realizations}\label{sec:supp_graph}

In the simulation study, each replicate uses a randomly generated connected graph as the adjacency structure. Table~\ref{tab:graph_summary} summarizes the degree distribution (i.e., the number of neighbors per spatial unit) across all $100$ graphs generated for each value of $n$. The median degree is 2 across all settings, with the range of degrees increasing with $n$. Figure~\ref{fig:random_graphs} displays two example graphs for each value of $n$. Figure~\ref{fig:sim_theta} shows example realizations of the spatial random effects on a $10\times10$ grid, illustrating how $\rho$ and $\kappa$ affect their spatial pattern and magnitude. Increasing $\rho$ produces smoother spatial patterns, while increasing $\kappa$ increases their magnitude.

\clearpage 

\begin{table}[htbp]
\centering
\caption{Summary of the number of neighbors per spatial unit across $100$ randomly generated connected graphs for each $n$}
\begin{tabular}{lcccc}
$n$ & Mean & Median & Min & Max \\
25  & 1.920 & 2 & 1 & 7  \\
100 & 1.980 & 2 & 1 & 11 \\
400 & 1.995 & 2 & 1 & 13 \\
\end{tabular}
\label{tab:graph_summary}
\end{table}
\clearpage 

\begin{figure}[p]
\centering
\includegraphics[ width=\textwidth, height=0.90\textheight, keepaspectratio ]{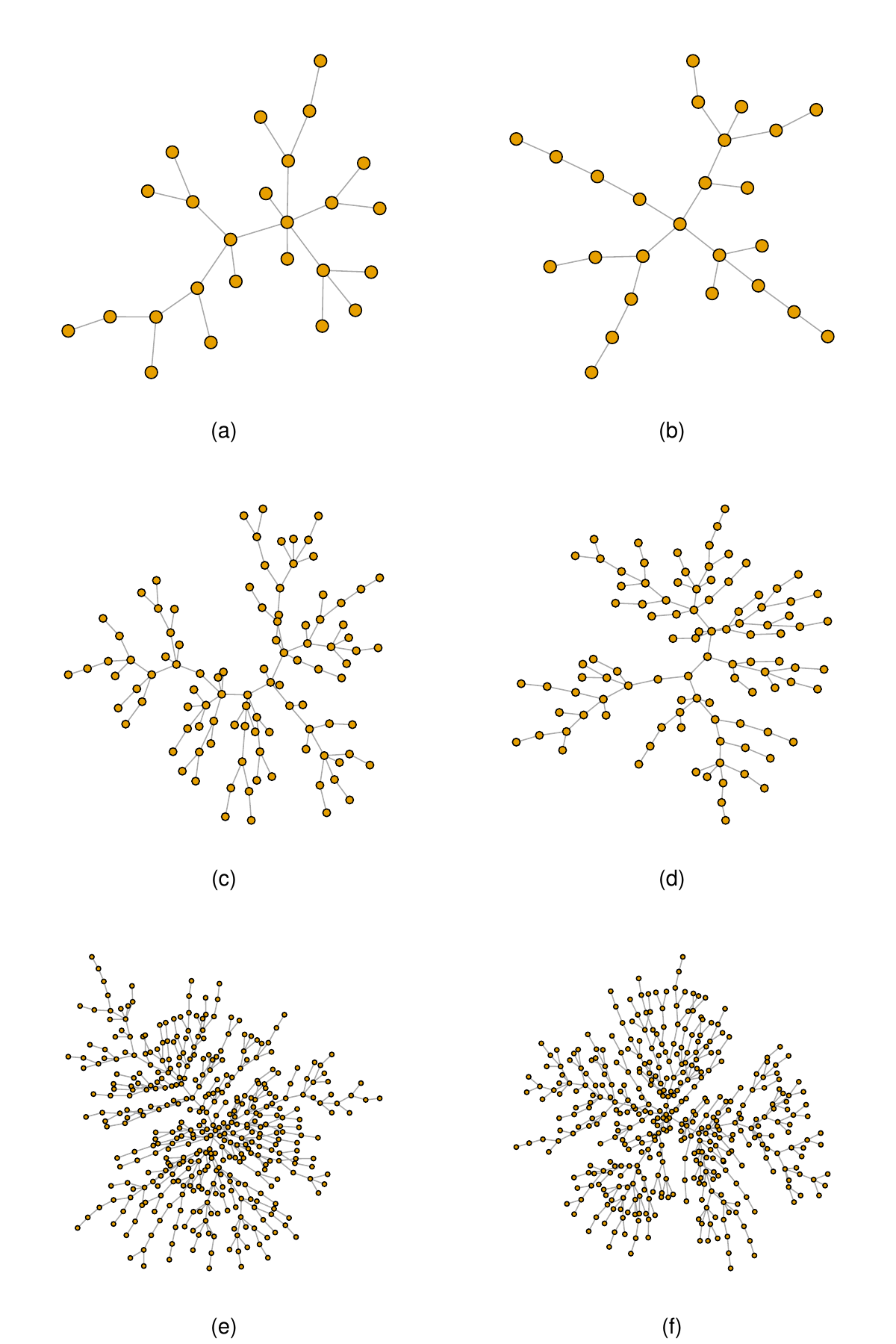}
\caption{Two randomly generated connected graphs used in the simulation study for (a)-(b) $n = 25$, (c)-(d) $n = 100$, and (e)-(f) $n = 400$ spatial units.}
\label{fig:random_graphs}
\end{figure}

\clearpage

\begin{figure}[htbp]
\centering
\includegraphics[width=\linewidth]{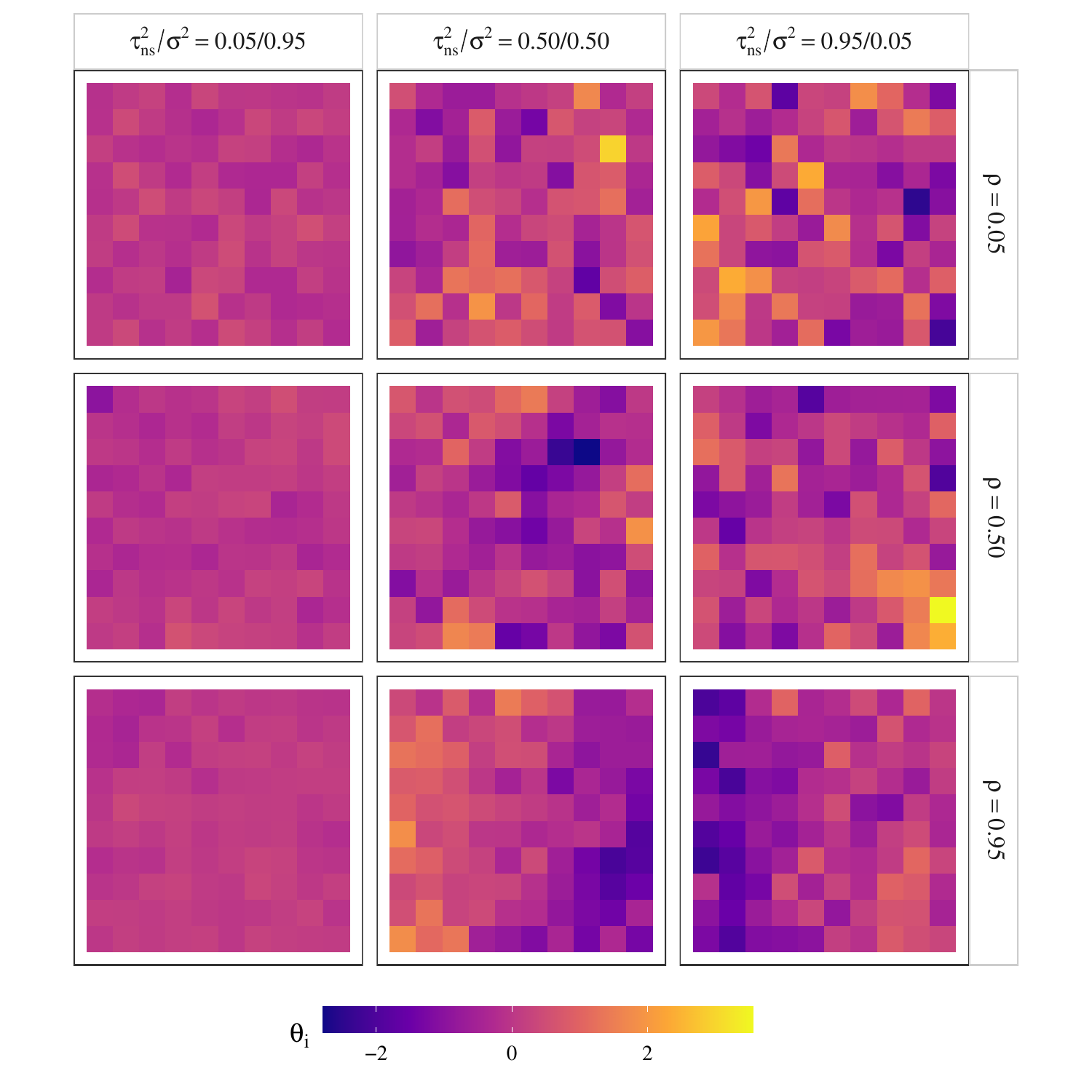}
\caption{Example realizations of the spatial random effects $\boldsymbol{\theta}$ across different levels of spatial correlation $\rho$ and variance ratio $\kappa=\tau_{ns}^2/\sigma^2$, generated on a $10\times10$ grid with neighbors defined by shared borders or vertices. The grid is used for illustration; the simulation study uses the random connected graphs summarized in Table~\ref{tab:graph_summary}.}
\label{fig:sim_theta}
\end{figure}
\end{APPENDICES}

\end{document}